\documentclass[twocolumn]{aastex701}

\definecolor{amaranth}{rgb}{0.9, 0.17, 0.31}
\newcommand{\teff}{T_{\rm eff}}
\newcommand{\logg}{{\rm log}(g)}

\definecolor{britishracinggreen}{rgb}{0.0, 0.26, 0.15}
\definecolor{caribbeangreen}{rgb}{0.0, 0.8, 0.6}
\newcommand{\bacchus}{\texttt{BACCHUS}}

\usepackage{xcolor}

\usepackage{float}
\usepackage{graphicx}
\usepackage{amsmath}
\usepackage{enumitem} 
\shorttitle{FGK Hosts of Wide-Orbit Brown Dwarfs}
\received{April 2, 2026}
\revised{July 6, 2026}
\accepted{July 8, 2026}
\begin{document}

\title{Benchmark Brown Dwarf Systems I: Chemical Abundance Analysis of FGK Stars with Wide-Separation Brown Dwarf Companions Using PEPSI}

\author[orcid=0000-0001-5610-5328,sname='CLP']{Caprice L. Phillips}
\altaffiliation{NASA Sagan Fellow}
\affiliation{Department of Astronomy \& Astrophysics, University of California, Santa Cruz, CA 95064, USA}
\email[show]{cphillips@ucsc.edu}  
\correspondingauthor{Caprice L. Phillips}

\author[0000-0001-9907-7742]{Megan Bedell}
\affiliation{Center for Computational Astrophysics, Flatiron Institute, 162 Fifth Avenue, New York, NY 10010, USA}
\email[]{mbedell@flatironinstitute.org}

\author[0000-0002-0900-6076]{Catherine Manea}
\affiliation{Department of Physics \& Astronomy, University of Utah, Salt Lake City, UT 84112, USA}
\email[]{catherinemanea@gmail.com}

\author[0000-0002-4531-6899]{Alison Duck}
\affil{Department of Astronomy, The Ohio State University, 100 W 18th Ave, Columbus, OH 43210 USA}
\email[]{Aduck@jpl.nasa.gov}

\author[0000-0002-8823-8237]{Anusha Pai Asnodkar}
\affil{Department of Astronomy, California Institute of Technology, Pasadena, CA 91125, USA}
\email[]{apaiasno@caltech.edu}

\author[0000-0001-9345-9977]{Emily J. Griffith}
\altaffiliation{NASA Hubble Fellow}
\affil{Center for Astrophysics and Space Astronomy, Department of Astrophysical and Planetary Sciences,
University of Colorado, 389 UCB, Boulder, CO 80309-0389, USA}
\email[]{Emily.Griffith-1@colorado.edu}

%--co-I PEPSI
\author[0000-0003-4083-9962]{Austin 
Rothermich}
\affiliation{Department of Astrophysics, American Museum of Natural History, Central Park West at 79th Street, NY 10024, USA}
\affil{Department of Physics, Graduate Center, City University of New York, 365 5th Avenue, New York, NY 10016, USA}
\affil{Department of Physics and Astronomy, Hunter College, City University of New York, 695 Park Avenue, New York, NY 10065, USA}
\email[]{apaiasno@caltech.edu}

\author[0000-0002-2682-0790]{Emily Calamari}
\affil{Department of Astrophysics, American Museum of Natural History, New York, NY 10024, USA}
\email[]{ecalamari@amnh.org}

\author[0000-0001-6251-0573]{Jacqueline K. Faherty}
\affiliation{Department of Astrophysics, American Museum of Natural History, New York, NY 10024, USA}
\email[]{jfahery@amnh.org}

\author[0000-0003-4600-5627]{Ben Burningham}
\affil{Centre for Astrophysics Research, School of Physics, Astronomy and Mathematics, University of Hertfordshire, Hatfield AL10 9AB}
\email[]{b.burningham@herts.ac.uk}

\author[0000-0001-6627-6067]{Channon Visscher}
\affiliation{Chemistry \& Planetary Sciences, Dordt University, Sioux Center IA 51250}
\affil{Center for Exoplanetary Systems, Space Science Institute, Boulder, CO 80301}
\email[]{channon.visscher@dordt.edu}

\author[0000-0003-4636-6676]{Eileen C. Gonzales}  
\affiliation{Department of Physics \& Astronomy, San Francisco State University, San Francisco, CA, 94132, USA }
\email[]{egonzales@sfsu.edu}
\author[0000-0002-0551-046X]{Ilya Ilyin}
\affil{Leibniz-Institute for Astrophysics Potsdam (AIP), An der Sternwarte 16, D-14482 Potsdam, Germany}
\email[]{lyin@aip.de}

\author[0000-0002-6192-6494]{Klaus Strassmeier}
\affil{Leibniz-Institute for Astrophysics Potsdam (AIP), An der Sternwarte 16, D-14482 Potsdam, Germany}
\email[]{kstrassmeier@aip.de}

\author[0000-0002-4361-8885]{Ji Wang}
\affiliation{Department of Astronomy, The Ohio State University, 100 W 18th Ave, Columbus, OH 43210 USA}
\email[]{wang.12220@osu.edu}

%% Use the \collaboration command to identify collaborations. This command
%% takes an optional argument that is either a number or the word "all"
%% which tells the compiler how many of the authors above the command to
%% show. For example "\collaboration[all]{(DELVE Collaboration)}" wil include
%% all the authors above this command.
%%
%% Mark off the abstract in the ``abstract'' environment. 
\begin{abstract}
We present results from a spectroscopic survey of 32 FGK stars hosting brown dwarfs, using high-resolution optical spectra (R = 130,000 and 50,000) obtained with the PEPSI spectrograph on the Large Binocular Telescope. The primary goal of this survey is to determine precise stellar parameters and abundances for 11 elements (C, O, Mg, Si, Ca, Al, Ti, Fe, Y, S, and N) in these systems.
We employ spectral synthesis within the \texttt{BACCHUS} framework to derive precise stellar properties and elemental abundance ratios. For our average S/N $>$ 200 data, we achieve a typical error of 42 K in $\teff$ and $\sim$0.03 dex for [Fe/H]. We observe a significant dispersion from a solar C/O ratio among the sample of brown dwarf host stars that host primarily wide-orbit brown dwarfs. Using established theoretical chemical frameworks, we discuss the implications of the observed Mg/Si and Ca/Al ratios for cloud properties in the brown dwarf companions.  Finally, we evaluate the applicability of the [Y/Mg] stellar clock for our sample and discuss the broader implications of our results. This work provides a timely and uniform abundance analysis of host stars, supporting extended wavelength brown dwarf observations in the era of JWST.

\end{abstract}

%% Keywords should appear after the \end{abstract} command. 
%% The AAS Journals now uses Unified Astronomy Thesaurus (UAT) concepts:
%% https://astrothesaurus.org
%% You will be asked to selected these concepts during the submission process
%% but this old "keyword" functionality is maintained in case authors want
%% to include these concepts in their preprints.
%%
%% You can use the \uat command to link your UAT concepts back its source.
% \keywords{\uat{Galaxies}{573} {\textemdash} \uat{Cosmology}{343} {\textemdash} \uat{High Energy astrophysics}{739} {\textemdash} \uat{Interstellar medium}{847} {\textemdash} \uat{Stellar astronomy}{1583} {\textemdash} \uat{Solar physics}{1476}}


%% From the front matter, we move on to the body of the paper.
%% Sections are demarcated by \section and \subsection, respectively.
%% Observe the use of the LaTeX \label
%% command after the \subsection to give a symbolic KEY to the
%% subsection for cross-referencing in a \ref command.
%% You can use LaTeX's \ref and \label commands to keep track of
%% cross-references to sections, equations, tables, and figures.
%% That way, if you change the order of any elements, LaTeX will
%% automatically renumber them.
\section{Introduction} 
\label{sec:intro}
Brown dwarfs are a category of astronomical objects with initial mass ($\leq$0.072 $M_{\sun}$) insufficient to maintain stable hydrogen fusion. As a result, these objects invariably contract and cool throughout their lifetimes \citep{Chabrier1997} and progress through their spectral classification sequence (M, L, T, and Y)\citep{Kirkpatrick2005}. This cooling provides a challenge in understanding formation history and evolution as it results in a mass-temperature-age degeneracy~\citep{Canty2013,Crepp2016}. One path forward to break this degeneracy is through the study of benchmark systems. Benchmark brown dwarfs can include systems in which properties such as mass, age, and chemistry may be determined independently and precisely~\citep{Zhang2025_bencharm,Rickman2024,Brandt_2021,Phillips2024,Allers2016,Rothermich2024}. Age-benchmarked brown dwarfs include those that are members of young moving groups, associations, or stellar clusters, where the co-eval members have known ages from age-dating mechanisms that can include isochrone fitting, kinematic traceback, and chromospheric activity or X-ray emission~\citep{Oretega2007, Gagne_2024,Lee2024}. Dynamical-mass provides the most direct and model-independent measurement of the mass of brown dwarfs~\citep{Brandt_2021,Franson2023,Dupuy2014}. As a result, accurate dynamical masses enable stringent tests of substellar evolutionary and atmospheric models, which can improve our understanding of brown dwarf formation and evolution~\citep{Maire2024,Franson2022}.

\par
In recent works, \cite{Calamari_2024} focused on compositional benchmarks, defined as ``benchmark brown dwarfs belonging to a binary or multiple system with the specification that the primary star in the system is of spectral type F, G, or K". By studying a sample of solar-type (FGK) stars hosting a brown dwarf companion, we can place system constraints on atomic and molecular abundances in order to inform our subsequent companion studies. \citet{Calamari_2024} established the underlying theoretical framework and derived the chemical relationships linking stellar and substellar compositions, providing a foundation for interpreting benchmark systems. FGK-type stars are ideal candidates for detailed compositional analysis, allowing for more accurate determination of elemental abundances, such as carbon, nitrogen, oxygen, magnesium, silicon, and other heavy elements, which will be valuable for future studies. Although such benchmark systems provide valuable constraints on brown dwarf compositions, they represent a relatively small and specialized subset of the overall brown dwarf population, as most brown dwarfs are typically found in isolation~\citep{Fontantive2023,Faherty2016}. Compositional benchmarks can include close companions and widely-separated companions to well-characterized stars. In this work, we focus on wide-companion systems, with projected separations $>$100 AU. Brown dwarf companions at wide separations are rare, with the frequency of wide-orbit systems estimated to be $\sim$1.3$\%$~\citep{Feng2022, Grieves2017, Kiefer2019}. Similarly, the detection frequency of brown dwarfs around solar-like stars is low, at 2.0 $\pm$ 0.5$\%$~\citep{Kiefer2019}. Given their rarity yet valuable power, it is essential to extract as much information as possible from these systems, including atmospheric composition, temperature, surface gravity, cloud properties, while also leveraging the host stars  abundances as a reference point for interpreting substellar compositions. 

\par

Stellar abundances are a quintessential component of our understanding of the planet and brown dwarf formation process and mechanisms~\citep[and references therein]{Teske_2024_Review}. While stellar compositions cannot be mapped directly into the bulk chemistry of planets, they establish the initial chemical reservoir from which planets form. In the substellar (brown dwarf) regime, stellar abundances more directly trace the formation environment of brown dwarfs, offering key constraints on their origins. In this work, we assume that brown dwarf companions retain the bulk elemental abundance patterns of their host stars, an assumption we revisit in Section \ref{sec:test_MgSi}. In many studies of brown dwarf and exoplanet atmospheres, host star abundances are often assumed to be solar for simplicity; however, this assumption can introduce significant biases in the interpretation of atmospheric retrievals and forward models~\citep{Baghel2025,Xuan_2024,Meynardie2025,Baburaj2025b}. This is particularly evident in the directly-imaged planets and planetary-mass companions such as $\beta$ Pictoris b and Ross 458C, where inferred formation interpretations are sensitive to the assumed composition of the host stars~\citep{Meynardie2025,Landman2024,Reggiani2024}. Stellar compositions can vary substantially across the Galactic population~\citep{Cabral2023,Teixeira2025}, with differences in metallicity and elemental ratios (e.g., C/O, Mg/Si, [Fe/H]), directly influencing the chemistry and cloud formation of their substellar companions, as explored in \citet{Calamari_2024}. A well-characterized host star thus serves as a compositional anchor by the following: enabling a more precise linkage between stellar and substellar chemistry, reducing degeneracies in retrieval analyses,  and ensuring interpretation of formation pathways is physically meaningful~\citep{Kolecki2022,daSilva2024,Polanski2022,Sun2025}.

\par
We present uniform abundance measurements for 11 elements for 32 stars with known brown dwarf companions. We provide diagnostic uses of the C/O, Mg/Si, Ca/Al, [Y/Mg], and S/N ratios for useful information regarding formation, age, metallicity for this critical sample. In Section \ref{sec:data}, we describe the observing campaign for our data. In Section \ref{sec:methods}, we describe our methodology and analysis technique using the~\bacchus~framework. In Section \ref{sec:validating_results} we validate our results with the Sun and comparison to literature values.  We provide discussions and implications for our work in Section \ref{sec:discussion} and \ref{sec:age_YtoMg}, and conclude in Section \ref{sec:conclusion}.

%{\textemdash}--Section Data
\section{Data}  \label{sec:data}
\subsection{Sample Selection}

We collected 32 high-resolution spectra of FGK hosts of primarily wide-orbit brown dwarfs~($>$ 100 AU) -- with the exception of HD 89744 (63 AU) and BD+01 299~(35 $-$ 45 AU). These wide-orbit brown dwarfs can be observed without the interference of light from their host stars, that affects close-in orbits. 
The host stars for the companion systems are known nearby stars and some of our targets have been previously observed as part of exoplanet and stellar observing campaigns~(e.g. SPFOT, Luck, SONS surveys; \citealt{Stonkute_2020, Luck_2018, Holland2017}).  Abundances for these stars have not necessarily been investigated or reported in a uniform catalog or dataset. We highlight the importance of a uniform analysis for our sample, as \cite{Hinkel_2016} found that choice of data sets, methodology, and analysis can lead to discrepancies between stellar parameters and abundances. We first compile our stellar sample from a previous literature collection of compositional benchmark brown dwarfs from \cite{Calamari_2024}. We prioritize stars that have brown dwarf targets observed as part of separate JWST programs: GO \#3670 (PI: Burningham; \cite{SinkingSilicates}, GO \#8140 (PI: Zhang; \cite{Zhang2025}. Our sample also consists of newly discovered brown dwarf companions in the solar neighborhood, identified through the citizen science project, Backyard Worlds: Planet 9~\citep{Rothermich2024}. An overview of the systems selected for this study is presented in Table \ref{tab:obs} and \ref{tab:sample}

\par

\subsection{Observations} 
\label{sec:observations}
We collect spectra from the Potsdam Echelle Polarimetric and Spectroscopic Instrument (PEPSI; \citealt{Strassmeier2015}) on the Large Binocular Telescope (LBT) between 2023 February 08 and 2024 January 16. PEPSI is a high-resolution echelle spectrograph with two arms (red and blue) that cover the wavelength range of 383 -- 914 nm across six different cross dispersers (CD). The primary CDs used in this work are the CD III~(4800 $-$ 5441 $\AA$) and CD VI~(7419 $-$ 9140 $\AA$). These CDs were chosen as they house many key lines, including the \ion{O}{1} triplet needed to calculate the [O/H] abundance and thus the C/O ratio. Most of our observations were taken with the 200 $\mu$m fiber yielding a two-pixel sampled resolution of R = 130,000. In our initial observation campaign, five systems~(HD 46588, HD 126054, GJ 417, HD 116012, BD+60 1417), were observed with the 300 $\mu$m fiber, resulting in a resolution of R = 50,000 instead of R = 130,000. We discuss how this difference in resolution affects our results in Section \ref{sec:abundance_determination}. Full observation details are provided in Table \ref{tab:obs}.

\begin{figure}[H]
\centering
\includegraphics[width=1.05\linewidth]{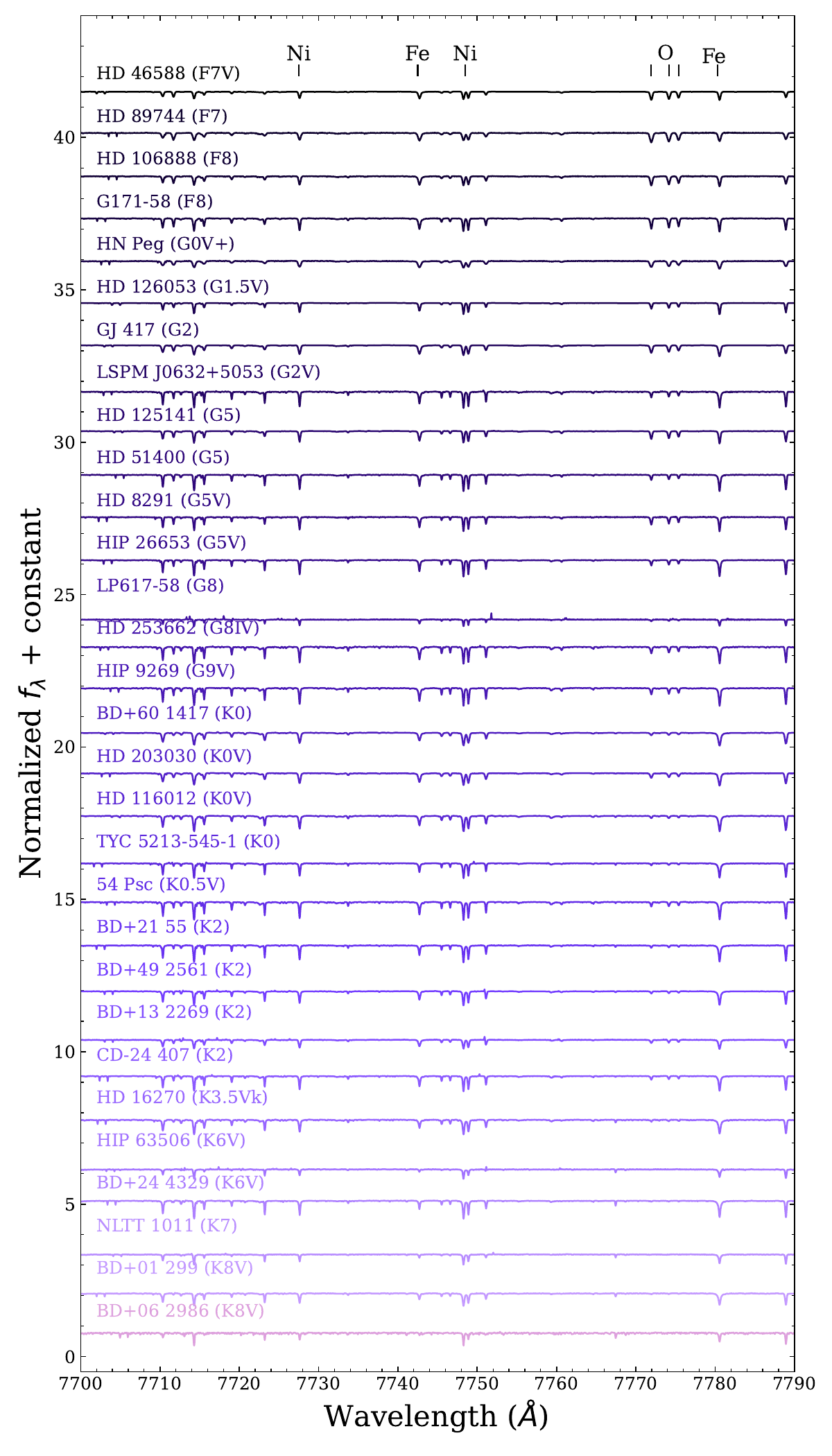}
    \caption{Subsets of our sample spectra in order of spectral class, from dark purple to light purple. Prominent absorption lines from oxygen triplet and iron (Fe) lines are labeled. Some nickel (Ni) lines are also labeled but are not used in this analysis.}
    \label{fig:pepsi_spectra}
\end{figure}

\par
We employ the spectroscopic data systems for PEPSI pipeline (SDS4PEPSI) to reduce the spectra of our targets, as described in \citet{Strassmeier2018}. In short, this pipeline applies bias and flat-field corrections and estimates the photon-noise contribution to the spectra. It removes contamination from scattered light as well as cosmic rays. The pipeline conducts flux extraction and wavelength calibrates the spectra against Th-Ar lines, as well as corrects for the echelle's blaze function, fringing, and vignetting effects. Finally, the pipeline corrects for the star's radial velocity shifts to bring the spectra to rest-frame wavelengths and applies a continuum normalization by fitting a 2D smoothing spline in the cross-dispersion and dispersion directions. While the pipeline's default continuum normalization removes much of the continuum variation, we found that our abundance measurements were sensitive to small continuum deviations that were not fully corrected by the pipeline. To fix this, we perform a second step of continuum normalization with a low-order polynomial fit individually to each of the two dispersers. Specifically, we perform iterative one-sided sigma-clipping to select continuum pixels in each wavelength region, then fit a third-order polynomial (for CD III) or a fifth-order polynomial (for CD VI) to these continuum pixels and divide the data by the best-fit polynomial to continuum normalize. A portion of the normalized spectra for our sample is shown in Figure \ref{fig:pepsi_spectra}.

%{\textemdash}-observation details
\begin{deluxetable*}{lrrrrrrr}
\tablecaption{Observing Details
\label{tab:obs}}
\tablehead{
\colhead{Object} & \colhead{V} & \colhead{Date} & 
\colhead{Exp (s)} & 
\colhead{SNR$_{\text{CD II}}$} & 
\colhead{SNR$_{\text{CD III}}$} &  
\colhead{SNR$_{\text{CD VI}}$} & 
\colhead{Resolution (R)}}
\startdata
HD 89744 & 5.72 & 2023-11-03 & 15.0 & {\textemdash} & 249 & 285 & 130,000 \\
HD 46588 & 5.41 & 2023-02-08 & 50.0 & {\textemdash} & 657 & 889 & 50,000 \\
$*$54 Psc & 5.87 & 2023-11-02 & 34.0 & {\textemdash} & 406 & 532 & 130,000 \\
V$^{*}$ HN Peg & 5.95 & 2021-10-22 & 50.0 & {\textemdash} & 375 & 298 & 130,000 \\
HD 126053 & 6.24 & 2023-02-08 & 70.0 & {\textemdash} & 467 & 664 & 50,000 \\
GJ 417 & 6.40 & 2023-02-08 & 120.0 & {\textemdash} & 887 & 1149 & 50,000 \\
HIP 9269 & 7.14 & 2023-10-22 & 170.0 & {\textemdash} & 533 & 519 & 130,000 \\
G171-58 & 7.55 & 2024-01-16 & 150.0 & {\textemdash} & 370 & 361 & 130,000 \\
HIP 26653 & 7.82 & 2023-11-04 & 180.0 & {\textemdash} & 370 & 459 & 130,000 \\
HD 106888 & 8.15 & 2024-01-16 & 160.0 & {\textemdash} & 256 & 312 & 130,000 \\
HD 51400 & 8.35 & 2023-11-04 & 170.0 & {\textemdash} & 225 & 380 & 130,000 \\
HD 203030 & 8.43 & 2023-10-22 & 513.0 & {\textemdash} & 461 & 481 & 130,000 \\
HD 8291 & 8.60 & 2023-11-06 & 180.0 & {\textemdash} & 239 & 272 & 130,000 \\
HD 16270 & 8.32 & 2023-11-06 & 210.0 & {\textemdash} & 243 & 291 & 130,000 \\
HD 116012 & 8.59 & 2023-02-08 & 660.0 & {\textemdash} & 763 & 1124 & 50,000 \\
BD+21 55 & 9.19 & 2023-10-22 & 1113.0 & {\textemdash} & 514 & 601 & 130,000 \\
BD+60 1417 & 9.43 & 2023-02-08 & 1260.0 & {\textemdash} & 517 & 795 & 50,000 \\
StKM 2-1777 & 9.29 & 2023-10-22 & 1500.0 & {\textemdash} & 560 & 869 & 130,000 \\
HD 253662 & 9.91 & 2023-11-02 & 600.0 & {\textemdash} & 265 & 297 & 130,000 \\
CD-24 407 & 9.94 & 2023-09-15 & 2000.0 & {\textemdash} & $\cdots$ & $\cdots$ & 130,000 \\
BD+13 2269 & 10.36 & 2024-01-14 & 3420.0 & {\textemdash} & 234 & 311 & 130,000 \\
BD+01 299 & 10.16 & 2023-11-06 & 1200.0 & {\textemdash} & 316 & 474 & 130,000 \\
StKM 1-1526 & 10.41 & 2023-09-15 & 4140.0 & {\textemdash} & $\cdots$ & $\cdots$ & 130,000 \\
TYC 5213-545-1 & 10.94 & 2023-09-15 & 4920.0 & {\textemdash} & $\cdots$ & $\cdots$ & 130,000 \\
LP 617-58 & 11.23 & 2024-01-16 & 1919.0 & {\textemdash} & 91 & 148 & 130,000 \\
HIP 63506 & 11.24 & 2024-01-14 & 4007.0 & {\textemdash} & 175 & 329 & 130,000 \\
LSPM J0632+5053 & 9.97 & 2023-11-02 & 590.0 & {\textemdash} & 237 & 288 & 130,000 \\
NLTT 1011 & 10.89 & 2023-11-02 & 1792.0 & {\textemdash} & 240 & 374 & 130,000 \\
HD 125141 (G200-28) & 7.63 & 2023-02-08 & 90.0 & {\textemdash} & 230 & 611 & 130,000 \\
BD+06 2986 & 9.77 & 2023-05-01 & 5000.0 & 186 & {\textemdash} & 817 & 130,000 \\
BD+24 4329 & 9.81 & 2023-06-19 & 3800.0 & 217 & {\textemdash} & 680 & 130,000 \\
BD+49 2561 & 10.09 & 2023-05-01 & 5000 & 218 & {\textemdash} & 650\tablenotemark{a} & 130,000 \\
\enddata
\tablecomments{V magnitude, observing date, exposure time, and SNR per pixel in all objects.}
\tablenotetext{a}{We use observations from 2023-05-01 for this target for CD VI.}
\end{deluxetable*}

%System Details for Host and Companion
\begin{deluxetable*}{lrrrcrr}
\tablecaption{Host Star Information \& Companion Information Organized by Host Spectral Type 
\label{tab:sample}}
\tablehead{
\colhead{Star Name} & \colhead{SpT} & \colhead{Companion} & \colhead{SpT} & \colhead{d$_\mathrm{{Host Star}}$} & \colhead{Proj. Sep} & \colhead{Ref} \\
\colhead{} & \colhead{Primary} & \colhead{} & \colhead{Secondary} & \colhead{(pc)} & \colhead{(AU)} & \colhead{}}
\startdata
HD 89744 & F7 & HD 89744 B$^{\dagger}$ & L0.5 & 38.68 $\pm$ 0.11 & 63 & 1,5 \\
HD 46588 & F7V & HD 46588 B$^{\dagger}$ & L9 & 18.21 $\pm$ 0.04 & 1420 &1,6 \\
G171-58 (HD 2057) & F8 & 2MASS J00250365+4759191 & L4+L4 & 54.01 $\pm$ 0.40 & 8800 & 7 \\
HD 106888 & F8 & 2MASS J12173646+1427119 & L1 & 67.18 $\pm$ 0.57 & 2170 & 1,8  \\
V$^{*}$ HN Peg & G0V+ & HN Peg B$^{\dagger}$ & T2.5 & 18.13 $\pm$ 0.02 & 795 &1,10  \\
HD 126053 (BD+01 2920) & G1.5V & BD+01 2920B & T8 & 17.44 $\pm$ 0.01& 2630 & 1,2 \\
GJ 417 & G2 & GJ 417 BC$^\diamond$ & L4.5+L6 & 22.65 $\pm$ 0.02 & 2000 & 1,10 \\
LSPM J0632+5053 & G2V & 2MASS J06324849+5053351$^{\dagger}$ & L1.5 & 82.58 $\pm$ 0.13 & 4499 & 11 \\
HD 51400 & G5 & CWISE J065752.45+163350.2 & L6$\beta$ & 37.08 $\pm$ 0.78 & 2300 & 1,3 \\
HD 8291 & G5V & HD 8291 B & L1+T3 & 50.38 $\pm$ 0.35 & 2570 & 1, 11 \\
HD 125141~(G200-28) & G5 & 2MASS J14165987+5006258$^{\dagger}$ & L4.5 & 47.11 $\pm$ 0.06 & 26951 &1,12 \\
HIP 26653 (HD 37216) & G5V & HD 37216 B$^{\dagger}$ & L5 & 28.08 $\pm$ 0.04 & 753 & 1,11 \\
HD 253662 & G8IV & 2MASS J06135342+1514062$^\diamond$ & L0.5 & 86.46 $\pm$ 0.34 & $>$1252 & 1,11\\
LP 617-58 & G8 & CWISE J132539.70+022309.4 & M9 & 118 $\pm$ 0.40 & 3600 & 3 \\
HIP 9269~(HD 12051) & G9V & HIP 9269B$^{\dagger}$ & L6 & 24.77 $\pm$ 0.02 & 1300 &1,11 \\
$*$ 54 Psc (HD 3651) & K0.5V & HD 3651 B & T7.5 & 11.14 $\pm$ 0.01 & 480 &1,11,9 \\
HD 203030 & K0V & HD 202030 B$^\diamond$ & L7.5 & 39.29 $\pm$ 0.09 & 487 & 1,11 \\
HD 116012 & K0V & 2MASS J13204427+0409045 & L5 & 30.31 $\pm$ 0.046 & 9708 &1,13 \\
BD+60 1417 & K0 & WISE J124332.17+600126.6$^\diamond$ & L8$\gamma$ & 44.96 $\pm$ 0.03 & 1662 &1,14,15 \\
TYC 5213-545-1 & K0 & CWISE J214129.80-024623.6 & L4 & 78.99 $\pm$ 0.06 & 9000 & 3 \\
BD+21 55 & K2 & 2MASS J00302476+2244492 & L0.5 & 55.21 $\pm$ 0.11 & 3970 &1, 11 \\
CD-24 407 (HIP 4417) & K2 & CWISE J005635.48–240401.9 & L8 & 67.60 $\pm$ 0.08 & 6924 & 1,3 \\
BD+49 2561 & K2 & CWISE J165325.10+490439.7 & L5 & 57.06 $\pm$ 0.03 & 3000 & 3 \\
BD+13 2269 & K2 & ULAS J103131.49+123736.4 & L3 & 99.0 $\pm$ 0.20 & 4100 & 3,4 \\
HD 16270 & K2.5Vk & HD 16270 B$^{\dagger}$ & L1.5 & 21.22 $\pm$ 0.02 & 250 &1,16 \\
StKM 1-1526 & K4/5 & CWISE J174509.03+380733.2 & L4$\beta$ & 57.20 $\pm$ 0.03 & 3900 & 3 \\
StKM 2-1777 (BD+29 5007) & K5V &WIS 395 B & L5.5 & 23.63 $\pm$ 0.02 &  22086 &1,17 \\
HIP 63506 (BD+42 2363) & K6V & 2MASS J13005061+4214473 & L1 & 44.15 $\pm$ 0.06 & 5640 &1,11 \\
BD+24 4329 & K6V & CWISE J210640.16+250729.0$^{\dagger}$ & T2 & 34.12 $\pm$ 0.02 & 38000 & 3 \\
NLTT 1011 (LP 192-58) & K7 & 2MASS J00193275+4018576$^{\dagger}$ & L2 & 55.21 $\pm$ 0.11& 3990 &1,11 \\
BD+01 299 & K8V & ULAS J014016.91+015054.7 & T5 & 38.56 $\pm$ 0.03 & 35-45 & 4,16 \\
BD+06 2986 & K8V & ULAS J150457.65+053800.8 & T6 & 19.02 $\pm$ 0.02 & 1230 &1,11 \\
\enddata
\tablerefs{1. \citet{Gaiadr3} 2. \citet{Pinfield2012}, 3. \citet{Rothermich2024}, 4. \citet{Skrzypek2016}, 5. \citet{Schneider2014}, 6. \citet{Loutrel2011}, 7. \citet{Reid_2006}, 8. \citet{Marocco2017}, 9. \citet{Faherty_2010}, 10. \citet{Kirkpatrick_2000}, 11. \citet{Deacon_2014}, 12. \citet{Chiu_2006}, 13. \citet{Gomes2013}, 14. \citet{Faherty2021}, 15. \citet{Phillips2024}, 16. \citet{Burningham_2018}, 17. \cite{Baig2024}}

\tablecomments{ $^{\dagger}$These targets were observed as a part of a JWST program with NIRSpec and MIRI instruments (GO \#3670)}
\tablecomments{$^{\diamond}$ These targets are observed and/or scheduled to be observed as a part of JWST programs with NIRSpec and/or MIRI instruments (GO \#8140)} 

\end{deluxetable*}

\section{Methods}
\label{sec:methods}

%{\textemdash}-Photometric Determination
\subsection{Photometric Stellar Parameters}
\label{sec:photometric_values}
The photometric stellar properties were estimated from spectral energy distributions~(SED) with \texttt{EXOFASTv2}~\citep{exofast1,exofastv2}. These photometric stellar properties were then used as starting values for our analysis of stellar properties from the PEPSI spectra. \texttt{EXOFASTv2} is based on a differential evolution Markov Chain Monte Carlo (MCMC) method to simultaneously characterize exoplanets and their host stars. Here, we use the MIST isochrones \citep{mist} fit jointly with an SED to characterize our host stars. We place no prior on the metallicity or temperature of these stars. We use broadband photometry measurements from Gaia DR3 RGB magnitudes ~\citep{Gaiadr3}, 2MASS $JHK_s$ magnitudes~\citep{cutri_2003}, and WISE 1,2,3, and 4 magnitudes \citep{wise} to construct the SED in combination with the parallax from Gaia. By default, \texttt{EXOFASTv2} requires a Gelman-Rubin  score of 1.01 for models to be considered converged. Thus the stellar evolution information from the MIST isochrones and the SED observations constrain estimates for stellar effective temperature and metallicity to be used as starting values for the analysis.

\par
\begin{figure}[hp!]
\gridline{\fig{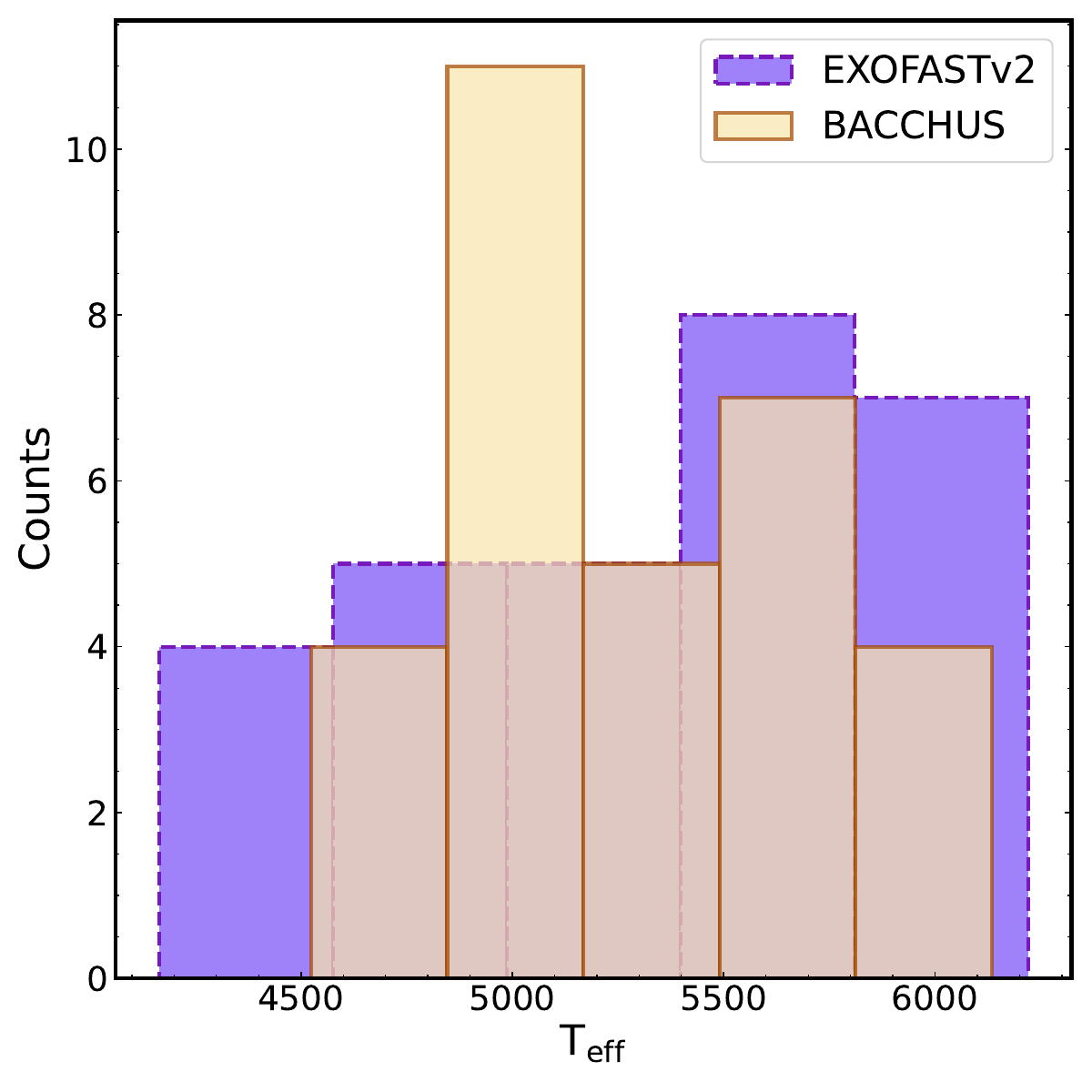}{0.37\textwidth}{\small(a)}}
\gridline{\fig{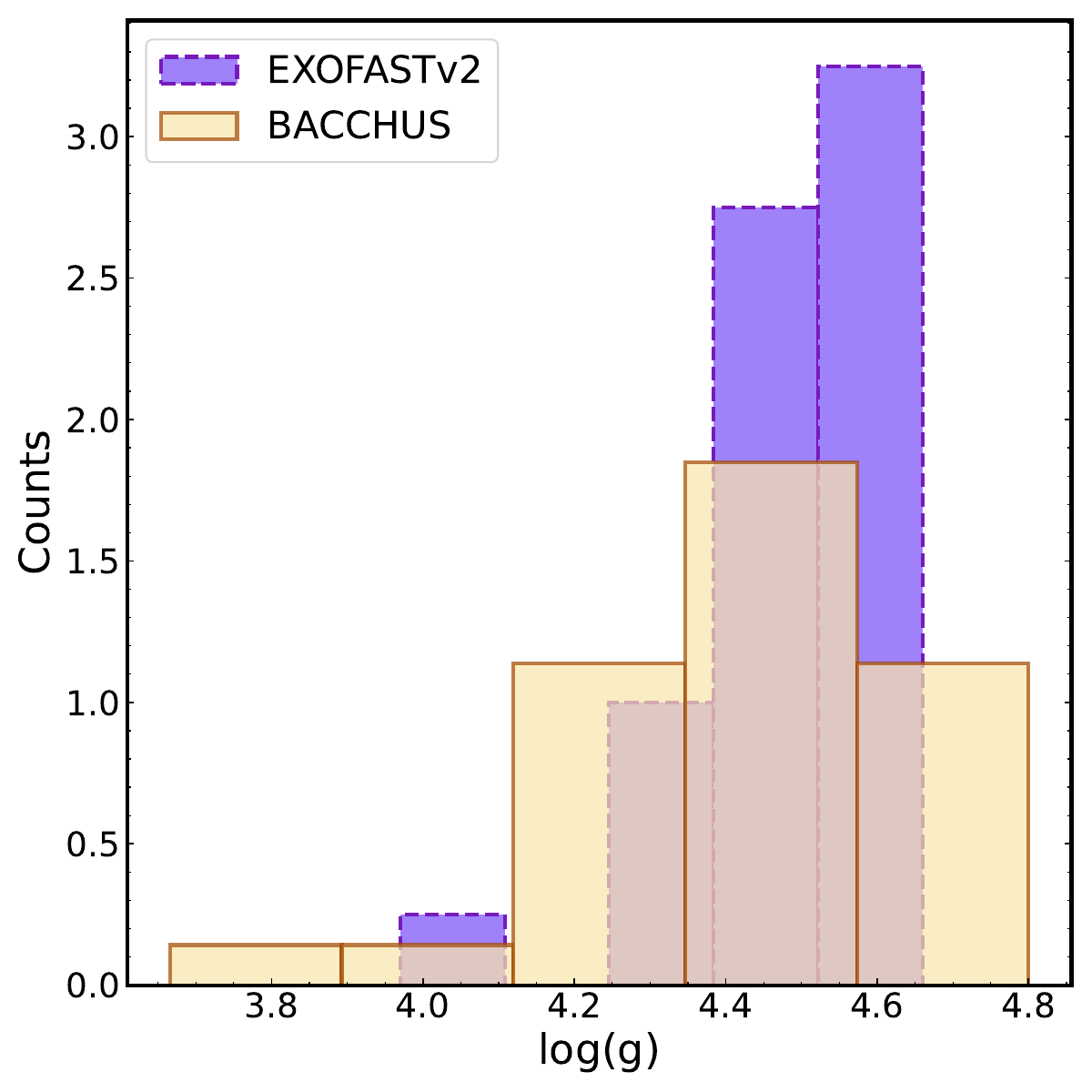}{0.37\textwidth}{\small(b)}}
\gridline{\fig{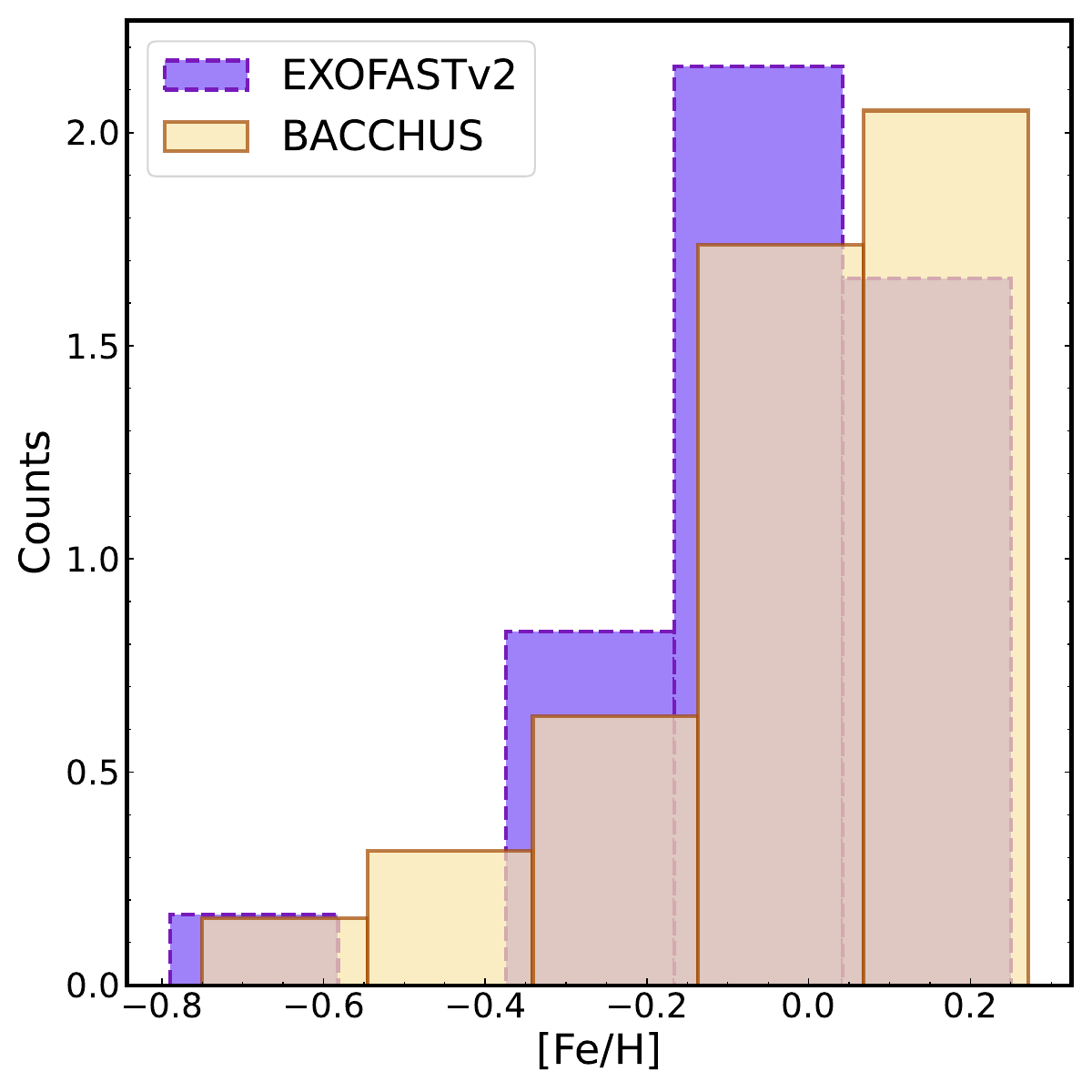}{0.37\textwidth}{\small(c)}}
 \caption{Histograms of the sample’s (a) $\teff$, (b) $\logg$, and (c) [Fe/H] distributions for parameters from spectroscopic parameters from \bacchus~(orange, this work) and photometric parameters from \texttt{EXOFASTv2} (purple)}
\label{fig:photo_vs_spec_params}
% \vspace{0.5cm}
\end{figure}

We note that the photometric stellar parameters used for initialization are not consistent with our derived spectroscopic parameters~(Figure \ref{fig:photo_vs_spec_params}). Although the photometric values of $T_{\mathrm{eff}}$ and $\log g$ are obtained from observational data and serve as the basis for estimating physical properties such as stellar radius and mass, because we have magnitudes, SEDs, and parallaxes, their use prevents us from achieving ionization and excitation equilibrium between Fe\,\textsc{i} and Fe\,\textsc{ii} abundances. Therefore, we adopt the spectroscopic parameters for subsequent abundance analyses, as they are optimized for this purpose. However, we emphasize that these spectroscopic parameters should alone not be used to infer the stellar mass or radius, as uncertainties in log(g) propagates into radius estimates, limiting the reliability of this approach~\citep{Jofre2019, Brucalassi2022, Torres2010}.

%{\textemdash}-Spectroscopic with BACCHUS

%{\textemdash}-Spectroscopic Parameters
\subsection{Spectroscopic  Parameters \& Abundances}
\label{sec:spectroscopic_parameters}

%{\textemdash}BACCHUS code section

\subsubsection{BACCHUS Code}
\label{sec:bacchus}
We use the Brussels Automatic Code for Characterizing High accUracy Spectra (\bacchus, \citealt{Masseron2013}) to derive spectroscopic parameters and elemental abundances for our sample. \bacchus~computes synthetic spectra for
a range of abundances and compares these syntheses to
observational data on a line-by-line basis, deriving abundances from different methods (e.g., using equivalent width, line depth, or $\chi^{2}$). The synthetic spectra are calculated using the 1D LTE \texttt{Turbospectrum} radiative transfer code~\citep{Alvarez1998,Pelz2012}, and the MARCS model atmosphere grids~\citep{Gustafsson2008}. \bacchus~uses four methods to compare the observed and synthetic
spectra and provides an abundance
measurement and associated quality flag for each method. The four methods are summarized as follows from \cite{Hays2022}:

\begin{enumerate}
    \item A$_{\chi^{2}}$ -- determines an abundance by minimizing the squared differences between synthetic and observed spectra
    \item A$_\mathrm{syn}$ --  looks for the abundance that makes the difference between the synthetic and the observed points zero
    \item A$_\mathrm{EW}$ -- determines the abundance needed to match the equivalent widths of the synthetic spectra to the observations.
    \item A$_\mathrm{int}$ -- measures abundances by matching the line core in the synthetic and observed spectra.
\end{enumerate}    
\par
We use the reduced PEPSI spectra to determine
stellar parameters and abundances with \bacchus. The fundamental  stellar parameters are determined by \bacchus~using the excitation-ionization balance method. We start with our initial photometric estimates from \texttt{EXOFASTv2}, and \bacchus~iteratively adjusts the parameters until it identifies the optimal combination of $\teff$, $\logg$, [M/H], and microturbulent velocity ($v_{\mathrm{mic}}$) that yields consistent abundances derived from both Fe\,\textsc{i} and Fe\,\textsc{ii} lines from excitation and ionization equilibrium. 
For one of our coolest targets, BD+06 2986, the microturbulent velocity did not converge, as a result we adopt a $v_{\mathrm{mic}}$ = 1.00 for this target. Our final spectroscopic parameters via \bacchus~are shown in Table \ref{tab:params}. The parameter uncertainties are estimated from the line-to-line scatter in Fe abundances, as discussed in \cite{Hays2022}
\begin{deluxetable*}{lccccccccccccccc}
\singlespace
\tablecaption{Spectroscopic and Photometric Stellar Parameters. \label{tab:params}}
\tablehead{
\multicolumn{1}{c}{} & \multicolumn{8}{c}{Spectroscopic} & \multicolumn{7}{c}{Photometric} \\
\cline{2-9}
\cline{11-16}
\colhead{Object} & \colhead{$\teff$ (K)}  &\colhead{$\sigma_{T}$}& \colhead{$\logg$} &\colhead{$\sigma_{log(g)}$}  &\colhead{$v_{\text{micro}}$}&\colhead{$\sigma\nu_{\text{micro}}$}&\colhead{[Fe/H]}&\colhead{$\sigma_{\text{[Fe/H]}}$}&\colhead{}&\colhead{$\teff$ (K)}  &\colhead{$\sigma_{T}$}& \colhead{$\logg$} &\colhead{$\sigma_{log(g)}$}&\colhead{[M/H]}&\colhead{$\sigma_{\text{[M/H]}}$}}
\startdata
HD 89744&6134&67&3.66&0.13&1.64&0.05&0.19&0.32&&6210&219&3.97&0.07&0.23&0.11\\
HD 46588&6093&53&4.29&0.31&1.33&0.07&$-$0.15&0.03&&6220&219&4.34&0.08&0.01&0.10\\
$*$ 54 Psc&5141&14&4.38&0.20&0.74&0.02&0.16&0.03&&5283&133&4.52&0.05&0.16&0.11\\
V$^{*}$ HN Peg&5927&1&4.35&0.07&1.21&0.03&$-$0.00&0.01&&6000&205&4.43&0.07&$-$0.01&0.10\\
HD 126053&5612&42&4.30&0.40&1.17&0.05&$-$0.38&0.02&&5830&191&4.44&0.08&--0.28&0.16\\
GJ 417&5779&42&4.42&0.28&1.19&0.03&0.05&0.03&&5490&156&4.43&0.07&0.20&0.14\\
HIP 9269&5493&111&4.41&0.60&1.42&0.07&0.16&0.07&&6050&304&4.25&0.09&0.17&0.14\\
G171-58&5798&47&4.17&0.34&1.35&0.03&0.09&0.03&&5720&262&4.46&0.06&-0.27&0.13\\
HIP 26653&5469&51&4.48&0.31&1.37&0.03&$-$0.06&0.03&&5860&205&4.30&0.09&-0.03&0.12\\
HD 106888&5922&28&4.48&0.24&0.92&0.04&0.02&0.11&&5480&170&4.50&0.06&0.00&0.11 \\
HD 51400&5415&19&4.53&0.25&1.05&0.02&0.05&0.01&&5570&170&4.52&0.05&0.04&0.12\\
HD 203030&5524&70&4.57&0.33&1.52&0.04&0.01&0.09&&5750&170&4.46&0.06&-0.05&0.09\\
HD 8291&5686&10&4.61&0.12&0.86&0.02&$-$0.07&0.01&&4910&255&4.62&0.04&-0.31&0.05\\
HD 16270&4719&29&4.27&0.48&0.62&0.04&0.24&0.09&&4982&122&4.56&0.05&0.16&0.14\\
HD 116012&4990&39&4.69&0.62&1.06&0.03&0.08&0.10&&4942&117&4.57&0.05&-0.07&0.14\\
BD+21 55&4820&28&4.36&0.07&0.21&0.05&$-$0.13&0.10&&4960&124&4.56&0.04&0.25&0.28\\
BD+60 1417&4907&49&4.66&0.51&1.12&0.03&0.11&0.11&&4452&105&4.62&0.04&-0.01&0.09\\
StKM 2-1777&4719&76&4.48&  $\cdots$   &1.12&0.071&$-$0.07&0.03&&5610&416&4.48&0.06&0.23&0.53\\
HD 253662&5280&23&4.35&0.06&0.51&0.03&0.28&0.07&&5160&141&4.51&0.05&0.11&0.28\\
CD-24 407&5035&15&4.48&0.41&0.31&0.03&$-$0.01&0.10&&5010&177&4.57&0.04&-0.17&0.37\\
BD+13 2269&5480&66&4.80&0.43&1.22&0.04&0.08&0.10&&4466&113&4.62&0.04&0.24&0.33\\
BD+01 299&4524&37&4.26&0.85&0.78&1.10&0.12&0.18&&4741&125&4.56&0.04&0.18&0.35\\
StKM 1-1526&4998&56&4.58&$\cdots$&0.95&0.07&0.13&0.02&&5613&141&4.58&0.05&-0.79&0.14\\
TYC 5213-5451&4829&20&4.70&0.17&0.67&0.03&0.04&0.03&&4164&96&4.66&0.04&-0.04&0.35\\
LP 617-58$^{\bigstar}$&5111&20&4.26&0.10&0.59&0.03&$-$0.39&0.01&&5349&${}^{+7.05}_{-7.34}$&4.52&${}^{+0.01}_{-0.00}$&$-$0.23&0.00\\
HIP 63506&4600&97&4.54&$\cdots$&1.06&0.08&$-$0.59&0.04&&5289&147&4.54&0.05&-0.14&0.20\\
LSPMJ0632+5053&5405&5&4.74&0.19&0.93&0.02&0.25&0.01&&4370&102&4.65&0.04&-0.02&0.07\\
NLTT 1011&5006&84&4.30&$\cdots$&1.62&0.06&$-$0.48&0.12&&5850&212&4.30&0.09&-0.03&0.13\\
HD 125141&5749&78&4.49&0.57&1.28&0.04&$-$0.08&0.10&&5570&234&4.53&0.05&-0.03&0.10\\
BD+06 2986&4468&106&4.61&$\cdots$& 1.00&$\cdots$&$-$0.80&0.06&&4543&135&4.61&0.04&0.33&0.28\\
BD+24 4329&4939&89&4.48&0.88&1.34&0.05&$-$0.17&0.09&&5420&178&4.50&0.06&0.21&0.34\\
BD+49 2561&4959&31&4.71&0.49&0.84&0.05&0.02&0.08&&4194&106&4.61&0.04&0.38&0.26\\
\enddata
\tablenotetext{\bigstar}{Not converged through \texttt{EXOFASTv2}. Initial photometric parameters taken from \textit{Gaia} DR3}
\tablecomments{Spectroscopic [Fe/H] $\approx$ [M/H] is assumed when determining best-fit parameters}
\end{deluxetable*}

\par
\subsubsection{Photometric vs Spectroscopic Surface Gravity}
In our initial determination of the fundamental stellar parameters for the sample, we adopt the photometric surface gravity derived from \texttt{EXOFASTv2} as an initial guess and solve for all parameters (effective temperature, surface gravity, metallicity, and microturbulence) with \bacchus. For several cooler K-type stars, we find large uncertainties in the spectroscopic surface gravity (exceeding 1 dex), likely due to weak Fe \textsc{ii} lines at lower temperatures and also the increased presence of molecular absorption features in these stars, both of which increase the line-to-line Fe \textsc{ii} abundance scatter.  For the following K dwarf targets, StKM~2-1777, StKM~1-1526, HIP~63506, NLTT~1011, and BD+06~2986, we therefore fix $\log g$ to the photometric value. Spectroscopic surface gravities in cool dwarfs are often poorly constrained due to the limited number and weakness of Fe\,\textsc{ii} lines~\citep{Neeves2009, Tsantaki2013}. As a result, enforcing ionization equilibrium can lead to unstable solutions and inflated uncertainties in the derived parameters. By fixing $\log g$ to the photometric value we can reduce these degeneracies and obtain  estimates of $T_{\mathrm{eff}}$ and $[\mathrm{Fe}/\mathrm{H}]$~\citep{Torres2010,Huber2013,Mortier2014}. We rerun the analysis for these stars while fixing the surface gravity to the photometrically determined value. We find that fixing the surface gravity with the photometric value alleviates larger errors on other parameters as well, so we proceed to adopt this method for these stars.

%{\textemdash}-Abundance Determination
\subsection{Abundance Determination}
\label{sec:abundance_determination}

Once the stellar parameters were established, we used \bacchus~to measure abundances for 11 elements. For each spectral line, the code synthesizes five model spectra with abundances spanning $--$0.6 to +0.6 dex around the expected abundance, assuming a solar-scaled composition adjusted to the star’s metallicity. Abundances are then determined using four independent approaches (see details in \S \ref{sec:bacchus}). The final abundance for each element is taken as the mean of all lines yielding consistent (unflagged) results across these methods.

We determine abundance uncertainties by first remeasuring each star's abundance in each element eight times, each time fixing all parameters to those reported in Table \ref{tab:params} except for one, which we modify by first adding and then subtracting the associated uncertainty in that parameter.  This process results in eight new abundances per element (four that correspond to adding the uncertainty to the modified parameter, four subtracting) that reflect how each adopted stellar parameter affects the resulting abundance.  The final abundance uncertainty we report is the quadrature sum of the average abundance difference induced by each stellar parameter's uncertainty (four terms) and also the standard error on the mean line-by-line abundance (the fifth term)~\citep{Bensby2014}.

We use the notation outlined in \citet{Brewer2016_CO} to convert from our reported $[X/H]$, the ${\rm log}_{10}$ of the solar relative number abundance of an element with respect to hydrogen, to abundance ratios for any given two elements:

\begin{equation}
    [X/H] = A(X)_{\star} - A(X)_{\odot},
\end{equation}
where X denotes the element.

\begin{equation}
    X_1/X_2 = 10^{\rm ([X_1/H] + (X_1/H)_{\odot})-([X_2/H] + (X_2/H)_{\odot})}
\end{equation}

We adopt solar abundances from \cite{Grevesse2007} (Table \ref{tab:Greeves_abund_photo}). 

\begin{table}[H]
    \centering
    \begin{tabular}{c|c}
        Element& Photospheric \\ & Abundance ($A(X)_\odot$) \\
        \hline
         C & 8.43\\
         O & 8.66 \\
         Mg & 7.53 \\
         Si & 7.51 \\
         Ca & 6.31 \\
         Al &6.37 \\
         Fe & 7.45\\
         Y & 2.21 \\
         N & 7.78 \\
         S & 7.14 \\
         \hline
         \end{tabular}
    \caption{Element abundances in the solar photosphere as measured from \cite{Grevesse2007}}
    \label{tab:Greeves_abund_photo}
\end{table}

We use the following formula from \cite{Hinkel2022} to determine the error for our abundance ratios:

\begin{equation}
    \delta  \frac{\mathrm{X_1}}{ \mathrm{X_2}} = \ln(10) \sqrt{\delta[\mathrm{X_1}/\mathrm{H}]_*^2 + \delta [\mathrm{X_2}/\textrm{H}]_*^2 }
\end{equation}

To propagate the error for [Y/Mg], we use the following formula

$$    \sigma_{[Y/Mg]} = \sqrt{\sigma^{2}_{[Y/H]} + \sigma^{2}_{[Mg/H]} } $$

We report the resulting ratios in Table \ref{tab:abundance_ratio}.

Our different resolution modes~(R = 50,000 or R = 130,000) affect our parameters, abundances, and abundance ratio precision. We find the following precision for R = 130,000 data: 0.03 dex for [Fe/H] and 0.11 for C/O ratios. For the 50,000 data, we find the following: 0.03 dex for [Fe/H] and 0.08 for C/O. We find a higher precision for the R = 50,000 data for the C/O abundance ratio, which we attribute to the higher average $\langle S/N \rangle > 500$ for this resolution~\citep{Bedell_2014, Nissen2015}.

\subsubsection{Line Selection}

We adopt version 5 of the \textit{Gaia}-ESO
linelist for our atomic transition data~\citep{Heiter2021} and combine molecular transition data from numerous
sources. Our line selections are initially from the NIST database~\citep{NIST_ASD, Ralchenko2020}. We follow \cite{Kolecki2022} and restrict our line choices to those that have a transition strength accuracy value of ``C" or better. We have a final line selection from referencing linelists from Gaia-ESO~\citep{Heiter2021}, the solar spectrum~\citep{SolarData}, and  \citet{Mack2018,Kolecki2022, Phillips2024}. We use different line selections curated for our two observing mode configurations: (A) CD III \& CD VI and (B) CD II \& CD VI~(Table \ref{tab:obs}). The main observing mode for our sample is with (A) CD III and CD VI, which covers wavelengths 4800 $-$ 5441 $\AA$ and 7410 $-$ 9140 $\AA$ respectively. Our line list for (A) is comprised of the following: 18 \ion{Ti}{1} lines, 16 \ion{C}{1} \& C$_{2}$ lines, 3 \ion{O}{1} lines, 20 \ion{Mg}{1} lines, 27 \ion{Si}{1} lines,  7 CN lines (for Nitrogen), 6 \ion{Ca}{1} lines, 4 \ion{Ca}{2} lines, 6 \ion{Al}{1} lines, 13 \ion{Y}{1} and \ion{Y}{2} lines, 2 \ion{S}{1} lines, and  111 \ion{Fe}{1} and \ion{Fe}{2} lines.

\par
Three targets were observed in configuration B, which covers wavelengths 4265 $-$ 4800 $\AA$ and 7410 $-$ 9140 $\AA$. Our line list for (A) is comprised of the following: 11 \ion{Ti}{1} lines, 6 \ion{C}{1} lines, 3 \ion{O}{1} lines, 21 \ion{Mg}{1} lines, 27 \ion{Si}{1} lines,   2 \ion{S}{1} lines, 4 CN lines (for Nitrogen), 4 \ion{Ca}{1} lines, 2  \ion{Ca}{2} lines, 6 \ion{Al}{1} lines, 7 \ion{Y}{1} and \ion{Y}{2} lines, 70 \ion{Fe}{1} and \ion{Fe}{2} lines.

\subsubsection{Carbon Abundance}
Our coolest K-type stars (StKM~2-1777, StKM~1-1526, HIP~63506, NLTT~1011, BD+06~2986, and BD+24~4329) are riddled with molecular features and have weak atomic carbon lines As a result of this, we turn to a set of C$_{2}$ swan lines to determine the carbon abundance of these stars~\citep{Brooke2013,Ryabchikova2015, Ryabchikova2022}. We use the following carbon \ion{C}{1} lines for our hotter stars: 4932.0, 5021.1, 5380.3, 8058.6, 8335.1, 8727.1, 9061.4 $\AA$. For the cooler stars we use the following C$_{2}$ lines: 5151.8, 5153.1, 5156.5, 5159.4, 5126.0,  5130.4,  5132.6,  5135.5,  5146.0,  5149.1 $\AA$.

The difficulty in measuring C and O makes determining these C/O ratios significantly challenging. Future high-resolution NIR spectroscopy of our coolest targets from IGRINS~(R$\sim$ 45,000) would prove beneficial for accessing CO and OH features to calculate C/O ratios~\citep{Hejazi_2024}.

\subsubsection{Nitrogen Abundance}
Measuring N from optical spectra is difficult because there are no reliable atomic N lines in this wavelength regime \citep[see discussion in, e.g.,][]{gaiaeso, Casali2019}.  As such, using CN molecules is our only option.  To determine the nitrogen abundance from CN molecular features, the carbon abundance was determined using the methods described in the previous subsection and then fixed for each star prior to the nitrogen determination. For stars that have uncertain carbon measurements and resulting upper limits (see \S \ref{sec:upperlimits}), we omit nitrogen abundance determination.

\subsubsection{Upper Limits on Abundances and Abundance Ratios}
\label{sec:upperlimits}

For some of our coolest K-dwarfs with weak lines, we are only able to report upper limits on abundances and resulting abundance ratios. To determine the upper limit on a given abundance, we first check that the \texttt{BACCHUS} fit is adequate and adopt the upper limit on A$_\mathrm{syn}$ for a given element~\citep[e.g.,][]{Hacksaw2025}. We describe our findings for NLTT 1011, HIP 63506, BD+06 2986, BD+24 4329, StKM 1-1526, StKM2-1777, and HD 116012. 
 
\par

For NLTT 1011 (K7), there is a O detection at the \ion{O}{1} 7771.9 $\AA$ line. We find an upper limit on [O/H] = $-0.85$, which leads to a reported C/O $>$ 0.76. There are two Ca detections at \ion{Ca}{1} 4847.3 \& 5260.4 $\AA$, which leads to a reported Ca/Al $<$ 1.8. There are 2 Y, detections at \ion{Y}{2} 4900.1 \& 5087.4 $\AA$. We find an average upper limit of [Y/H] = $-0.77$ and [Y/Mg] $<$ $-0.314$. 
\par
We find that for HIP 63506 (K6V), there are no O or C detections, therefore we cannot place an upper limit on the C/O ratio. Similar for NLTT 1011, there are 2 Y detections at \ion{Y}{2} 4900.1 \& 5087.4 $\AA$. We find an average upper limit of [Y/H] = $-1.02$ and [Y/Mg] $<$ $ -0.54$. 
\par
% BD+06  2986 (K8)
Same as HIP 63506, there are no usable O and C lines for BD+06 2986 so we cannot constrain or place an upper limit on the C/O ratio. There are two Y detections at \ion{Y}{1} 4643.7 \& 4674.8 $\AA$. We find an average upper limit of [Y/H] = -1.11 and  [Y/Mg]$< -0.687$.

\par
%BD+24 4329
BD+24 4329 (K6) has one C detection at the \ion{C}{1} 8335.1 $\AA$. [C/H] has an upper limit of $-1.17$, with leads to a C/O $<$ 1.79. We use 2 Ca detections at \ion{Ca}{2} 8248.8 \& 8912.2 $\AA$ and find upper limit of [Ca/H] = $-0.79$, leading to Ca/Al $<$ 1.37.

\par
%StKM 1-1526
StKM 1-1526 (K4/5) has two C detections at C$_{2}$ 4932.0 \& 5380.3 $\AA$. This results in an upper limit on [C/H] = -1.94 and C/O $>$ 1.04 for StKM 1-1526. StKM 2-1777 (K5) has only one C detection at \ion{C}{1} 5149.1 $\AA$. We find an upper limit on [C/H] = $-2.67$, and a C/O $>$ 1.005. HD 116012 has two Y detections at \ion{Y}{2} at 4883.1,  5087.4, and 5289.8 $\AA$, with an upper limit average for [Y/H] = $-0.78$ , leading to [Y/Mg] $<$$ -0.94$.

%{\textemdash}-NLTE Oxygen
\subsubsection{NLTE Effects}

In this work, we calculate the 1D non-LTE (NLTE) oxygen abundances for our sample by applying correction factors. NLTE effects are known to be a cause of abundance offsets and can require abundance corrections $>$0.1 dex (e.g. \citealt{Jofre2019}). We use NLTE correction lookup tables from the Max Planck Institute for \ion{O}{1}~\citep{Sitnova2013}\footnote{\url{https://gemini-web.mpia.de/gui-siuAC_secEnew.php}}. We use the spectroscopic parameters~(Table \ref{tab:params}) found with \bacchus~as input stellar parameters and the plane-parallel 1D MAFAGS-OS. We find the mean NLTE oxygen for the O triplet lines and apply this to our [O/H] to determine the NLTE [O/H] abundance. We provide the oxygen corrections in Table \ref{tab:oxy_correction}.

%{\textemdash}Validating Results

\section{Validating Results}
\label{sec:validating_results}
%{\textemdash}-Calibration with the Sun
\subsection{Testing BACCHUS with a Solar Spectrum from PEPSI}
\label{sec:test_sun_pepsi}
To calibrate and test our line list, derived stellar parameters, and methodology~(\S \ref{sec:abundance_determination}), we anchor our analysis with the Sun as seen from PEPSI. We test the Sun to derive the following ratios: [O/H], [C/H], [S/H], [N/H], [Mg/H], [Si/H], [Ca/H], [Y/H], and derive C/O and Mg/Si ratios to compare to literature values~\citep{Asplund2009,Grevesse2007}.
\bacchus~has been used in recent abundance and parameter estimation for stellar hosts of directly-imaged planets and brown dwarfs~(\citealp{Zhang2023,Zhang2025_elpis2}, Rothermich et al, submitted). To test the validity of our method and the \bacchus~code with data from PEPSI we use observations of the Sun taken with PEPSI from \cite{Strassmeier2018_sun}. These Solar observations were taken in R = 250,000 mode, so we must convolve them down to the resolution of our observations before proceeding with ~\bacchus.  We use a 1D-Gaussian kernel to convolve down to the resolution of most of our data R = 130,000. In \cite{Strassmeier2018_sun}, the Sun was observed across all cross dispersers (CD I $-$ CD VI), we take the Sun's spectrum with PEPSI and reduce to just CD III and CD VI to test \bacchus, as this is the dominant cross-disperser configuration for our sample. 

\begin{table}[H]
    % \centering
    \begin{tabular}{c|cc}
        Parameter&Sun$_{\odot}$ & Sun$_\mathrm{\bacchus}$ \\
        \hline
         $\teff$&5774&5710 $\pm$ 11\\ 
         $\logg$&4.44 & 4.41 $\pm$ 0.22\\
         $[Fe/H]$ &0.00 &$-$0.00 $\pm$ 0.01\\
         \end{tabular}
    \caption{Known Sun parameters versus spectroscopic parameters found for the Sun from \bacchus~using PEPSI data.}
    \label{tab:sun_PEPSI}
\end{table}

Generally we find good agreement between our spectroscopic parameters found for the Sun with \bacchus. We note that the largest discrepancy appears between $\teff$, where we find a lower $\teff$ by about $\sim$ 50K~(Table \ref{tab:sun_PEPSI}).

\begin{deluxetable}{lr}[h!]
\tabletypesize{\small}
\tablecolumns{2}
\tablecaption{Stellar abundances and ratios for the Sun from \bacchus}
\tablewidth{0pt}
\setlength{\tabcolsep}{0.3in}
\label{tab:sun_parameters}
\tablehead{\colhead{Property}  & \colhead{Sun$_\mathrm{\bacchus}$}}
\startdata
\\\hline
$[{\rm Fe/H}]$      &  $-0.00 \pm 0.01$ dex  \\ 
$[{\rm C/H}]$       &   $0.08 \pm 0.09$ dex  \\ 
$[{\rm O/H}]_{\rm NLTE}$ & $0.18 \pm 0.06$ dex \\
$[{\rm Mg/H}]$      &   $0.01 \pm 0.02$ dex   \\ 
$[{\rm Si/H}]$      &   $-0.03 \pm 0.07$ dex   \\ 
$[{\rm Ca/H}]$      &   $0.11 \pm 0.09$ dex   \\ 
$[{\rm Al/H}]$      &   $-0.01 \pm 0.03$ dex   \\ 
$[{\rm N/H}]$      &   0.10 $\pm$ 0.03 dex   \\ 
$[{\rm Y/H}]$      &   $-0.22 \pm 0.11$ dex   \\ 
$[{\rm Ti/H}]$      &   $-0.07 \pm 0.05$ dex   \\ 
$[{\rm S/H}]$      &   $0.20 \pm 0.19$ dex   \\ 
\hline
C/O       &   $0.53$  \\ 
Mg/Si     &   $1.14$  \\ 
\enddata
\end{deluxetable}

From our line selection, we find non-zero systemic offsets between expected and measured [X/H] values for a few elements, with the biggest discrepancy in yttrium, oxygen, and sulfur for the Sun when compared to \cite{Grevesse2007}~(Table \ref{tab:Greeves_abund_photo}). In general, \bacchus~achieves good fits to the PEPSI solar spectrum (see Figure \ref{fig:solar_fits_bacchus} in the Appendix), so it is likely that the discrepancies come from issues with the line list or the spectral synthesis method rather than from data quality. Systematic abundance differences between surveys are common and remain an open issue in high-resolution spectroscopy. These offsets can arise from differences in adopted atomic data (e.g., laboratory values versus empirically tuned oscillator strengths), unidentified or imperfectly modeled blends, the use of 1D rather than 3D model atmospheres, and residual non-LTE effects~\citep{Socas2015, Brewer2016,Bergemann2021, Zhang2024, Pietrow2026}.

When reporting \bacchus~ abundances for other stars in the format [X/H], we are implicitly normalizing the abundances to the Solar values. Therefore, we can use the observed offsets in Solar abundances from Table \ref{tab:sun_parameters} to correct the \bacchus-reported stellar abundances and bring them in line with the \bacchus-determined Solar abundance values~(Equation \ref{eq:pepsi_norm}). We calculate a [Fe/H] = $-0.00 \pm 0.01$ for the Sun so we do not apply a correction for the sample for [Fe/H]. In what follows, we subtract these Solar correction factors from the \bacchus~ abundances in Table \ref{tab:sun_parameters} to determine final Solar-normalized [X/H] abundances in Table \ref{tab:sampl_abundances}.

\begin{equation}
\label{eq:pepsi_norm}
    [X/H] = \big((A(X)_{\star} - A(X)_{\odot}\big) - [X/H]_{\odot, \textrm{PEPSI}}
\end{equation}
where X denotes the element.

\par

\begin{deluxetable*}{lccccccccccc}
\tablecaption{Stellar Abundances and Uncertainties for PEPSI Sample 
\label{tab:sampl_abundances}}
\tabletypesize{\footnotesize}
\tablehead{
\colhead{Object} & \colhead{[C/H]} & \colhead{[O/H]} & \colhead{$[\mathrm{O/H}]_\mathrm{NLTE}$} & \colhead{[Ca/H]}& \colhead{[Al/H]} & \colhead{[Mg/H]} & \colhead{[Si/H]} & \colhead{[Y/H]} & \colhead{[S/H]} & \colhead{[N/H]}  & \colhead{[Ti/H]} }
\startdata
HD 89744 & 0.04 $\pm$ 0.09 & 0.31 $\pm$ 0.13 & $-0.00$ $\pm$ 0.11 & 0.09 $\pm$ 0.07 & 0.18 $\pm$ 0.03 & 0.11 $\pm$ 0.44 & 0.24 $\pm$ 0.19 & 0.08 $\pm$ 0.05 & 0.17 $\pm$ 0.18 & 0.18 $\pm$ 0.03 & 0.12 $\pm$ 0.06 \\
HD 46588 & $-0.07$ $\pm$ 0.10 & $-0.00$ $\pm$ 0.13 & $-0.10$ $\pm$ 0.13 & $-0.18$ $\pm$ 0.06 & $-0.13$ $\pm$ 0.05 & $-0.13$ $\pm$ 0.03 & $-0.12$ $\pm$ 0.05 & $-0.14$ $\pm$ 0.14 & $-0.10$ $\pm$ 0.19 & $-0.11 \pm 0.07$  & $-0.10$ $\pm$ 0.08  \\
$\ast$ 54 Psc & 0.13 $\pm$ 0.13 & 0.14 $\pm$ 0.07 & 0.11 $\pm$ 0.07 & 0.13 $\pm$ 0.07 & 0.23 $\pm$ 0.05 & 0.23 $\pm$ 0.02 & 0.14 $\pm$ 0.19 & 0.06 $\pm$ 0.09 & $-0.02 \pm 0.09$ & $0.43 \pm 0.17$ &  0.23 $\pm$ 0.06   \\
$\mathrm{V}^{\ast}$ HN Peg & $-0.01$ $\pm$ 0.08 & 0.00 $\pm$ 0.03 & $-0.09$ $\pm$ 0.03 & 0.04 $\pm$ 0.06 & $-0.13$ $\pm$ 0.02 & $-0.08$ $\pm$ 0.03 & $-0.19$ $\pm$ 0.05 & 0.03 $\pm$ 0.06 & $-0.07 \pm 0.13$ & $0.01 \pm 0.01$ & 0.02 $\pm$ 0.02 \\
HD 126053 & $-0.37$ $\pm$ 0.13  & $-0.27$ $\pm$ 0.12 & $-0.23$ $\pm$ 0.12 & $-0.33$ $\pm$ 0.07 & $-0.23$ $\pm$ 0.03 & $-0.28$ $\pm$ 0.03 & $-0.30$ $\pm$ 0.06 & $-0.47$ $\pm$ 0.18 & $-0.30 \pm 0.23$ & $-0.43 \pm 0.04$ & $-0.28$ $\pm$ 0.08 \\
GJ 417 & 0.05 $\pm$ 0.09 & 0.20 $\pm$ 0.14 & 0.13 $\pm$ 0.14 & 0.13 $\pm$ 0.07  & 0.09 $\pm$ 0.05 & 0.00 $\pm$ 0.03 & 0.11 $\pm$ 0.04 & 0.13 $\pm$ 0.13 & $-0.10 \pm 0.08$ & $0.01 \pm 0.00$ & 0.12 $\pm$ 0.07 \\
HIP 9269 & 0.07 $\pm$ 0.27 & $-0.02$ $\pm$ 0.28 & $-0.07$ $\pm$ 0.28 & 0.06 $\pm$ 0.19 & 0.28 $\pm$ 0.09 & 0.21 $\pm$ 0.07 & 0.07 $\pm$ 0.13 & $-0.06$ $\pm$ 0.17 & $-0.12 \pm 0.27$ & $0.33 \pm 0.09$ & 0.18 $\pm$ 0.16   \\
G171-58 & 0.23 $\pm$ 0.13 & 0.33 $\pm$ 0.17 & 0.22 $\pm$ 0.17 & 0.22 $\pm$ 0.15 & 0.21 $\pm$ 0.03 & 0.15 $\pm$ 0.03 & 0.10 $\pm$ 0.07 & 0.03 $\pm$ 0.12 & $0.17 \pm 0.09$ & $0.32 \pm 0.08$ & 0.07 $\pm$ 0.08   \\
HIP 26653 & $-0.22$ $\pm$ 0.25 & $-0.21$ $\pm$ 0.14 & $-0.18$ $\pm$ 0.14 & $-0.12$ $\pm$ 0.08 & $-0.07$ $\pm$ 0.04 & $-0.05$ $\pm$ 0.02 & $-0.17$ $\pm$ 0.09 & $-0.12$ $\pm$ 0.13 & $-0.40 \pm 0.16$ & $-0.01 \pm 0.06$ & $-0.03$ $\pm$ 0.07  \\
HD 106888 & 0.16 $\pm$ 0.10  & 0.41 $\pm$ 0.07 & 0.33 $\pm$ 0.07 & 0.00 $\pm$ 0.07 & 0.06 $\pm$ 0.08 & $-0.01$ $\pm$ 0.03 & 0.03 $\pm$ 0.05 & 0.12 $\pm$ 0.10 & $0.05 \pm 0.05$ & $0.05 \pm 0.04$ & 0.11 $\pm$ 0.07  \\
HD 51400 & $-0.11$ $\pm$ 0.18 & $-0.08$ $\pm$ 0.09 & $-0.11$ $\pm$ 0.09 & 0.06 $\pm$ 0.08 & 0.09 $\pm$ 0.06 & 0.00 $\pm$ 0.02 & 0.00 $\pm$ 0.07 & 0.04 $\pm$ 0.1 & $0.06 \pm 0.25$ & $0.16 \pm 0.02$ & 0.08 $\pm$ 0.06 \\
HD 203030 & $-0.15$ $\pm$ 0.22  & $-0.06$ $\pm$ 0.12 & $-0.10$ $\pm$ 0.12 & 0.05 $\pm$ 0.04 & $-0.02$ $\pm$ 0.05 & $0.00$ $\pm$ 0.03 & 0.00 $\pm$ 0.07 & 0.10 $\pm$ 0.16 & $-0.20 \pm 0.16$ & $-0.07 \pm 0.17$ & 0.03 $\pm$ 0.08  \\
HD 8291 & $-0.11$ $\pm$ 0.08 & $-0.02$ $\pm$ 0.04 & $-0.06$ $\pm$ 0.04 & 0.00 $\pm$ 0.09 & $-0.08$ $\pm$ 0.05 & $-0.12$ $\pm$ 0.03 & $-0.07$ $\pm$ 0.06 & 0.04 $\pm$ 0.06 & $-0.03 \pm 0.08$ & $-0.03 \pm 0.05$ & $-0.02$ $\pm$ 0.03    \\
HD 16270 & 0.26 $\pm$ 0.08 & 0.17 $\pm$ 0.16 & 0.15 $\pm$ 0.16 & 0.15 $\pm$ 0.20 & 0.27 $\pm$ 0.07 & $0.18 \pm 0.05$ & $0.16 \pm 0.17$ & $0.19 \pm 0.2$ & \nodata & $0.28 \pm 0.19$ & $0.24 \pm 0.18$  \\
HD 116012 & 0.16 $\pm$ 0.07 & $-0.06 \pm 0.23$ & $-0.07 \pm 0.23$ & $0.03 \pm 0.14$ & $0.22 \pm 0.07$ & $0.16 \pm 0.03$ & $0.09 \pm 0.18$ & $-0.78$\tablenotemark{a} & $-0.16 \pm 0.29$ & $-0.04 \pm 0.17$ & $0.09 \pm 0.16$ \\
BD+21 55 & $-0.29 \pm 0.18$ & $0.00 \pm 0.05$ & $-0.00 \pm 0.05$ & $-0.11 \pm 0.08$ & $-0.02 \pm 0.06$ & $-0.08 \pm 0.03$ & $-0.08 \pm 0.05$ & $-0.19 \pm 0.75$ & $-0.04 \pm 0.04$ & $0.14 \pm 0.04$ & $-0.04 \pm 0.03$ \\
BD+60 1417 & $0.17 \pm 0.06$ & $0.34 \pm 0.2$ & $0.33 \pm 0.2$ & $0.12 \pm 0.12$ & $0.08 \pm 0.07$ & $0.07 \pm 0.09$ & $0.17 \pm 0.14$ & $0.21 \pm 0.24$ & \nodata & $-0.22 \pm 0.03$  & $0.06 \pm 0.08$ \\
StKM 2-1777 & $-2.57$\tablenotemark{a} & $-0.29 \pm 0.16$ & $-0.31 \pm 0.16$ & $-0.51 \pm 0.11$ & $0.09 \pm 0.04$ & $-0.11 \pm 0.02$ & $-0.33 \pm 0.09$ & $0.03 \pm 0.07$ & \nodata & \nodata & $-0.03 \pm 0.10$  \\
HD 253662 & $0.21 \pm 0.04$ & $0.46 \pm 0.03$ & $0.43 \pm 0.03$ & $0.36 \pm 0.10$ & $0.34 \pm 0.04$ & $0.31 \pm 0.02$ & $0.36 \pm 0.06$ & $0.31 \pm 0.01$ & $0.33 \pm 0.03$ & $0.35 \pm 0.03$ & $0.36 \pm 0.05$  \\
CD-24 407 & $-0.05 \pm 0.23$ & $0.12 \pm 0.18$ & $0.10 \pm 0.18$ & $0.03 \pm 0.12$ & $0.08 \pm 0.06$ & $0.04 \pm 0.03$ & $0.11 \pm 0.12$ & $-0.03 \pm 0.22$ & $0.12 \pm 0.19$ & $0.10 \pm 0.10$ & $0.09 \pm 0.13$ \\
BD+13 2269 & $-0.14 \pm 0.23$ & $0.00 \pm 0.15$ & $-0.02 \pm 0.15$ & $0.19 \pm 0.09$ & $0.08 \pm 0.06$ & $0.07 \pm 0.04$ & $0.11 \pm 0.07$ & $0.19 \pm 0.16$ & $0.01 \pm 0.27$ & $0.21 \pm 0.10$ & $0.13 \pm 0.06$ \\
BD+01 299 & $0.03 \pm 0.17$ & $-0.12 \pm 0.18$ & $-0.13 \pm 0.18$ & $0.02 \pm 0.3$ & $0.09 \pm 0.06$ & $-0.06 \pm 0.23$ & $-0.08 \pm 0.34$ & $0.00 \pm 0.22$ & \nodata & $0.19 \pm 0.40$ & $0.21 \pm 0.24$ \\
StKM 1-1526 & $-1.94$\tablenotemark{a} & $-0.40 \pm 0.08$ & $-0.43 \pm 0.08$ & $-0.31 \pm 0.10$ & $-0.39 \pm 0.04$ & $0.18 \pm 0.02$ & $0.03 \pm 0.08$ & $0.12 \pm 0.02$ & \nodata & \nodata & $0.11 \pm 0.07$ \\
TYC 5213-545-1 & $0.12 \pm 0.05$ & $0.05 \pm 0.06$ & $0.04 \pm 0.06$ & $-0.02 \pm 0.14$ & $-0.04 \pm 0.06$ & $0.03 \pm 0.04$ & $0.05 \pm 0.08$ & $0.04 \pm 0.11$ & \nodata & $-0.15 \pm 0.07$ & $0.12 \pm 0.06$ \\
LP 617-58 & $-0.28 \pm 0.06$ & $-0.15 \pm 0.04$ & $-0.17 \pm 0.04$ & $-0.28 \pm 0.07$ & $-0.09 \pm 0.03$ & $-0.05 \pm 0.01$ & $-0.21 \pm 0.05$ & $-0.48 \pm 0.07$ & \nodata & $0.83 \pm 0.05$ & $-0.08 \pm 0.04$  \\
HIP 63506 & \nodata & \nodata & \nodata & $-0.02 \pm 0.08$ & $-0.29 \pm 0.04$ & $-0.47 \pm 0.05$ & $-0.64 \pm 0.09$ & $-1.02$\tablenotemark{a} & \nodata & \nodata & $-0.45 \pm 0.15$  \\
LSPMJ0632+5053 & $0.17 \pm 0.16$ & $0.17 \pm 0.06$ & $0.14 \pm 0.06$ & $0.23 \pm 0.07$ & $0.25 \pm 0.06$ & $0.19 \pm 0.02$ & $0.27 \pm 0.06$ & $0.34 \pm 0.09$ & $0.26 \pm 0.19$ & $0.27 \pm 0.04$ & $0.34 \pm 0.09$  \\
NLTT 1011 & $0.25 \pm 0.22$ & $-0.86$\tablenotemark{a} & $-0.85$\tablenotemark{a} & $-0.82$\tablenotemark{a} & $-0.00 \pm 0.06$ & $-0.45 \pm 0.02$ & $-0.77$\tablenotemark{a} & \nodata & \nodata & \nodata & $-0.65 \pm 0.32$ \\
HD 125141 & 0.00 $\pm$ 0.18 & $0.03 \pm 0.23$ & $-0.01 \pm 0.23$ & $0.00 \pm 0.11$ & $0.06 \pm 0.04$ & $-0.10 \pm 0.05$ & $0.02 \pm 0.05$ & $-0.13 \pm 0.23$ & $-0.27 \pm 0.21$ & $-0.00 \pm 0.09$ & $-0.07 \pm 0.13$ \\
BD+06 2986 & \nodata & \nodata & \nodata & $0.22 \pm 0.16$ & $-0.27 \pm 0.08$ & $-0.41 \pm 0.08$ & $-0.72 \pm 0.17$ & $-0.40 \pm 0.05$ & \nodata & $-0.64 \pm 0.13$ & $-0.50 \pm 0.10$ \\
BD+24 4329 & $-1.17$\tablenotemark{a} & $-0.84 \pm 0.46$ & $-0.85 \pm 0.46$ & $-0.79$\tablenotemark{a} & $0.10 \pm 0.07$ & $-0.17 \pm 0.14$ & $-0.38 \pm 0.36$ & $-0.26 \pm 0.4$ & \nodata & \nodata & $0.40 \pm 0.16$ \\
BD+49 2561 & $0.02 \pm 0.15$ & $0.02 \pm 0.20$ & $0.01 \pm 0.20$ & $-0.00 \pm 0.09$ & $0.04 \pm 0.06$ & $0.01 \pm 0.07$ & $0.05 \pm 0.18$ & $0.06 \pm 0.17$ & $-0.11 \pm 0.34$ & $-0.23 \pm 0.15$ & $0.15 \pm 0.15$ \\
\enddata
\tablenotetext{a}{Upper Limit}
\end{deluxetable*}

%{\textemdash}--Abundance Ratios
\begin{deluxetable}{l|ccccc}
\tablecaption{Key Abundance Ratios and Errors for the PEPSI Sample 
\label{tab:abundance_ratio}}
\tablehead{
\colhead{Object} &    \colhead{C/O} & \colhead{Mg/Si} &\colhead{Ca/Al} & \colhead{[Y/Mg]} & \colhead{S/N}}
\startdata
HD 89744 & 0.56 $\pm$ 0.09 & 0.77 $\pm$ 0.37 & 0.72 $\pm$ 0.06 & $-0.02$ $\pm$ 0.44 & 0.41 $\pm$ 0.07   \\
HD 46588 & 0.57 $\pm$ 0.09 & 1.02 $\pm$ 0.06 & 0.77 $\pm$ 0.06 & 0.00 $\pm$ 0.14 & 0.44 $\pm$ 0.09 \\
$*$ 54 Psc & 0.56 $\pm$ 0.08  & 1.14 $\pm$ 0.08 & 0.69 $\pm$ 0.05 & $-0.16$ $\pm$ 0.09 & 0.15 $\pm$ 0.02  \\
V$^{*}$ HN Peg & 0.52 $\pm$ 0.04 & 0.90 $\pm$ 0.05 & 1.29 $\pm$ 0.09 & 0.11 $\pm$ 0.07 & 0.34 $\pm$ 0.04   \\
HD 126053 & 0.46 $\pm$ 0.08 & 1.09 $\pm$ 0.07 & 0.69 $\pm$ 0.05 & $-0.19$ $\pm$ 0.18 & 0.58 $\pm$ 0.14 \\
GJ 417 & 0.45 $\pm$ 0.07 & 0.81 $\pm$ 0.04 & 0.94 $\pm$ 0.08  & 0.12 $\pm$ 0.13 & 0.32 $\pm$ 0.02\\
HIP 9269 & 0.76 $\pm$ 0.29  & 1.44 $\pm$ 0.21 & 0.52 $\pm$ 0.11 & $-0.26$ $\pm$ 0.18 & 0.14 $\pm$ 0.04 \\ 
G171-58 & 0.54 $\pm$ 0.11 & 1.17 $\pm$ 0.09 & 0.89 $\pm$ 0.13 & $-0.11$ $\pm$ 0.12 & 0.30 $\pm$ 0.03   \\
HIP 26653 (HD 37216) & 0.56 $\pm$ 0.16 & 1.37 $\pm$ 0.12 & 0.77 $\pm$ 0.07 & $-0.06$ $\pm$ 0.13 & 0.17 $\pm$ 0.03 \\
HD 106888 & 0.38 $\pm$ 0.04 & 0.93 $\pm$ 0.05 & 0.74 $\pm$ 0.08 & 0.13 $\pm$ 0.10 &0.42 $\pm$ 0.03 \\
HD 51400 & 0.54 $\pm$ 0.11  & 1.04 $\pm$ 0.08 & 0.80 $\pm$ 0.08 & 0.04 $\pm$ 0.10 & 0.34 $\pm$ 0.08  \\
HD 203030 & 0.47 $\pm$ 0.11  & 1.04 $\pm$ 0.08 & 1.03 $\pm$ 0.07 & 0.09 $\pm$ 0.16 & 0.31 $\pm$ 0.07 \\
HD 8291 & 0.47 $\pm$ 0.04  & 0.93 $\pm$ 0.06 & 1.06 $\pm$ 0.11 & 0.15 $\pm$ 0.06 & 0.42 $\pm$ 0.04  \\
HD 16270 & 0.69 $\pm$ 0.12  & 1.09 $\pm$ 0.19 & 0.66 $\pm$ 0.14 & 0.01 $\pm$ 0.20 & \nodata \\
HD 116012 & 0.91 $\pm$ 0.21  & 1.20 $\pm$ 0.22 & 0.57 $\pm$ 0.8 & $<$ $-0.94$  & 0.31 $\pm$ 0.10  \\
BD+21 55 & 0.27 $\pm$ 0.05  & 1.05 $\pm$ 0.07 & 0.71 $\pm$ 0.07 & $-0.10$ $\pm$ 0.75  & 0.27 $\pm$ 0.01  \\
BD+60 1417 &  0.37 $\pm$ 0.07 & 0.84 $\pm$ 0.14 & 0.93 $\pm$ 0.13 & 0.14 $\pm$0.26 & \nodata    \\
StKM 2-1777 & $<$1.00  & 1.73 $\pm$ 0.37 & 0.21 $\pm$ 0.05 & $-0.06$ $\pm$ 0.03 &\nodata \\
HD 253662 & 0.32 $\pm$ 0.01 & 0.93 $\pm$ 0.05 & 0.89 $\pm$ 0.09 &0.08 $\pm$ 0.03 & 0.40 $\pm$ 0.01\\
CD-24 407 & 0.37 $\pm$ 0.11  & 0.90 $\pm$ 0.11 & 0.79 $\pm$ 0.10& $-0.08$ $\pm$ 0.22 & 0.44 $\pm$ 0.09 \\
BD+13 2269 & 0.41 $\pm$ 0.11 & 0.96 $\pm$ 0.08 & 1.10 $\pm$ 0.12 & 0.11 $\pm$ 0.16  & 0.26 $\pm$ 0.07 \\
BD+01 299 & 0.78 $\pm$ 0.19  & 1.24 $\pm$ 0.5 & 0.73 $\pm$ 0.22 & 0.01 $\pm$ 0.31 & \nodata \\
StKM 1-1526 & $<$ 1.04 & 1.47 $\pm$ 0.28 &1.04 $\pm$ 0.11 & $-0.06$ $\pm$ 0.03 & \nodata \\
TYC 5213-545-1 & 0.63 $\pm$ 0.05 & 0.99 $\pm$ 0.09 & 0.90 $\pm$ 0.13 & 0.01 $\pm$ 0.11 & \nodata \\
LP 617-58 & 0.41 $\pm$ 0.03  & 1.48 $\pm$ 0.20 & 0.56 $\pm$ 0.11 & $-0.42$ $\pm$ 0.07 & \nodata \\
HIP 63506 & \nodata  & 1.54 $\pm$ 0.40 & 1.59 $\pm$ 0.34 & $<$ $ -0.54$ & \nodata  \\
LSPM J0632+5053 & 0.57 $\pm$ 0.09 & 0.87 $\pm$ 0.05 & 0.83 $\pm$ 0.08 & 0.14 $\pm$ 0.10 & 0.41 $\pm$ 0.08   \\
NLTT 1011 & $>$0.94  & 2.64 $\pm$ 0.20 & $<-$0.88 &$<$$-0.314$ & \nodata \\
HD 125141 (G200–28) & 0.56 $\pm$ 0.16 & 0.95 $\pm$ 0.07  & 0.76 $\pm$ 0.09 & $-0.11$ $\pm$ 0.23  &  0.22 $\pm$ 0.05 \\
BD+06 2986 & \nodata &1.47 $\pm$ 0.29& 2.70 $\pm$ 0.50 & $< -0.687$ & \nodata   \\
BD+24 4329 & $< 1.79$ & 1.69 $\pm$ 0.65 & $<$1.37 &  $-0.08$ $\pm$ 0.42 & \nodata  \\
BD+49 2561 & 0.54 $\pm$ 0.13 & 0.97 $\pm$ 0.19 & 0.77 $\pm$ 0.09 &0.04 $\pm$ 0.19 & 0.55 $\pm$ 0.21\\
\enddata
\end{deluxetable}

%{\textemdash}{\textemdash}--comparision to Brewer
\subsection{Comparison to Brewer Catalog and \cite{Rice2020}}
\label{sec:brewer_compare}
Five stars in our sample~(HD 46588, HD 203030, HD 126053,  GJ 417, and HD 37216) were previously observed as part of the \cite{Brewer2016_CO} study within the California Planet Search (CPS) program (hereinafter: Brewer Catalog) and one star $^{*}$54 Psc from \cite{Rice2020}. Observations were carried out using the HIRES spectrograph on the Keck~I telescope (R $\sim$ 70,000) with a typical signal-to-noise ratio of approximately 100. The resulting spectra were analyzed in a uniform manner employing the \textit{Spectroscopy Made Easy} (SME; \citealt{Valenti1996}) package, which provided precise determinations of fundamental stellar parameters (e.g., $\teff$, $\logg$, and \( v \sin i \)) and elemental abundances for 15 species, utilizing atomic data from the Vienna Atomic Line Database (VALD-3). Species measured in the Brewer Catalog include: C, N, O, Na, Mg, Al, Si, Ca, Ti, V, Cr, Mn, Fe, Ni, and Y. Of cross interest between this work and the Brewer Catalog include the following species: C, N, O, Mg, Al, Si, Ca, Ti, Fe, and Y (see Section \ref{sec:discussion}).

\par
\begin{figure*}[h!]
    \centering
\includegraphics[width=0.85\linewidth]{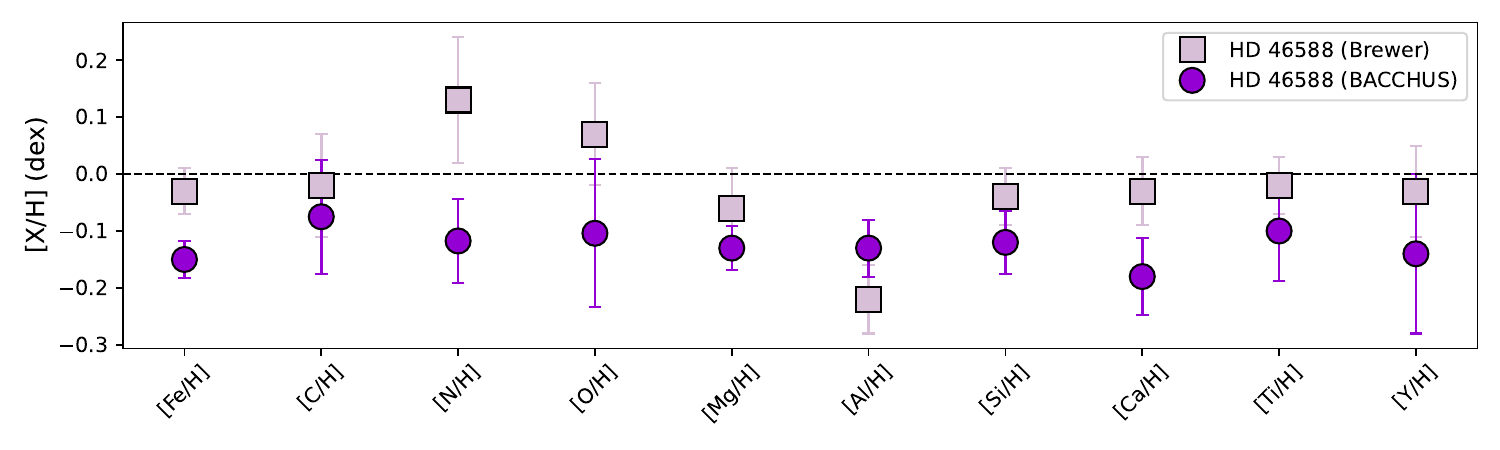}
      \includegraphics[width=0.85\linewidth]{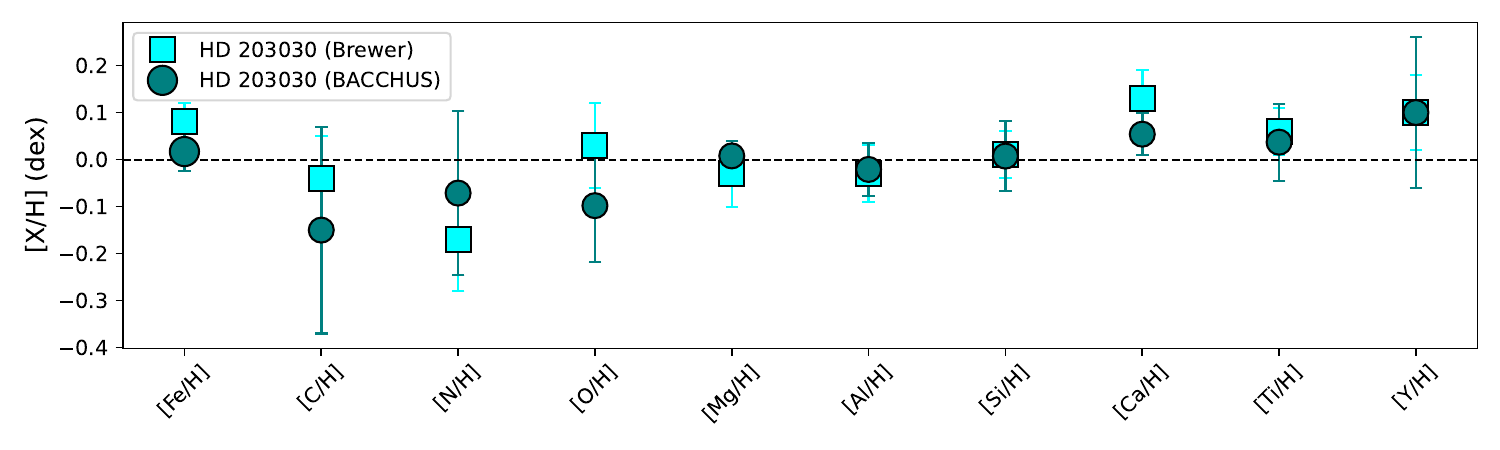}
\includegraphics[width=0.85\linewidth]{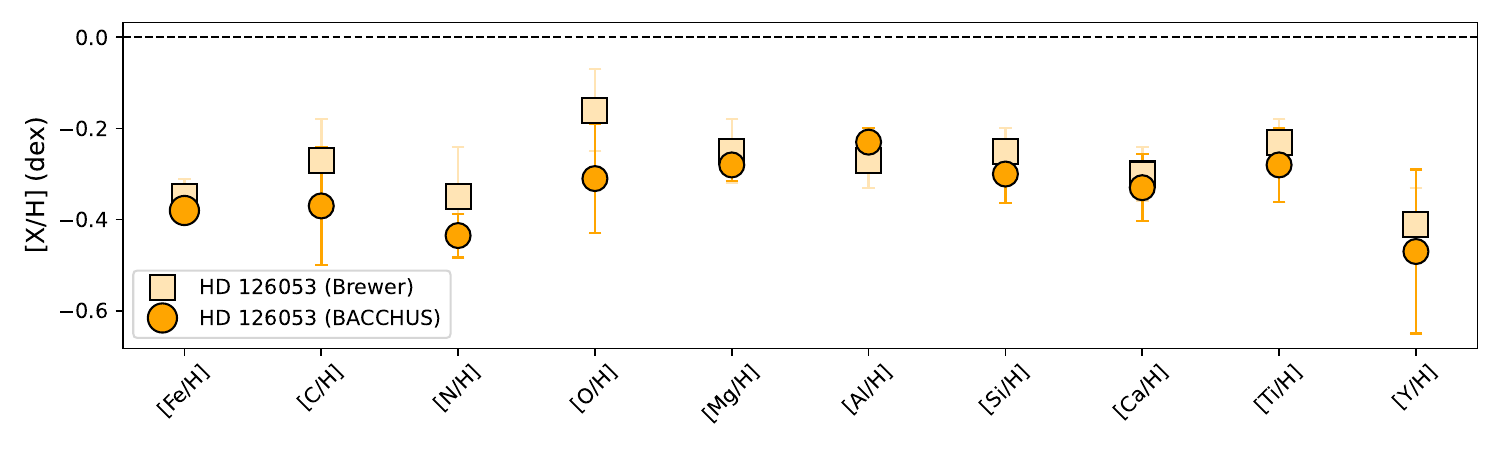}
\includegraphics[width=0.85\linewidth]{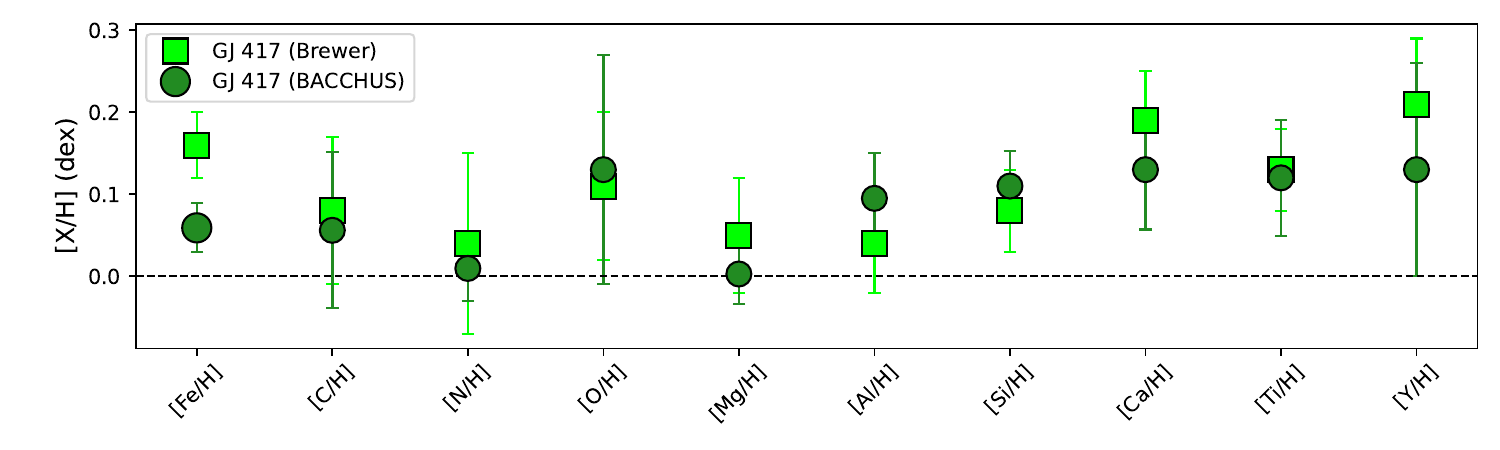}
\includegraphics[width=0.85\linewidth]{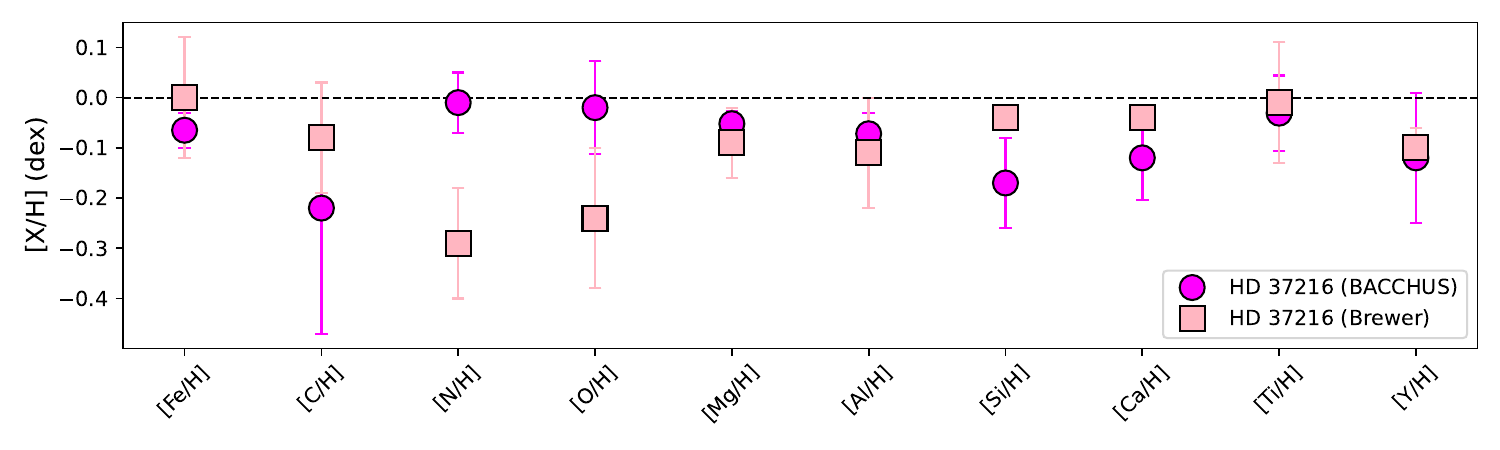}
    \caption{Elemental abundances for HD 46588 (purple) from \bacchus~(circles), and from \cite{Brewer2016}~(squares) and their associated errors. Same for HD 203030 (blue), HD 126053 (orange), GJ 417~(green), and HD 37216 (pink). }
    \label{fig:brewer_comparision}
\end{figure*}

\begin{figure*}[h!]
    \centering
    
    \includegraphics[width=0.85\linewidth]{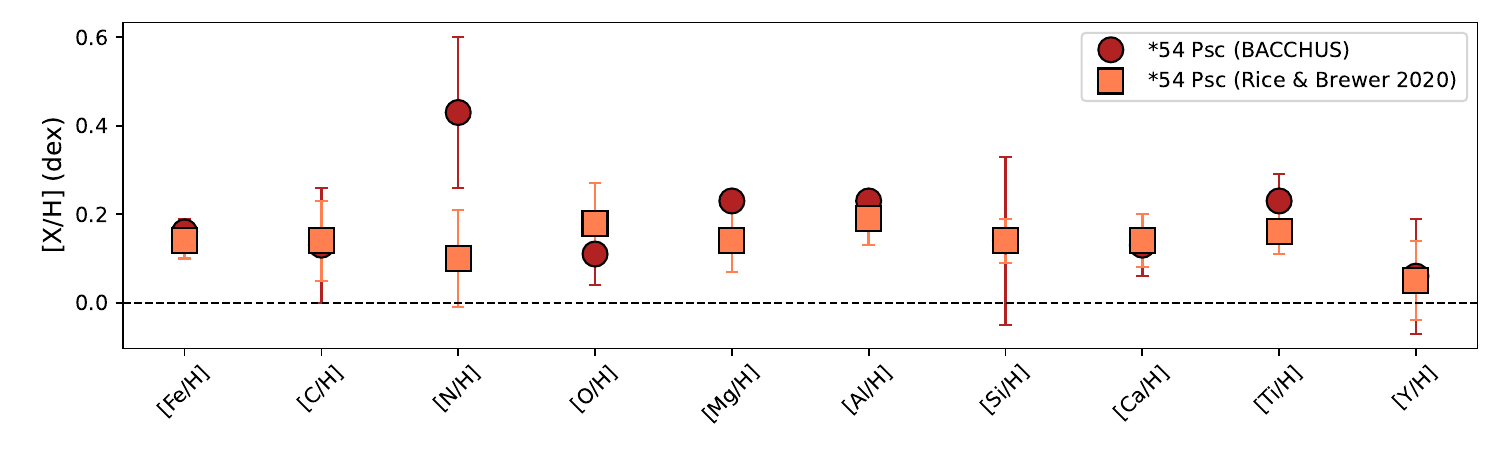}
    \caption{Same as Figure \ref{fig:brewer_comparision} abundances for $^*$54 Psc from \bacchus~(red circle) and \citet{Rice2020}~(light red squares)}
    \label{fig:mrice_2020_54Psc}
\end{figure*} 

For the five stars in \cite{Brewer2016}, we find generally good agreement between our abundances derived from \bacchus~and the Brewer Catalog~(Figure \ref{fig:brewer_comparision} and \ref{fig:mrice_2020_54Psc}). We note a systematic offset of [O/H] and [Y/H]. In \cite{Brewer2016}, NLTE corrections were not applied to their sample. However, we note that the original discrepancy for [O/H] and [Y/H] is caused by an offset from the PEPSI data, as discussed in \S \ref{sec:test_sun_pepsi}. Two of our stars, HD 46588 and HD 37216 have the biggest offset in [N/H] and [O/H], where our values are sub-solar compared to the near-solar values from \cite{Brewer2016}. For $^*$54 Psc, we find close agreement between our abundance values and \cite{Rice2020}, with the most notable offset for the [N/H] abundance~(Figure \ref{fig:mrice_2020_54Psc}). 

\par
\cite{Brewer2016} uses 12 OH and \ion{O}{1} lines and 7 nitrogen~(\textit{private communication}) lines in addition to the molecular lines (CN), and does not use any NLTE corrections for oxygen. The Brewer Catalog and \cite{Rice2020} differs from \bacchus~in several ways, including atomic data source, determination of $\teff$, $\log$ and [Fe/H], microturbulence determination and abundance determination methodology. These various factors can impact our final abundance determination.

\par

Typical uncertainties in the stellar parameters derived from the \citet{Brewer2016} catalog are on the order of $\sim$60~K in effective temperature, $\sim$0.15~dex in surface gravity, and $\sim$0.06~dex in metallicity.
For HD 203030, the $\teff$ = $5441 \pm 60$ K, $\logg$ = $4.51 \pm 0.15$ and [Fe/H] = $0.08 \pm 0.06$~\citep{Brewer2016}. Our values for HD 203030 are $\teff$ = $5524 \pm 70$ K, $logg$ = $4.57 \pm 0.33$, and [Fe/H] = $0.01 \pm 0.09$. For HD 46588 from the Brewer Catalog, the $\teff$ = $6145 \pm 60$ K, $\logg$ = $4.30 \pm 0.15$ and [Fe/H] = $-0.05 \pm 0.06$. Our values for HD 46588 are $\teff$ = $6093 \pm 53$ K, $logg$ = $4.29 \pm 0.31$, and [Fe/H] = $-0.03 \pm 0.03$. HD 126053 has a $\teff$ = $5714 \pm 60$ K, $\logg = 4.54 \pm0.15$, and [Fe/H] = $-0.35 \pm 0.06$ from the Brewer Catalog and our values are $\teff = 5612 \pm 42$ K, $\logg = 4.30 \pm 0.40$, and [Fe/H] = $-0.38 \pm 0.02$. Lastly, from the Brewer Catalog for GJ 417, $\teff = 5905 \pm 60$ K, $\logg = 4.47 \pm 0.15$ and [Fe/H] = 0.16 $\pm$ 0.06 and our values are $\teff = 5779 \pm 42$ K, $\logg = 4.42 \pm 0.28$, and [Fe/H] = $0.05 \pm 0.03$. 

\par

\begin{figure*}
    \centering
    \includegraphics[width=\linewidth]{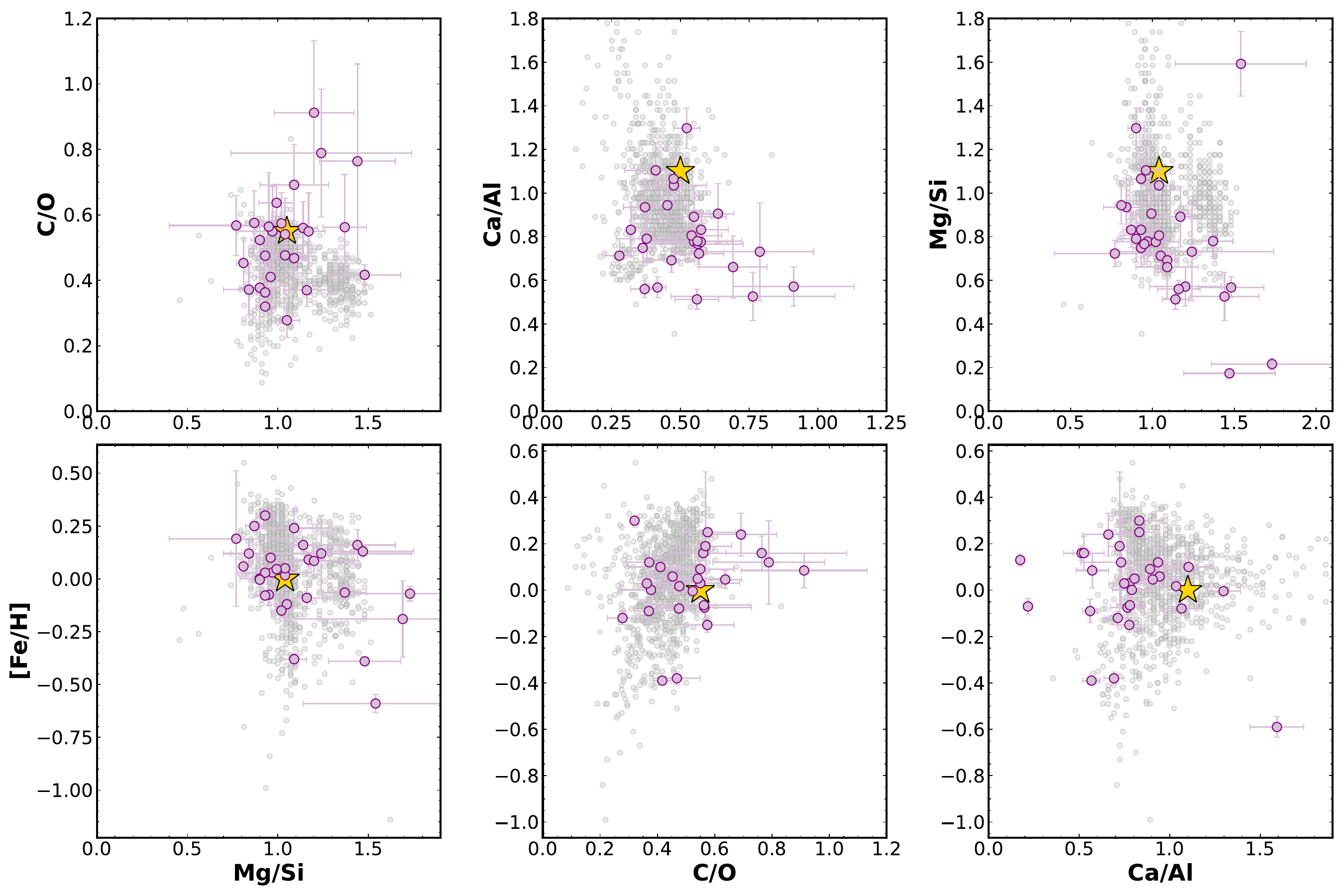}
    \caption{Abundance ratios of stars in the solar neighborhood from \cite{Brewer2016,Brewer2016_CO}, plotted following Figure 1 from \cite{Calamari_2024} with our PEPSI sample (purple points) overlaid from Table \ref{tab:obs}. The Sun's values are marked with a gold star~\citep{Grevesse2007}. }
    \label{fig:ratio_compare_solarneighborhood}
\end{figure*}

In Figure \ref{fig:ratio_compare_solarneighborhood}, we show abundance ratios for C/O, Mg/Si, and Ca/Al, and [Fe/H]~(metallicity) for our host stars and the local solar neighborhood from \cite{Brewer2016}, following restrictions in \cite{Calamari_2024}. This comparison places our sample in the broader chemical context of nearby FGK stars analyzed in a homogeneous manner. This result highlights the spread in abundance ratios for our sample and the solar neighborhood. The distributions demonstrate that our host stars span a range in chemical abundances comparable to that of the solar-neighborhood population, encompassing both sub-solar and super-solar metallicities, alongside a wide dispersion in C/O and Mg/Si ratios. Our host stars predominantly occupy the central region of the solar-neighborhood distribution, suggesting that they are chemically typical of nearby stars rather than representing extreme deviations or chemically peculiar set. We note that when our sample deviates from other solar neighborhood values of \cite{Brewer2016}, we begin to see an increase in the error bars on our ratios., mainly C/O and Mg/Si ratios. Our determined Mg/Si and C/O are closer to solar values than our Ca/Al values. 

\par

\subsubsection{C/O and Mg/Si from Brewer Catalog}

\par
We use uncertainties and parameters found from \cite{Brewer2016}, where uncertainties in the C/O values are $\pm$10\% and Mg/Si values are $\pm$3.3\%. We find the following in  \cite{Brewer2016_CO}: HD 203030 has a reported C/O = 0.45 $\pm$ 0.04, and Mg/Si = 0.95 $\pm$ 0.03,  HD 126053 has a C/O = 0.41 $\pm$ 0.04 and Mg/Si = 1.04 $\pm$ 0.03, and GJ 417 (HD 97334) has a C/O = 0.50 $\pm$ 0.05 and Mg/Si = 0.98 $\pm$ 0.03. From our analysis  our Mg/Si ratio for HD 203030 is within 1$\sigma$ of the Brewer Catalog (1.14 $\pm$ 0.08 vs 0.98). We find that HD 126053 has a Mg/Si = 1.09 $\pm$ 0.07 which is very close to the Brewer Catalog value (1.04 $\pm$ 0.03). Similarly, our derived Mg/Si for GJ 417 agrees within uncertainties. Our derived C/O ratios for HD 203030, HD 126053, and GJ 417 is within 1$\sigma$ of the Brewer Catalog.

\subsection{Comparison to Literature}

We examine the literature for our sample that was not examined in \citet{Brewer2016_CO, Rice2020}. We focus on stars with previously reported C/O and Mg/Si measurements and below we highlight a few examples. 

\par

In \cite{Luck_2018, Luck2017}, abundances for carbon, oxygen, magnesium and silicon are presented for V$*$ HN Peg and HIP 9269 (HD 12051) with the Sandiford spectrograph and the Hobby–Eberly telescope at the McDonald Observatory. The spectral lines used to determine C/O were C\textsc{I} at 5052\,\AA\ and 5380\,\AA, C$_2$, O\,\textsc{I} at 6155\,\AA, and [O\,\textsc{I}] at 6300\,\AA -- where LTE is assumed. For V$*$ HN Peg, [C/H] =$-0.10 \pm 0.09$, and [O/H] = $0.07 \pm 0.09$, which leads to a C/O = 0.37 $\pm$ 0.03 using solar normalizations taken from \citet{Asplund2009}. This value is lower than our C/O of $0.52\pm0.04$ from \bacchus. The [Mg/H] and [Si/H] are $0.10\pm0.02$ and $0.05\pm0.01$ respectively., leading to Mg/Si = $1.38\pm0.22$, which is higher than our Mg/Si =  0.90 $\pm$ 0.05. The [Fe/H] of 0.01 $\pm$ 0.01 matches our values of $0.00 \pm 0.01$. The $\teff$ = 5972.0 K from \cite{Luck2017}, is close to our value of $5927 \pm 1$ K.

HIP 9269 has a C/O value of 0.40, from $<C> = 8.58$ and $<O> = 8.98$  compared to our C/O value of 0.76 $\pm$ 0.29, a 2$\sigma$ difference.  The [Mg/H] = $0.37\pm0.07$ and [Si/H] = $0.30 \pm 0.06$, which leads to a Mg/Si = $1.44 \pm  0.13$~\citep{Luck2017}. This is in line with our Mg/Si value for HIP 9269 of 1.44 $\pm$ 0.21. The [Fe/H]  = $0.24\pm0.04$~\citep{Luck2017} is within 1$\sigma$ of our values for [Fe/H] which is $0.16 \pm 0.07$.

% the carbon abundance $<C> = 8.26$ and mean oxygen abundance $<O> = 	8.69$, with normalization taken from \cite{Asplund2009}, this leads to a C/O ratio of 0.37, which is lower than our C/O of $0.52\pm0.04$. 

\par 

HD 89744 has multiple prior determinations of its atmospheric parameters reported in the literature along with C/O and Mg/Si ratios~\citep{Delgado2010}.  \citet{Delgado2010} find stellar parameters of $\teff$ = 6234 K, $\logg$ = 3.98, and [Fe/H] = 0.22. They also find [O/H] = 0.00, [C/H] = 0.17, [Mg/H] = 0.16 and [Si/H] = 0.26. The most recent findings  with \texttt{MOOG} in \cite{Morris_2019}, show $\teff$ =  6070 $\pm$ 110, $\logg$ 4.18 $\pm$ 0.05, and [Fe/H] = 0.26. Our stellar parameters are within 2$\sigma$ for $\teff$ and [Fe/H]. \cite{Delgado2010} report a C/O ratio of 0.98 and Mg/Si ratio of 0.85, a notable contrast to our derived C/O = 0.54 but within 1$\sigma$ for our derived Mg/Si = 0.77 $\pm$ 0.37. There is a lack of literature information on how these ratios were determined, making exact comparisons difficult.

\par
% \subsection{BD+13 2269}

BD+13 2269 has few literature measurements of its stellar parameters. The most recent include measurements are presented in \cite{Rothermich2024, Zong2020,Wang_2020}. $\teff$, $\logg$, and 11 elements were measured from the Large Sky Area Multi-Object Fiber Spectroscopic Telescope (LAMOST) Medium Resolution~(R $
\sim$ 7500) Survey data sets with a machine learning method. \cite{Wang_2020,Zong2020} found $\teff$ = 5542, $\logg$ = 4.60,  [Fe/H] = 0.22 $\pm$ 0.01, and C/O = 1.40 and Mg/Si = 0.79. The [Fe/H] is about 2$\sigma$ from our value, in addition our $\teff$ = 5480 $\pm$ 66 ($>$3$\sigma$), $\logg$ = 4.80 $\pm$ 0.33 (within 1$\sigma$). We find our Mg/Si value to be 0.96 $\pm$ 0.18 and C/O to be 0.41. The Mg/Si ratio is within 1$\sigma$ of the literature value, but our derived C/O ratio is $>$3$\sigma$ from literature value.
\par

Previous abundance measurements for BD+60 1417 (K0) use \texttt{iSpec}~\citep{BlancoCuaresma2019, Phillips2024} and report a substellar C/O = 0.23 $\pm$ 0.12 and enhanced Mg/Si = 1.41 $\pm$ 0.19. Compared to \cite{Phillips2024}, we find a C/O of 0.33 $\pm$ 0.07 compared to 0.23 $\pm$ 0.12 (within 1$\sigma$).  Contrarily, we do not report an enhanced Mg/Si ratio 0.84 $\pm$ 0.14 vs 1.41 $\pm$ 0.19. We investigate if our abundances discrepancy is due to different choices solar normalization of \cite{Asplund2009}(hereinafter;  ASP2009 vs \cite{Grevesse2007} (hereinafter; GR2007). We determine our abundances using ASP2009 and GR2007 solar abundances values. We find that our determined abundances are within 1$\sigma$ of each other from the two different normalizations. Thus we rule out discrepancy due to solar normalization choices.

\par

%{\textemdash}{\textemdash}Discussion Section
\section{Discussion}

With our parameters and abundances in hand, we can investigate a number of aspects of the benchmark brown dwarf systems. In this section, we discuss the implications of our results.
\label{sec:discussion}
\subsection{Metallicity}

The comparison between the metallicity of brown dwarfs and their host stars provides a powerful diagnostic of formation pathways~\citep{Xuan_2024, Meynardie2025,Ma2014,Zhang2023, Adams_2023}. If a brown dwarf forms through the star-like route of gravitational collapse and fragmentation of a molecular cloud, it should share the same [Fe/H] as its host star, reflecting a common origin in the same natal environment~\citep{Hsu2024, Hawkins2020, Sinha2026}. This expectation contrasts with the formation of planets, which exhibit a strong dependence on host star metallicity. It is now well established that giant planets are more commonly found orbiting metal-rich stars, known as the planet metallicity correlation~\citep{Fischer_2005}. From observational studies, brown dwarf host star-metallicity correlation is not as strongly correlated as planet-metallicity. Stellar hosts of giant planets are more metal-rich, in contrast, stars that host brown dwarfs have compositions resembling those of stars with low-mass planets; $\alpha$-elements ([$X_{\alpha}/Fe$], which include Mg, Si, Ca , and Ti), and iron-peak elements ([$X_{Fe}/Fe$], which include Cr, Mn, Co, and Ni) appear solar, compared to the $\sim$+0.15 dex peak in giant-planet hosts~\citep{Fischer_2005,Ma2014,Sahlmann2011,Maldonado2017,Mata2014}. Taken together, these results indicate the absence of a statistically significant metallicity correlation for brown dwarf host stars.

\par

\begin{figure*}
\includegraphics[width=\textwidth]{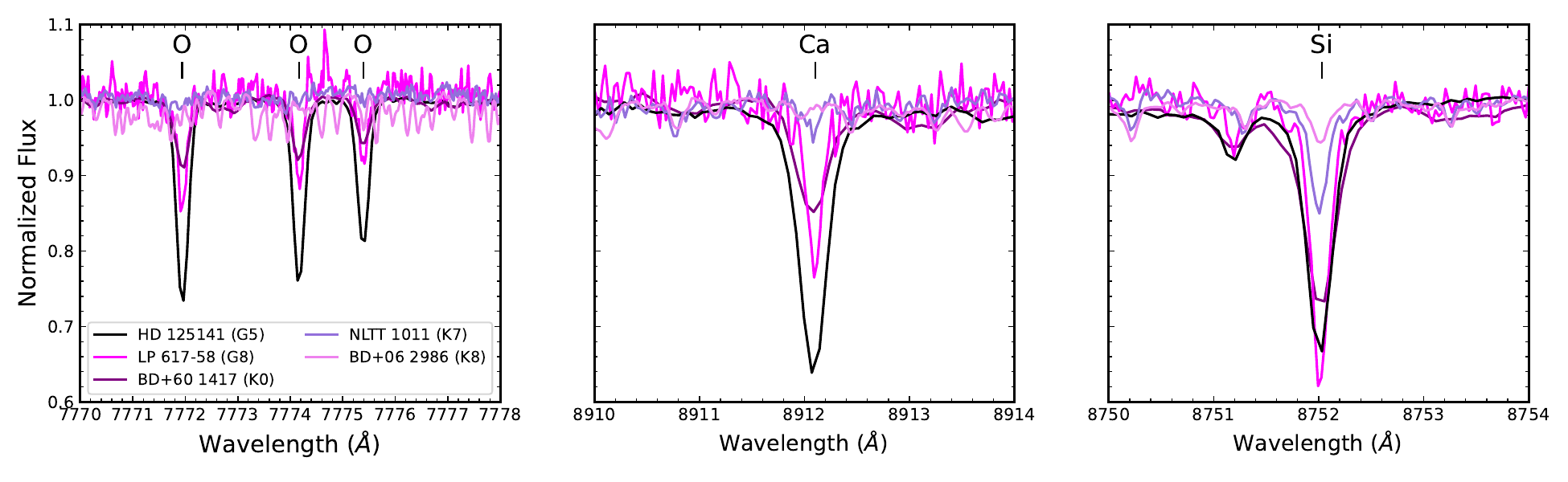}
    \caption{PEPSI spectra of stars of different\textbf{ spectral types},  HD 125141 (G5), LP 617-58 (G8), BD+60 1417 (K0), NLTT 1011 (K7) and BD+06 2986 (K8) in selected wavelength regions of 7770–7778 Å (left), 8910-8914 Å (center), and 8750-8754 Å (right). Line features used in the abundance analysis have been labeled.}
    \label{fig:compare_metallicity_odd}
\end{figure*}

For most of our stars, we find a near solar metallicity. However, compared to the current population of host stars of directly-imaged planets~\citep{Baburaj2025a,Baburaj2025b}, our spread in [Fe/H] for wider-orbit brown dwarfs is larger~(Figure \ref{fig:Distance_vs_co_metallicity}). The larger spread is in part due to a number of stars that are metal-poor: HIP 63506 (K6), LP 617-58 (G8), NLTT 1011 (K7), and BD+06 2986 (K8). In metal-poor stars, fewer heavy elements lead to weaker absorption features, making abundance determination difficult. For our metal poor stars we see this effect for a number of lines including \ion{O}{1}, \ion{Ca}{1}, and \ion{Si}{1}, as demonstrated in Figure \ref{fig:compare_metallicity_odd}. For a number of these metal-poor stars, we note enhanced Mg/Si ratios, in particular, NLTT 1011 and BD+06 2986 with Mg/Si = 2.47 $\pm$ 0.20 and 1.47 $\pm$ 0.29 respectively have silicon depletion, [Si/H] = $-0.77$ (upper limit) and $-0.72 \pm0.10$ respectively. 

\par

To investigate the validity of our finding that some of our stars are metal-poor, we examined previous measurements in the literature for both the host star and its companion. The companion T dwarf, ULAS J150457.65+053800.8, to BD+06 2986 is classified as a metallicity benchmark subdwarf with [M/H] = $-0.47^{+0.09}_{-0.10}$ \citep{Burgasser_2025,Zhang2019}. We find a host star metallicity ([Fe/H] = $-0.82 \pm 0.15$), along with previous literature measurements that support a sub-solar metallicity environment. From the MILES survey,  \cite{Garcia_Perez2021,Prugniel2011} find a [Fe/H] = $-1.32 \pm0.0$ and $-0.83 \pm 0.15$ respectively. Previous works, find a [Fe/H] $\approx$ $-0.38$, which is closer to the metallicity for the brown dwarf companion and supports a metal-poor environment~\citep{Arensten2019}. We report a sub-solar metallicity for LP 617-58 of [Fe/H] =  $-0.39 \pm 0.01$, which is in line with the spectroscopic literature value reported of [Fe/H] =  $-0.36$ $\pm$ 0.04~\citep{Xiang_2019} from low-resolution~(R$\sim$1800) spectra from LAMOST DR5. The photometric [Fe/H] from \textit{Gaia} DR3 for LP 617-58 is sub-solar, $-0.23\pm0.01$. Two of our other metal-poor stars, NLTT 1011 and HIP 63506, do not have reported literature measurements for comparison. 
\par

\subsection{Silicate Cloud Prediction from Host Star Chemistry}

As brown dwarfs cool and evolve from spectral classes M to Y, they experience dynamic changes in the atmospheric properties and subsequent cloud formation. L-dwarfs have a typical temperature regime of $\sim$ 1300 $-$ 2200 K, which favors the formation of atmospheric condensates~\citep{Allard2001}. These clouds are primarily composed of silicate species such as enstatite, forsterite, quartz, SiO, along with aluminum oxides, like corundum,  and Fe in the lower layers. 
\par
Both Ca and Al are highly refractory elements that condense into solid grains at high temperatures, primarily as Al$_2$O$_3$ (corundum) and CaTiO$_3$ (perovskite)~\citep{sun_caal,Wakeford2016}. Although their overall abundances are lower than more dominant refractories such as Mg, Si, and Fe, the relative abundance of Ca to Al can influence which condensates dominate at a given condensation temperature. A lower Ca/Al ratio favors Al-bearing condensates such as corundum, which act as efficient cloud seeds and can lead to thicker, more optically dense cloud decks~\citep{Allard2001,Helling2008}. 
In contrast, a higher Ca/Al ratio promotes the early formation of Ca-bearing species such as perovskite and Ca$_2$Al$_2$SiO$_7$ (gehlenite), which accelerate the depletion of Ti from the gas phase and thus influence the disappearance of TiO and VO bands during the spectral transition from M to L dwarfs~\citep{Lodders2002,Helling2008,Woitke2020}. While condensation temperature primarily determines the sequence in which species solidify, variations in Ca/Al ratios provide context for modeling early cloud formation, cloud opacity, and spectral diversity in brown dwarf atmospheres.

\par
From the theoretical chemical framework presented in \cite{Calamari_2024}, we can predict cloud species of brown dwarfs based on FGK host star chemical abundances for well-known silicates enstatite (MgSiO$_{3}$), forsterite (Mg$_{2}$SiO$_{4}$) and quartz (SiO$_{2}$). The relative Mg/Si abundance ratio and prediction of cloud species are as follows:
\begin{itemize}
    \item Mg/Si $\lesssim$ 0.9 : Enstatite + Quartz
    \item Mg/Si  $\sim$ 0.9 : Enstatite
    \item Mg/Si $\gtrsim$ 0.9 : Enstatite + Forsterite
\end{itemize}

We use the chemical network from \cite{Calamari_2024} to predict the formation of cloud species of our sample (Table \ref{tab:cloud_species}). Generally, we find that our Mg/Si ratios are roughly solar (1.05; \cite{Grevesse2007,Asplund2009}) or higher.
This would result in the production of mainly enstatite and forsterite clouds for our sample. There are, however, a number of systems with Mg/Si that predict SiO$_{2}$ or quartz clouds. Future retrieval analyzes could provide deeper insight into the potential presence of SiO$_{2}$ and contribute to the expanding catalog of substellar objects exhibiting SiO$_{2}$ clouds~\citep{Burningham2021}.

\par

\begin{figure*}
\includegraphics[width=\textwidth]{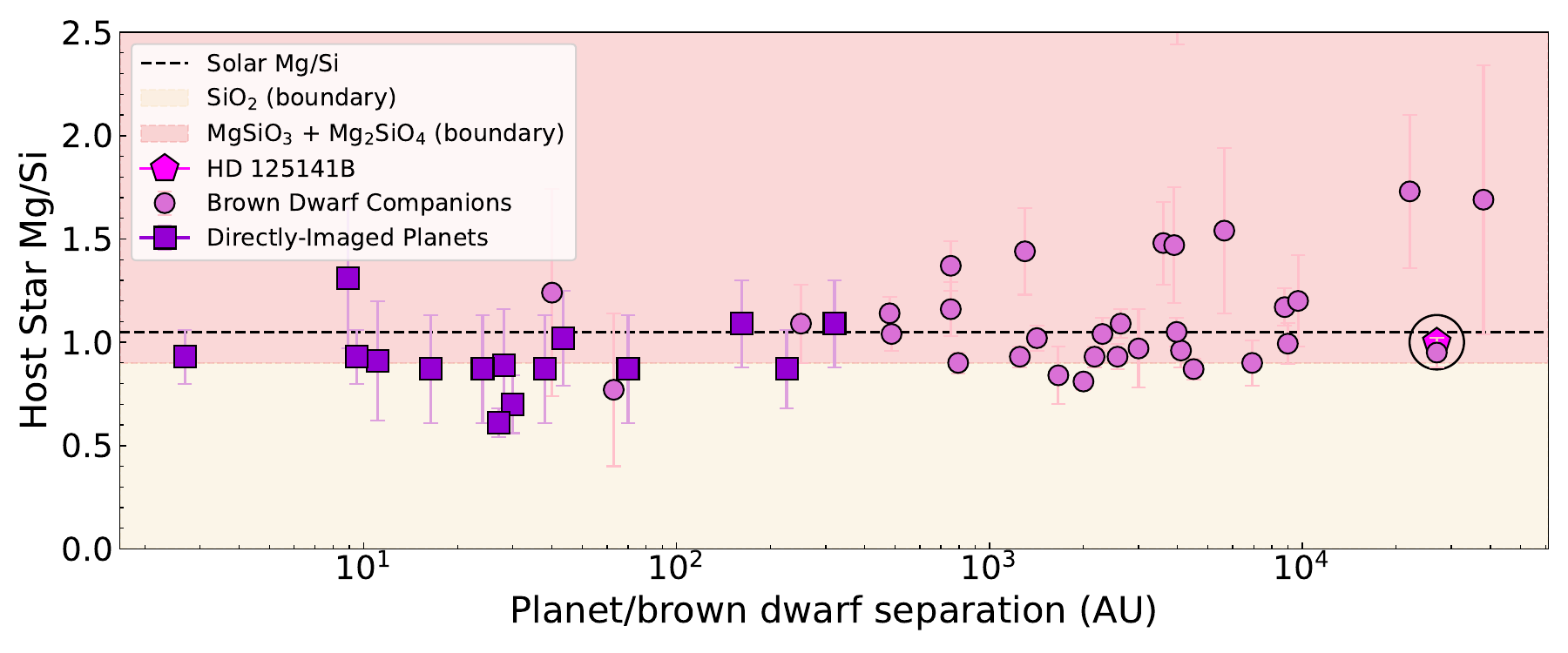}
    \caption{Host star Mg/Si ratios for directly-imaged planets (purple squares) \textbf{and} brown dwarf companions in this sample with FGK hosts. The dashed line represents solar Mg/Si ratio. We denote the region for SiO$_{2}$ cloud predictions and Mg$_{2}$SiO$_{4}$ + MgSiO$_{3}$ from \cite{Calamari_2024}. We highlight a recent retrieval results from  Kecskem\'ethy et al. in prep that highlights the similarities between host star Mg/Si and inferred Mg/Si of the brown dwarf companion to HD 125141. }
    \label{fig:compare_MgSi_host}
\end{figure*}

We compare the host star Mg/Si ratios for current wide-orbit directly-imaged planet host stars and our sample of host stars of brown dwarfs in Figure \ref{fig:compare_MgSi_host}. We find that the median Mg/Si ratio for our host star sample is $\approx$1.04, indicating the primary prediction that enstatite and forsterite clouds might form in the companion. We find that the average Mg/Si for the directly-imaged planets host stars is $\approx$ 0.89, which predicts enstatite + quartz clouds formation.

\subsubsection{Testing Host-Brown Dwarf Companion Mg/Si Connections}
\label{sec:test_MgSi}

In this work, we adopt the hypothesis that brown dwarf companions broadly reflect the elemental abundance patterns of their host stars, as predicted by phase-equilibrium condensation models~\citep{Lodders_2003, Lodders2002, Marley2002}. However, this assumption remains largely untested. The advent of JWST enables detailed atmospheric retrievals that can constrain cloud composition and refractory element ratios~\citep[e.g.,][]{Hoch2025, Zhang2025_elpis2}.   
Recent works have begun to explore the question of whether brown dwarf and giant exoplanets inherit their host star Mg/Si chemistry~\citep{Calamari_2024, Calamari2026}. Establishing observational constraints on refractory ratios in substellar atmospheres is therefore a critical step towards testing this phase equilibrium connection. \cite{Burningham2021} used the column density~($N$) retrieved from the magnesium and oxygen bearing cloud species to infer Mg/Si ratios of a brown dwarf companion.
Using the \textit{Brewster} framework~\citep{Burningham2021}, Kecskem\'ethy et al. in prep recently found that the companion to HD 125141, has an inferred Mg/Si ratio of 1.00, which is in close proximity to our determined Mg/Si ratio of 0.95 $\pm$ 0.07, which suggests chemical similarities between the host and the brown dwarf companion. 14/32 of our host star sample have current or incoming JWST NIRSpec/PRISM \& MIRI/LRS data for the brown dwarf companion, which will enable detailed studies of silicate-based cloud properties. Future retrieval analysis with extended wavelength coverage enabled by this JWST, will allow for tests of whether these brown dwarf systems show similar Mg/Si chemical composition with their host stars, as recently found with an ultra-hot Jupiter ~\citep{Sanchez2025}

\begin{deluxetable}{l|cc}[h!]
\tablewidth{\columnwidth}
\tablecaption{Prediction of cloud species to form in brown dwarf companion from Mg/Si ratio of host star.}
\label{tab:cloud_species}
\tablehead{
\colhead{Object} & \colhead{Mg/Si} &  \colhead{Cloud Species Predicted}
}
\startdata
$*$ 54 Psc & 1.14 & Enstatite + Forsterite \\
BD+01 299 & 1.24 & Enstatite + Forsterite  \\
BD+13 2269 &  0.96 & Enstatite + Forsterite  \\
BD+49 2561 &  0.97 & Enstatite + Forsterite\\
BD+ 60 1417$^{\diamond}$ & 0.84 &Enstatite + Quartz\\
BD+21 55 & 1.05 & Enstatite + Forsterite \\
CD-24 407 & 0.90  &  Enstatite \\
G171-58 & 1.17  & Enstatite + Forsterite\\
GJ 417$^{\diamond}$ &  0.81& Enstatite + Quartz  \\
HD 106888 & 0.93 & Enstatite + Forsterite\\
HD 116012 &  1.20 & Enstatite + Forsterite \\
HD 125141  & 0.95 & Enstatite + Forsterite  \\
HD 126053 & 1.09 &  Enstatite + Forsterite \\
HD 16270 & 1.09 & Enstatite + Forsterite \\
HD 203030$^{\diamond}$ & 1.04 & Enstatite + Forsterite\\
HD 253662$^{\diamond}$ & 0.93 & Enstatite + Forsterite \\
HD 46588$^{\dagger}$ & 1.02 &  Enstatite + Forsterite \\
HD 51400 & 1.04  & Enstatite + Forsterite \\
HD 8291 & 0.93 &Enstatite + Forsterite  \\
HD 89744$^{\dagger}$ & 0.77 &Enstatite + Quartz  \\
HIP 26653$^{\dagger}$ &  1.37 &Enstatite + Forsterite \\
HIP 63506 & 1.54 & Enstatite + Forsterite \\
V$^{*}$ HN Peg$^{\dagger}$ & 0.90 &Enstatite   \\
HIP 9269 &  1.44 &  Enstatite + Forsterite\\
StKM 2-1777 & 1.73 & Enstatite + Forsterite  \\
StKM 1-1526 &1.47  &  Enstatite + Forsterite   \\
TYC 5213-545-1 & 0.99 & Enstatite + Forsterite \\
LP 617-58 & 1.48 & Enstatite + Forsterite  \\
LSPM J0632+5053 & 0.87 & Enstatite + Quartz  \\
NLTT 1011$^{\dagger}$ &  2.64 & Enstatite + Forsterite   \\
BD+06 2986 &  1.47 & Enstatite + Forsterite \\
BD+24 4329 & 1.69 & Enstatite + Forsterite   \\
BD+49 2561 & 0.97 & Enstatite + Forsterite  \\
\enddata
\tablecomments{$^{\diamond}$ These targets are observed and/or scheduled to be observed as a part of JWST programs with NIRSpec and/or MIRI instruments (GO \#8140)}
\tablecomments{$^{\dagger}$ These targets are observed and/or scheduled to be observed as a part of JWST programs with NIRSpec and/or MIRI instruments (GO \#3670)}

\end{deluxetable}

%{\textemdash}C/O Ratio Discussion and Oxygen Sinks

\subsection{C/O Ratio: Formation and Oxygen Sinks}

The carbon-to-oxygen (C/O) ratio is widely considered a key diagnostic of substellar formation pathways~\citep{Oberg2011,Oberg2023}. Comparison of the C/O ratio of a brown dwarf to that of its host star can offer valuable insight into the origin of the companion~\citep{Xuan2022,Wang2022,Meynardie2025, Adams_2023}. If the brown dwarf formed via gravitational collapse of a molecular cloud or through gravitational instabilities within a protoplanetary disk, processes that involve well-mixed material, its C/O ratio is expected to resemble that of the host star~\citep{Ahmad_2025,Phillips_2024, Bate2002, Vowell2025}. Similarly, a stellar-like C/O ratio is anticipated for planets that form inside the snowline of both water and carbon grains. In contrast, a superstellar C/O ratio may suggest that the companion's atmosphere formed primarily from gas accreted beyond the water snowline, consistent with core-accretion scenarios. The formation location and possible migration history of the companion can be inferred by comparing its current position to the ice lines of key molecules such as water (H$_2$O), carbon dioxide (CO$_2$), and carbon monoxide (CO) in the protoplanetary disk~\citep{Oberg2011, Oberg2023}.

\par

Potential trends with the C/O ratio have been investigated in the literature, and at least for directly-imaged planets, the C/O ratio appears consistently near solar, and potential trends between C/O and separation have been identified~\citep{Hock2023}. Recent work has focused on the C/O ratio of brown dwarf companions, and current trends suggest near-solar composition~\citep{Costes2024,Hsu2024,Xuan_2024,Kammerer2025}. While some brown dwarfs have been found to have super-solar C/O ratios~\citep{Gaarn2023,Calamari2022}, these have recently been rectified with the discoveries of binarity for Gliese 229B (Gliese 229Ba \& Gliese 229Bb) and updated line lists for analysis of Ross 458c with JWST/NIRSpec~\citep{Xuan2024Nature,Meynardie2025}.

\begin{figure*}
    \centering    
    \includegraphics[width=\textwidth]{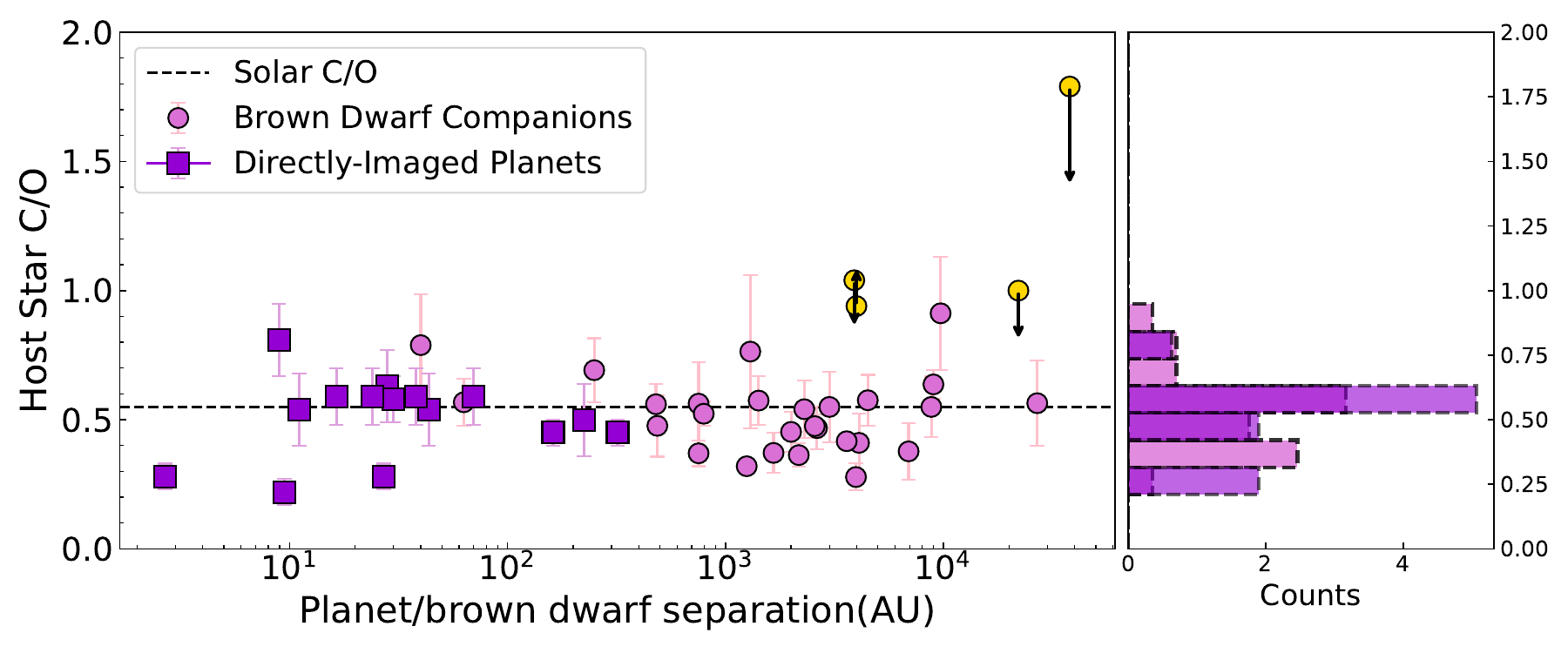}
    \includegraphics[width=\textwidth]{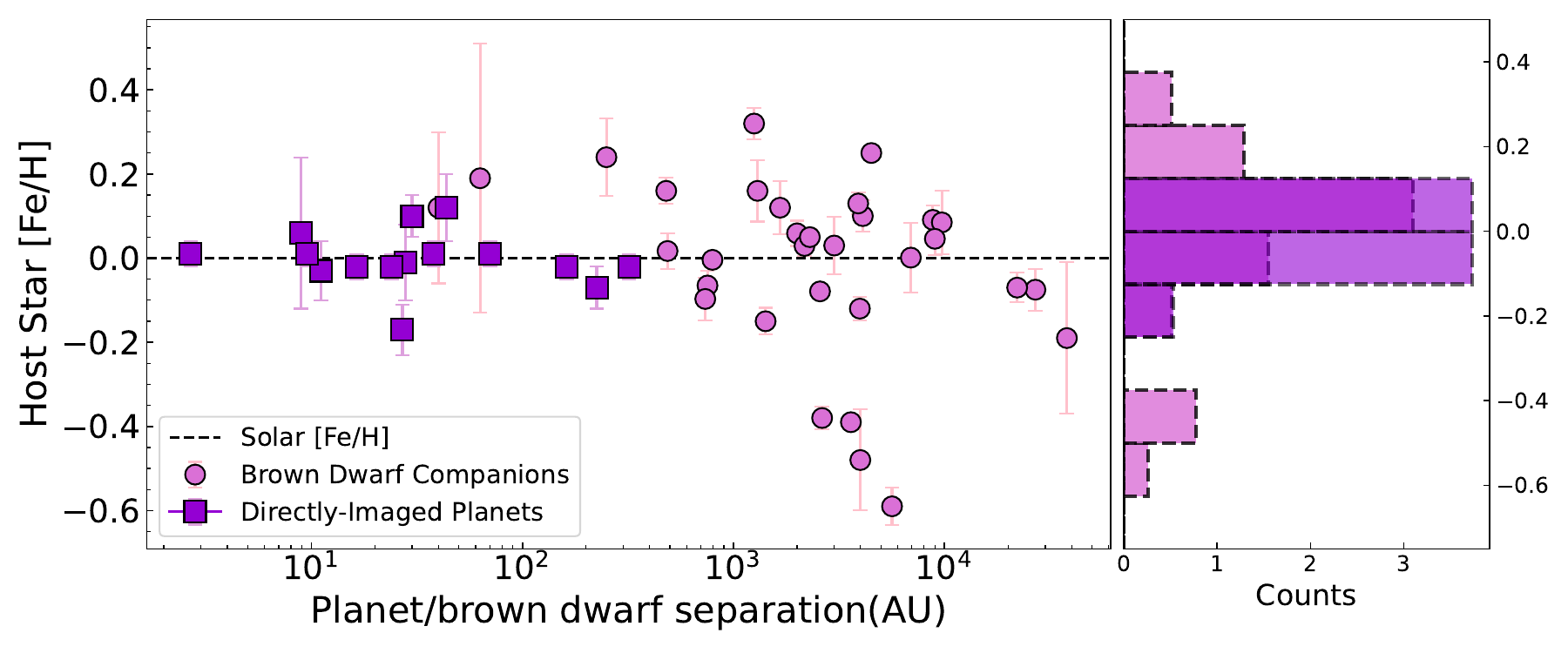}
    \caption{\textbf{Top:} Host star C/O ratios for directly-imaged planets (purple squares) versus brown dwarf companion (pink circles) in this sample with FGK hosts. Host stars of brown dwarf companions that only have an upper limit on the C/O ratio are shown in gold circles. The dashed black line indicates the solar C/O ratio. \textbf{Bottom}:  Host star metallicity ([Fe/H]) ratios for directly-imaged planets (purple squares) versus brown dwarf companion (pink circles) in this sample with FGK hosts. The dashed black line indicates the solar metallicity ratio. Directly-imaged planets are a compilation from \cite{Baburaj2025a,Baburaj2025b} }
    \label{fig:Distance_vs_co_metallicity}    
\end{figure*}

Historically, forward modeling and atmospheric retrieval analyses of brown dwarf companions have often assumed a solar C/O ratio, implicitly treating host stars as chemically homogeneous and representative of the solar neighborhood~\citep{Xuan_2024}. This assumption is particularly common for wide-separation brown dwarfs, where stellar-like formation pathways such as turbulent fragmentation or gravitational instability are expected to produce companions that inherit the elemental composition of their host stars~\citep{Boss1997,Cameron1979}. Recent efforts have established that there is a diversity of C/O ratios between directly imaged companion hosts~\citep{Baburaj2025a,Baburaj2025b}. Our determined host-star C/O ratios shown in Figure \ref{fig:Distance_vs_co_metallicity} demonstrate that the stellar environments in which brown dwarfs form exhibit similar chemical diversity. We implement the Anderson-Darling (AD) test to determine whether the C/O and [Fe/H] distributions of directly-imaged planet host stars and our PEPSI sample of brown dwarf host stars are statistically different. For the C/O ratios, we find an Anderson-Darling Statistic = $-$0.01 and \textit{p}-value of 0.25. This indicates that the two populations are similar, however with low confidence.  For the [Fe/H] abundances,  we find an Anderson-Darling Statistic = 1.41 and the \textit{p}-value is 0.08 that remains slightly above the standard \textit{p}-value = 0.05 significance threshold.
\par
\begin{figure*}
    \centering
    \includegraphics[width=\linewidth]{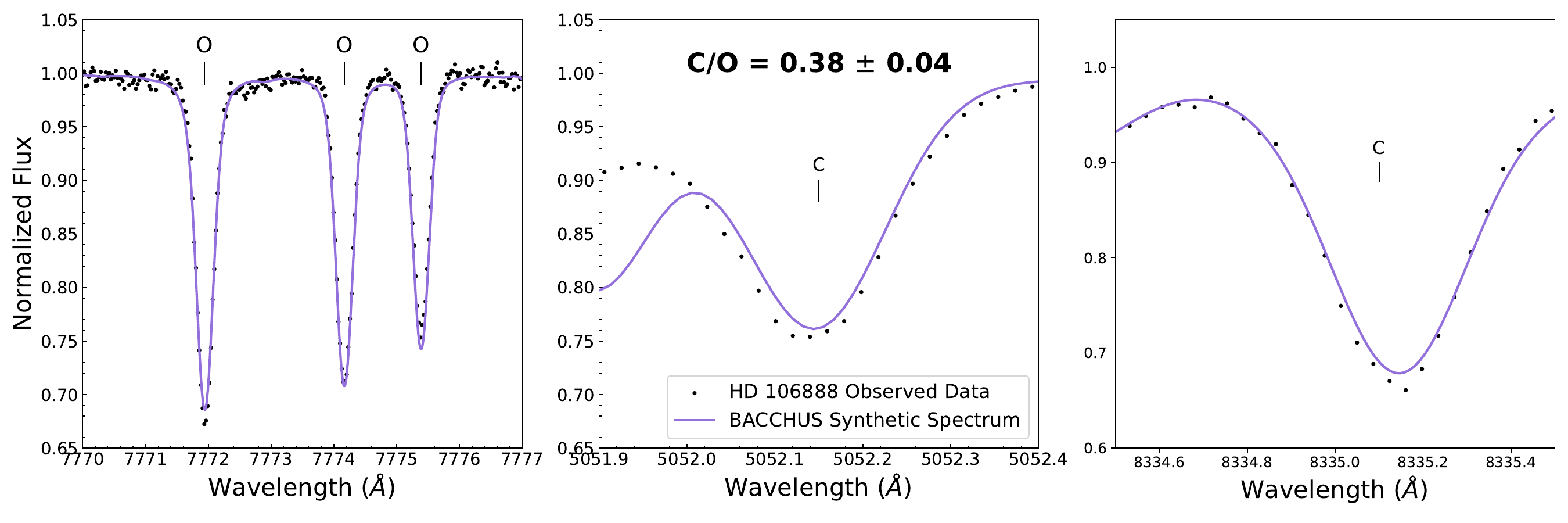}
    \includegraphics[width=\linewidth]{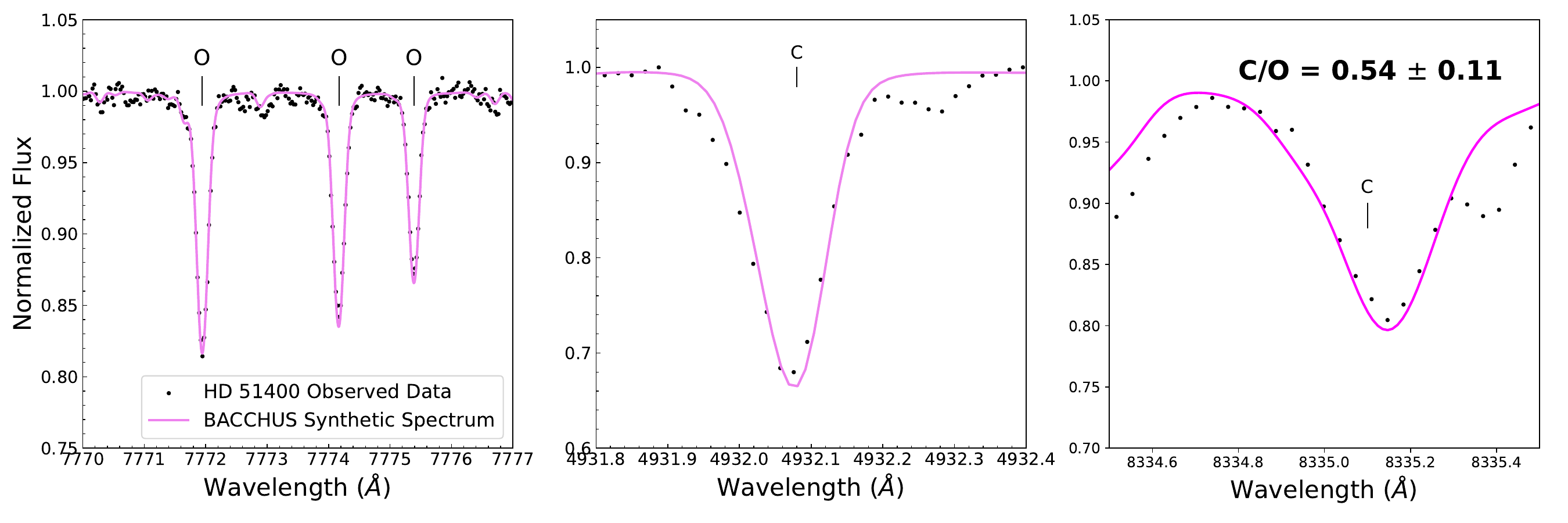}
    \caption{\textbf{Top:} We show the \ion{O}{1} triplet line and two different carbon lines used to calculate the C/O ratio for HD 106888 which yields a sub-solar C/O ratio and smaller error bars.  \textbf{Bottom:} \ion{O}{1} triplet line and two different carbon lines used to calculate the C/O ratio for HD 514000 which yields a near solar C/O ratio and larger error bars.}
    \label{fig:CtoO_fits_diff}
\end{figure*}

As our C/O ratios have a range of precision from 0.03 $-$ 0.29 for solar and sub-solar C/O ratios, we examine the \bacchus~fits. We compare the sub-solar C/O \bacchus~fits for HD 106888 and  near solar C/O fits HD 51400, with 0.04 and 0.11 error on the C/O respectively~(Figure \ref{fig:CtoO_fits_diff}). Both targets have similar precision for $\teff$, $\logg$, and [Fe/H], but HD 51400 has higher errors for [C/H] and [O/H], resulting in a higher propagated error in its C/O ratio.

\par

\subsubsection{Oxygen Sink Fraction}

Discrepancies between atmospheric and host-star C/O do not uniquely diagnose formation pathways and must be interpreted in the context of atmospheric chemistry and cloud physics. Condensation of oxygen-bearing species such as H$_2$O, silicates, and other metal oxides, followed by rainout, can remove oxygen from the observable atmosphere of a brown dwarf, yielding elevated C/O ratios even when the bulk composition matches that of the host star.

We use the formulas from \cite{Calamari_2024} to determine the relative percentage of oxygen lost in the brown dwarf companion from the host star oxygen budget.
\begin{equation}
    \Sigma \rm O_{cloud} = 2\Sigma Si + \Sigma Mg + \Sigma Ca + 1.5\Sigma Al + 2\Sigma Ti + \Sigma V
\end{equation}
\begin{equation} \label{eq: osink_fraction}
    \rm O_{sink} = \frac{\Sigma O_{cloud}}{\Sigma O}
\end{equation}

We convert metallicity measurements expressed in logarithmic (dex) notation to absolute molar abundances relative to hydrogen following \cite{Hinkel2022}.

%{\textemdash}Oxygen Sinks Table
\cite{Calamari_2024} found from compositional benchmarks in comparison to the local solar neighborhood population from \cite{Brewer2016}, the median oxygen sink is 17.8$^{+1.7}_{-2.3}$ \%. We find that on average our oxygen sink percentage lies at the higher end of the local neighborhood~(21.40 $\pm$ 5.39\%; Table \ref{tab:oxygen_sequestering}). In future modeling via retrievals or forward modeling, we would expect this to have implications on the retrieved versus bulk C/O ratio (e.g., Kecskem\'ethy et al. in prep). 

\begin{deluxetable}{l|c}[h!]
\tablewidth{\columnwidth}
\tablecaption{Estimation of percentage of oxygen lost from silicate-based clouds based on host star chemistry}
\label{tab:oxygen_sequestering}
\tablehead{
\colhead{Object} & \colhead{Oxygen Sink Percentage}   
}
\startdata
$*$ 54 Psc &  25.2 $\pm$ 4.7\% \\
BD+01 299 & 26.0 $\pm$ 16.5\%  \\
BD+13 2269 &  26.9 $\pm$ 9.7\%  \\
BD+49 2561 &  22.7 $\pm$ 12.1\%  \\
BD+ 60 1417 & 13.6 $\pm$ 6.9\%\\
BD+21 55 & 17.7 $\pm$ 2.7\% \\
CD-24 407 & 20.2 $\pm$ 9.1\% \\
G171-58 &  13.5 $\pm$ 5.4 \% \\
GJ 417 & 16.3 $\pm$ 5.4\%  \\
HD 106888 &  8.8 $\pm$ 1.6\%\\
HD 116012 &  33.1 $\pm$ 19.2\% \\
HD 125141  & 20.5 $\pm$ 11.0\%  \\
HD 126053 & 20.7 $\pm$ 6.0\% \\
HD 16270 & 21.8 $\pm$ 9.5\% \\
HD 203030 & 25.5 $\pm$ 7.5\%\\
HD 253662 & 17.8 $\pm$ 2.2\% \\
HD 46588 & 16.4 $\pm$ 5.1\% \\
HD 51400 & 26.7 $\pm$ 6.6\% \\
HD 8291 & 18.5 $\pm$ 2.4\%  \\
HD 89744 & 16.7 $\pm$ 8.4\%  \\
HIP 26653 & 26.5 $\pm$ 9.1\%  \\
HIP 63506 & 25.0 $\pm$ 11.9 \% \\
V$^{*}$ HN Peg & 19.6 $\pm$ 2.3\%  \\
HIP 9269 &  31.3 $\pm$ 20.1\% \\
StKM 2-1777 &  25.2 $\pm$ 9.6\%   \\
StKM 1-1526 & \nodata  \\
TYC 5213-545-1 & 21.0 $\pm$ 4.3\%  \\
LP 617-58 & 21.7 $\pm$ 2.5\%   \\
LSPM J0632+5053 & 25.7 $\pm$ 4.3\%   \\
NLTT 1011 & \nodata  \\
BD+06 2986 & \nodata  \\
BD+24 4329 & \nodata   \\
BD+49 2561 & 22.7 $\pm$ 12.0\%  \\
\enddata
\end{deluxetable}

\subsection{Formation and Migration Tracers}

Elemental abundance ratios (such as C/O, N/O, C/N, and S/N) in protoplanetary disks serve as powerful tracers of formation and migration histories \citep{Turrini_2021}. By comparing the atmospheric abundances of a substellar companion to its host star, shown as:
\begin{equation}
    X/Y^* = \frac{(X_\mathrm{comp}/Y_\mathrm{comp})}{(X_\mathrm{star}/Y_\mathrm{star})}
\end{equation}
following \citet{Kolecki2022}, multi-elemental analysis can break degeneracies between star-like cloud collapse and disk-born formation pathways.

While we do not currently possess the precise atmospheric abundance measurements for this brown dwarf sample to calculate these companion-to-star ratios, future high-signal-to-noise observations, particularly with \textit{JWST}, will enable these multi-elemental constraints by targeting key nitrogen- and sulfur-bearing species. 

Measuring nitrogen in brown dwarfs remains challenging because $\mathrm{N}_{2}$ is spectroscopically inactive. Instead, the primary tracers are $\mathrm{NH}_{3}$ in T-dwarfs, $\mathrm{HCN}$ in warmer objects, and potentially $\mathrm{NH}$ (imidogen) \citep{PerriNH2024}, though $\mathrm{NH}$ also serves as a theoretical temperature diagnostic in equilibrium chemistry \citep{Lodders2002}. Future or current \textit{JWST} observations of later-type T-dwarf companions can reveal $\mathrm{NH}_{3}$ features in the mid-infrared, significantly improving constraints on the total nitrogen budget and allowing for accurate determinations of $\mathrm{C/N}$ and $\mathrm{N/O}$ ratios. 

Complementarily, sulfur species play a key role in shaping the atmospheric chemistry and cloud structures of cooler brown dwarfs. In some of the T-dwarfs in our  sample, $\mathrm{H}_{2}\mathrm{S}$ is anticipated to be the dominant molecular carrier across a broad range of temperatures, eventually condensing into $\mathrm{MnS}$, $\mathrm{Na}_{2}\mathrm{S}$, $\mathrm{ZnS}$, and $\mathrm{NH}_{2}\mathrm{SH}$ cloud layers \citep{Visscher2006, Morley2012}. \textit{JWST}'s access to these mid-infrared molecular features \citep{Meynardie2025} can provide the full suite of volatile abundances required to directly test these formation frameworks and illuminate the evolutionary histories of these brown dwarf systems.

\begin{deluxetable}{l|cc}[h!]
\tablecaption{Abundance ratios and errors for the PEPSI sample related to additional formation and migration diagnostics from \cite{Turrini_2021}
\label{tab:turrini_abundances}}
\tablehead{
\colhead{Object} & \colhead{N/O} & \colhead{C/N}}
\startdata
HD 89744 & 0.19 $\pm$ 0.02 & 2.94 $\pm$ 0.29\\
HD 46588 & 0.12 $\pm$ 0.01 & 4.49 $\pm$ 0.55 \\
$*$ 54 Psc & 0.50 $\pm$ 0.09 & 2.03 $\pm$ 0.43 \\
V$^{*}$ HN Peg & 0.16 $\pm$ 0.01 & 3.14 $\pm$ 0.27\\ 
HD 126053 & 0.09 $\pm$ 0.01 & 4.74 $\pm$ 0.65 \\
GJ 417 & 0.09 $\pm$ 0.01 &4.53 $\pm$ 0.43 \\
HIP 9269 &  0.34 $\pm$ 0.10 & 2.22 $\pm$ 0.63\\
G 171-58 & 0.16 $\pm$ 0.03 &3.31 $\pm$ 0.50 \\
HIP 26653 & 0.22 $\pm$ 0.03 & 2.48 $\pm$ 0.63\\
HD 106888 & 0.06 $\pm$ 0.01 & 5.19 $\pm$ 0.56 \\
HD 51400 & 0.24 $\pm$ 0.02 & 2.18 $\pm$ 0.39\\
HD 203030 & 0.14 $\pm$ 0.02 & 3.39 $\pm$ 0.95 \\
HD 8291 & 0.14 $\pm$ 0.01 & 3.38 $\pm$ 0.33\\
HD 16270 & 0.18 $\pm$ 0.04 & 3.80 $\pm$ 0.80 \\
HD 116012 & 0.14 $\pm$ 0.04 & 6.46 $\pm$ 1.19 \\
BD+21 55 & 0.18 $\pm$ 0.01 & 1.48 $\pm$ 0.27 \\
BD+60 1417 & 0.03 $\pm$ 0.01 & 10.04 $\pm$ 0.73 \\
SkTKM 2-1777 & \nodata & \nodata \\
HD 253662 & 0.10 $\pm$ 0.01 & 2.95 $\pm$ 0.14 \\
CD-24 407 & 0.13 $\pm$ 0.02 & 2.83 $\pm$ 0.71\\
BD+13 2269 & 0.22 $\pm$ 0.04 &1.80 $\pm$ 0.45 \\
BD+01 299  &0.28 $\pm$ 0.12 & 2.78 $\pm$ 1.22\\
StKM 1-1526 & \nodata & \nodata \\
TYC 5213-545-1 & 0.08 $\pm$ 0.01 & 7.63 $\pm$ 0.72 \\
LP 617-58 & 1.33 $\pm$ 0.09 & 0.31 $\pm$ 0.02 \\
HIP 63506 & \nodata & \nodata\\
LSPM J0632+5053 & 0.18 $\pm$ 0.01 &  3.18 $\pm$ 0.53\\
NLTT 1011 & \nodata & \nodata \\
HD 125141 & 0.13 $\pm$ 0.03 & 4.14 $\pm$ 0.85 \\
BD+06 2986 & \nodata &\nodata \\
BD+24 4329& \nodata  & \nodata  \\
BD+49 2561 & 0.07 $\pm$ 0.01 & 7.27 $\pm$ 1.56 \\
\enddata
\end{deluxetable}

\begin{figure}
    \centering
    \includegraphics[width=1.05\linewidth]{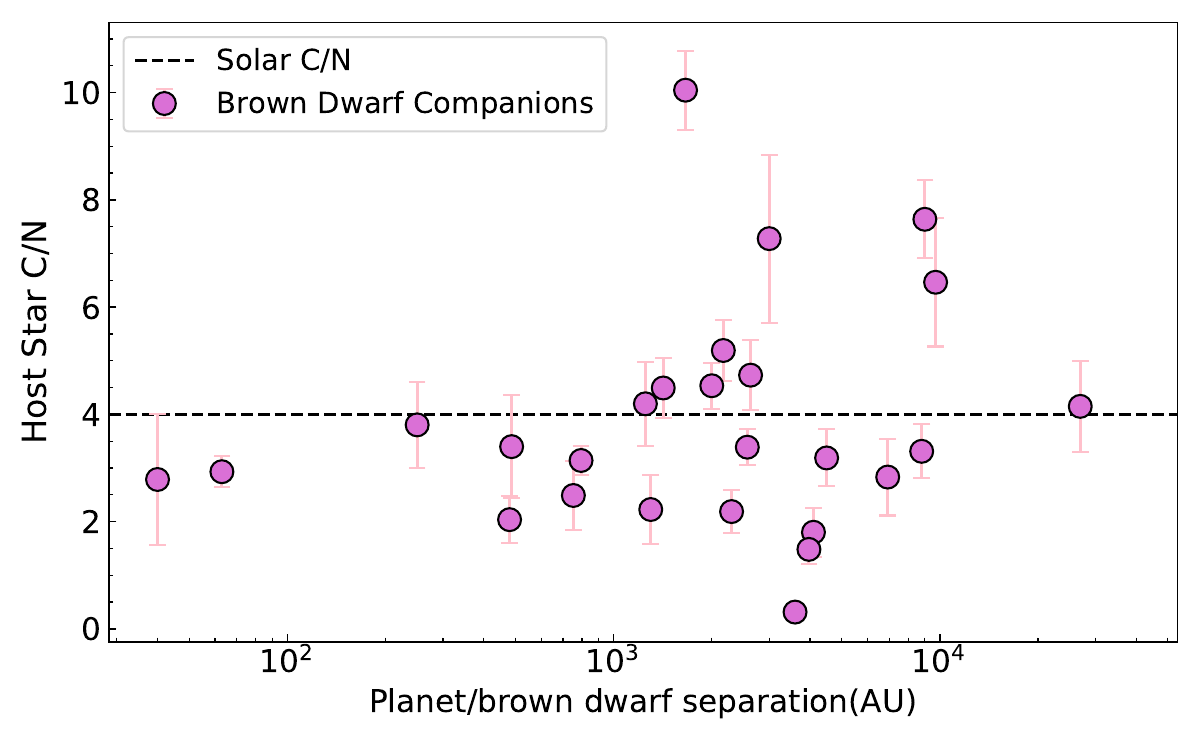}
    \includegraphics[width=1.05\linewidth]
    {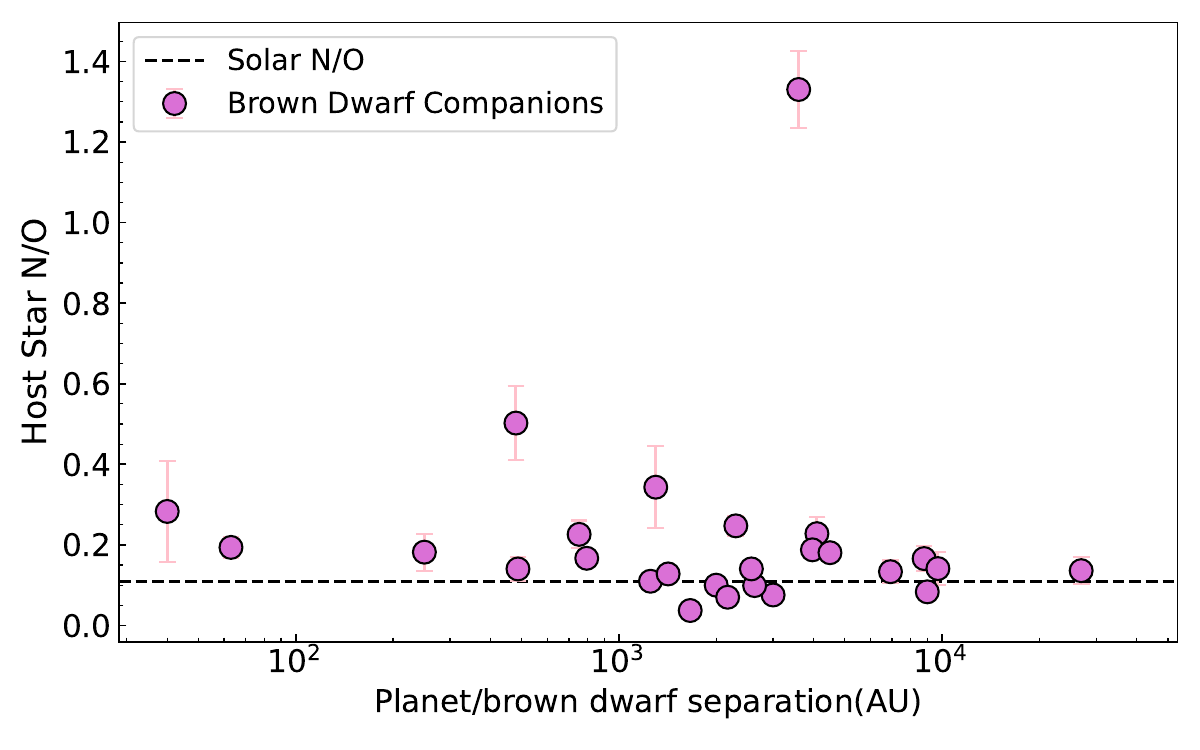}
    \includegraphics[width=1.05\linewidth]
    {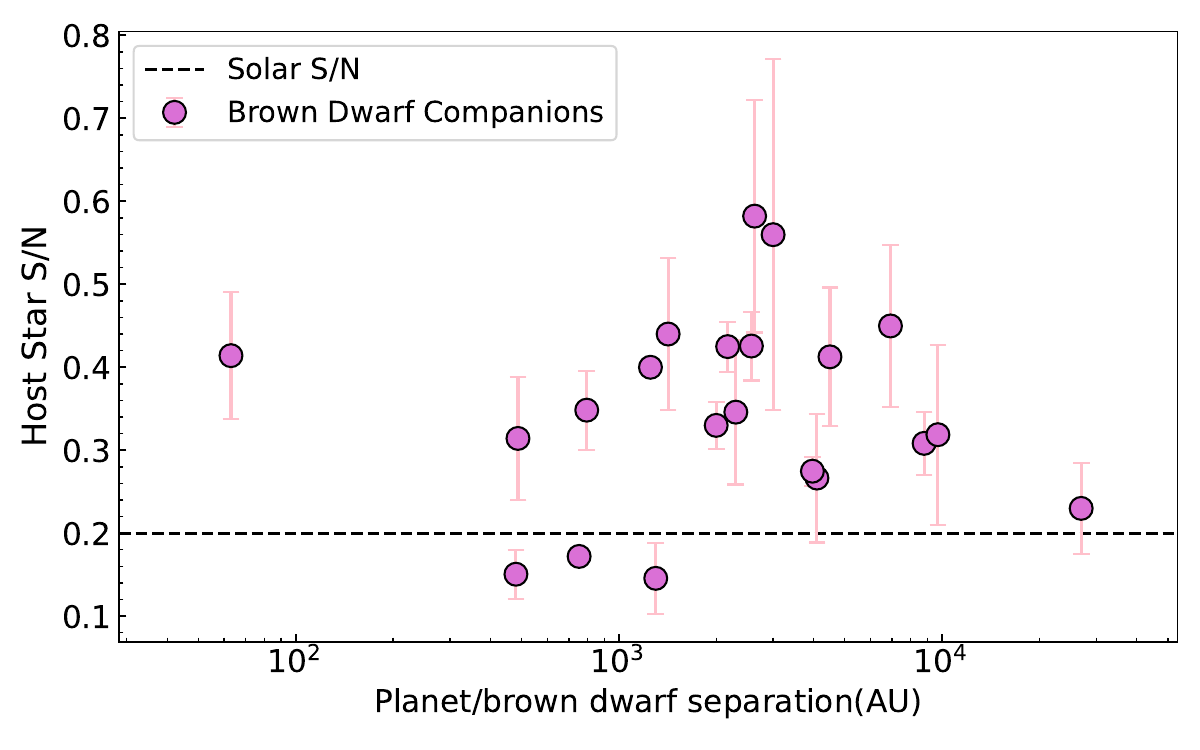}
     \caption{\textbf{Top:} Host star C/N vs separation for our sample,  \textbf{Middle:} Host star N/O vs separation, and \textbf{Bottom:} S/N vs separation ratios. }
    \label{fig:turrini_plots}
\end{figure}

%{\textemdash}{\textemdash}Prediction of Ages from Y/Mg
\section{Ages from Y/Mg}
\label{sec:age_YtoMg}

As brown dwarfs contract and cool over time, their apparent properties depend strongly on both mass and age. Without an independent age estimate, disentangling these parameters is fundamentally degenerate~\citep{Burrows2011,Burrows2001,Baraffe2002,Baraffe2015}. Observable quantities such as luminosity and effective temperature are sensitive to both age and mass, producing a well-known degeneracy among these parameters. Several approaches can help break this degeneracy, including membership in clusters or young moving groups, kinematic analyses, spectroscopic gravity indicators, the lithium test, or association with binary or multiple-star systems~\citep{Allers2016, Phillips2020,Gagne2023, Kirkpatrick2008}. In particular, binary or co-moving systems provide a unique opportunity: if the primary companion star can be characterized in detail, its properties can yield valuable age constraints for all components of the system.
\par

Recent studies have proposed the abundance ratio [Y/Mg] as a useful diagnostic of stellar age for solar-type stars~\citep{Nissen2015,Tucci2016, Katime2022,Ratcliffe2024,Shejeelammal2024}, and FGK stars~\citep{Berger_2022}. The [Y/Mg] ratio appears to be correlated with stellar age, reflecting the nucleosynthetic history of the Galactic disk~\citep{Shejeelammal2024}. Brown dwarfs orbiting solar-type stars provide an excellent opportunity to use the [Y/Mg] chemical clock. Because isolated field brown dwarfs typically have highly uncertain ages, determining the [Y/Mg] ratio of the host star may help to place a spectroscopic age constraint for the entire system.

\par

There are several methods used to estimate the ages of solar analogs. These include gyrochronology, which infers stellar ages from rotation rates of field stars; X-ray emission, where coronal activity serves as an age indicator because magnetic activity decreases as stars spin down over time; and chromospheric emission, arising from the cores of the Ca II H \& K lines, which has been observed to decline with stellar age. Lithium abundance is another diagnostic, since lithium is rapidly depleted in the cores of solar-type stars early in their evolution. Additionally, isochrone fitting allows ages to be determined by placing stars on theoretical Hertzsprung–Russell diagrams using observed parameters such as effective temperature (T$_\mathrm{eff}$), absolute magnitude (M$_v$), and metallicity ([Fe/H]), along with kinematic information. Some of these methods have previously been applied to different subsets of our sample, here we investigate an abundance based method that provides another independent check on system age.

\par

\cite{Berger_2022} show that the [Y/Mg] abundance ratio serves as a useful age indicator for FGK-type stars in general, but achieves its highest reliability and precision within solar twins and analogs. Here we investigate the age determination for our sample from the [Y/Mg] ratio using formulation from \cite{Berger_2022} and values for [Y/Mg]–Rotation Age best-fit relations from the \citealt{Brewer_2018} catalog. We present our results in Table \ref{tab:ages}. 

We use the following formulation to determine the age

\begin{equation}
    [Y/Mg] = m \times age +b,
    \label{eq:y_mg}
\end{equation}
where for \cite{Brewer_2018}, slope ($m$) = $-$0.228, intercept ($b$) = 0.121, and [Y/Mg] are values taken from [Y/H]$-$[Mg/H], and the ratio is presented in Table \ref{tab:abundance_ratio}.

To determine the error on the age, we use:

\begin{equation}
    \sigma_{\text{age}} = \text{Age} \times \sqrt{\frac{(\sigma_{\text{[Y/Mg]}})^2 + (\sigma_b)^2}{(\text{[Y/Mg]} - b)^2} + \left(\frac{\sigma_m}{m}\right)^2}
\end{equation}
where $\sigma_{b}$ = 0.016 and $\sigma_{m}$ = 0.0044 and $\sigma_{[Y/Mg]}$ is the propagation error of the values of [Y/Mg]. Due to the shallow slope ($m$), small variations in $\text{[Y/Mg]}$ near the intercept ($b$) sometimes produces  
negative ages. This occurs primarily for young stars with independent literature constraints (e.g., gyrochronology or isochrones). Because negative ages are unphysical, we truncate the distribution at $0\,\text{Gyr}$. We define their age constraints as an upper physical limit ranging from $0$ to $(\text{Age} + \sigma_{\text{age}})\,\text{Gyr}$.

\begin{figure}
    \centering
    \includegraphics[width=\columnwidth]{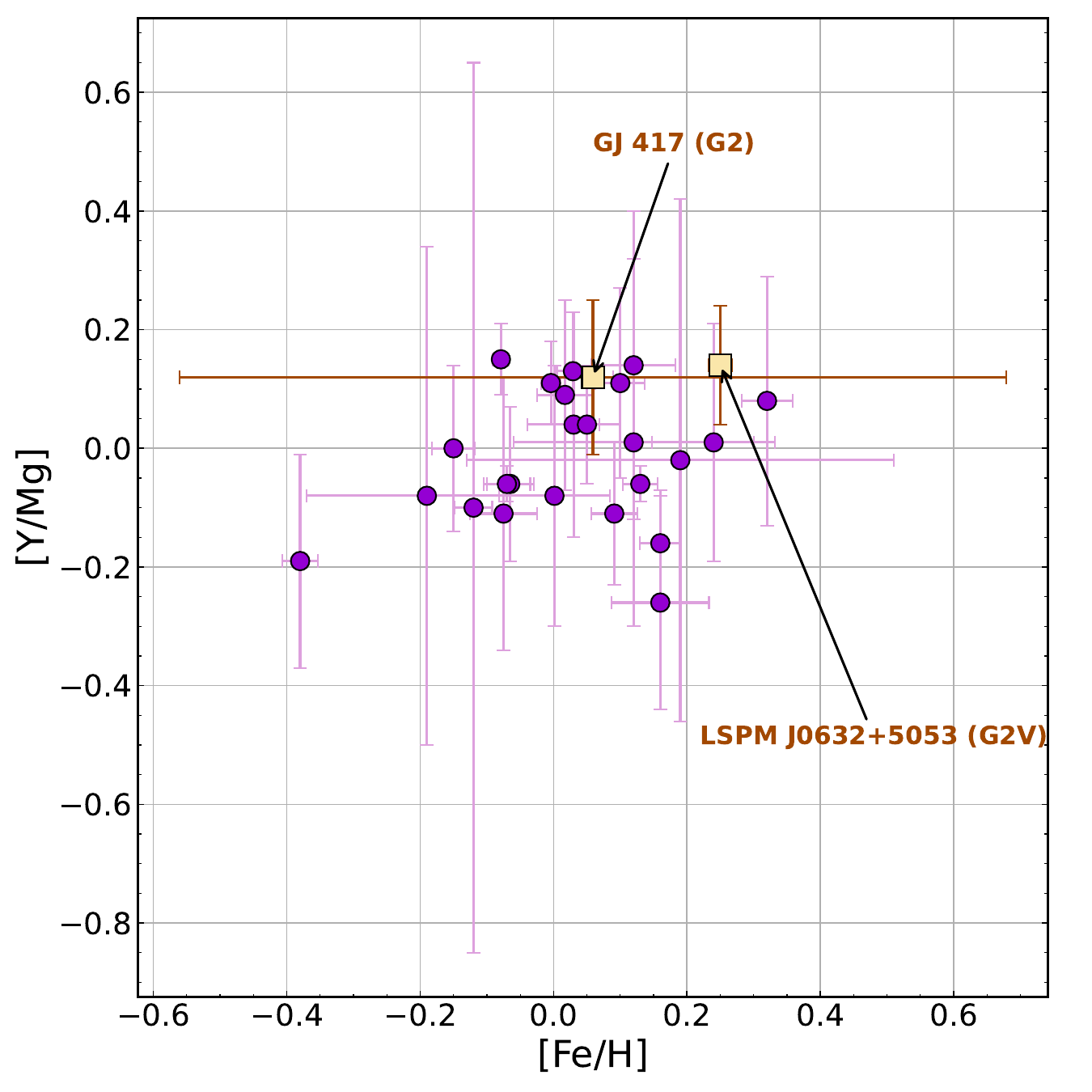}
    \caption{[Y/Mg] vs. [Fe/H]~(metallicity)  for our sample of FGK stars (purple circles). We highlight solar-twins (G2) $-$ GJ 417 and LSPM J0632+5053 in yellow squares.  }
    \label{fig:Y_Mg_plot}
\end{figure}

\par

Our sample has previous literature age estimates from a mix of isochrone fitting, chromospheric activity, x-ray emission, disk-like kinematics often give wide age estimates (e.g. 0.2 $-$ 10 Gyr). Generally we find a tighter-constraint on ages, but often higher estimates on the age from [Y/Mg] than literature estimates. We find that the age determined from [Y/Mg] for four of our systems~(HD 126053, HIP 9269, LP 617-58, $^*$54 Psc), provide implausible ages -- ages greater than the age of the universe~\citep{Spergel1997}. Notably, we find that one target HD 8291 (G5), provides a [Y/Mg]-derived age in line with previous age estimates, this is the only star with an age estimate from \textit{Gaia} photometry.
In Figure \ref{fig:Y_Mg_plot} we show the [Y/Mg] vs. [Fe/H] determined from Eq. \ref{eq:y_mg} for our sample.

%54 Psc = HD 3651
%HIP 26653 = HD 37216
%HIP 63506 - BD+42 2363e
%LP 192-58 = NLTT 1011
\begin{deluxetable*}{l|cccc}
\tablecaption{Comparison of Stellar Ages: Literature vs.\ [Y/Mg]-based Estimates
\label{tab:ages}}
\tablehead{
\colhead{Object} & \colhead{Previous Age (Gyr)$^{a}$} & \colhead{[Y/Mg] (Gyr)} & \colhead{Age Estimate Method}& \colhead{References}
}
\startdata
HD 89744\tablenotemark{c} &1.5 $-$ 3 & 6.19 $\pm$ 19.42 & Isochrone Fitting& \cite{Wilson_2001} \\
HD 46588 & 1.3$-$ 4.3 & 5.44 $\pm$ 6.43 &Isochrone Fitting&\cite{Casagrande2011}   \\
$*$ 54 Psc & 0.7 $-$ 4.7  & 12.55 $\pm$ 4.90 & Isochrone Fitting & \cite{Valenti_2005} \\
HN Peg & 0.1 $–$ 0.5 & 0.34 $\pm$ 3.34  & Chromospheric Activity & \cite{Gaidos1998}\\
HD 126053\tablenotemark{c} & 2.3 $-$ 14.4 & 13.70 $\pm$ 8.53 & Chromospheric Activity & \cite{Vican_2012}  \\
GJ 417 & 0.08 $–$ 0.3 & 0.00 $-$ 5.99 & X-ray luminosity & \cite{Kirkpatrick_2000}  \\
HIP 9269\tablenotemark{c} & 2.2 $–$ 10.2 & 17.12 $\pm$ 8.68 & Isochrone Fitting & \cite{Valenti_2005} \\
G171-58 & 1.8 -- 3.5 & 10.30 $\pm$ 5.75 & Isochrone Fitting & \cite{Faherty_2010}  \\
HIP 26653 & 1.1 $-$ 9.3   & 8.12 $\pm$ 5.99 & Isochrone Fitting & \cite{Valenti_2005}  \\
HD 106888 & 0.3 $-$ 2.5 & $0-4.02$ & Isochrone Fitting & \cite{Shevelev1993} \\
HD 51400 & 0.6 $-$ 0.7 & 3.50 $\pm$ 4.63 & Color - Period diagram & \cite{Rothermich2024}  \\
HD 203030 & 0.13 $-$ 0.4 & 1.23 $\pm$ 7.08 & X-ray Emission \& Chromospheric Emission & \cite{Metchev_2006} \\
HD 8291& 0.89$^{+1.30}_{-0.69}$& $0-1.46$ &  \textit{Gaia} Photometry & \cite{Baig2024}\\
HD 16270 & $<$ 1 & 4.49 $\pm$ 9.08 & X-ray Emission & \cite{Gizis_2001}   \\
HD 116012\tablenotemark{d} & 12 $-$ 14 & \nodata & Chromospheric Activity& \cite{Faherty_2010} \\
BD+21 55 & 0.5 $-$ 10 & 9.84 $\pm$ 33.02 & X-ray Emission 
\& Disk-like Kinematics & \cite{Deacon_2014} \\
BD+60 1417\tablenotemark{b} & 0.01 $-$ 0.15  & $0-10.4$ & Gyrochronology \& Color-Color/Period Diagrams & \cite{Faherty2021} \\
StKM 2-1777\tablenotemark{b} & \nodata & $0-2.61$  & Chromospheric Activity & \cite{Pace2013}  \\
HD 253662 & $<$ 10  & 1.63 $\pm$ 9.30 & Disk-like Kinematics & \cite{Deacon_2014}  \\
CD-24 407 & $>$1  & 8.81 $\pm$ 9.99 & Chromospheric Emission & \cite{Zerjal_2017}  \\
BD+13 2269\tablenotemark{b} & \nodata & 0.39 $\pm$ 7.30 & \nodata & {\textemdash}\\
BD+01 299\tablenotemark{b} & 6.5 $-$ 13.5 & 4.61$\pm$ 14.02 & \nodata & {\textemdash} \\
StKM 1-1526\tablenotemark{b} & \nodata & 8.18 $\pm$ 2.26 &  \nodata & {\textemdash}  \\
TYC 5213-545-1\tablenotemark{b} & 0.1 $-$ 1  & 4.63 $\pm$ 5.11 & Chromospheric Emission &\cite{Zerjal_2017}  \\
LP 617-58\tablenotemark{c} & \nodata  & 23.85 $\pm$ 5.72 & \nodata & {\textemdash}  \\
HIP 63506\tablenotemark{d} & 0.3 $-$ 10 & \nodata & Disk-like Kinematics & \cite{Shkolnik_2009} \\
LSPM J0632+5053 & 0.2 $-$ 10 & $0-3.33$ & X-ray flux \& Disk-like Kinematics &\cite{Deacon_2014} \\
NLTT 1011\tablenotemark{d} & 0.3 $-$ 10 & \nodata 
&Disk-like Kinematics &   \cite{Deacon_2014}  \\
HD 125141 & 8.5 $-$ 11 & 10.25 $\pm$ 10.51 & Isochrone Fitting &\cite{Holmberg2008}  \\
BD+06 2986\tablenotemark{d} & $>$ 1.6 & \nodata &  Chromospheric Activity & \cite{Murray2011}  \\
BD+24 4329\tablenotemark{b} & \nodata & 9.01 $\pm$ 18.72 & \nodata & {\textemdash}   \\
BD+49 2561\tablenotemark{b} & \nodata  & 3.23 $\pm$ 8.41 & \nodata & {\textemdash}  \\
\enddata
\tablecomments{Comparison between previously determined stellar ages and those derived from the [Y/Mg] abundance ratio.}
\tablenotetext{a}{Published age ranges found in the literature}
\tablenotetext{b}{Age determined using [Y/Mg]-Rotation age relations.}
\tablenotetext{c}{Implausible old age, as age is older than estimated age of universe $\approx$ 13.8 Gyr}
\tablenotetext{d}{Only upper limit found for [Y/Mg], so no [Y/Mg] reported age}

\end{deluxetable*}
\par

\subsection{Age Implications for Brown Dwarf Companions}
Stellar ages derived from the [Y/Mg] chemical clock can provide a valuable constraint on the evolutionary state of substellar companions and can therefore be used as robust priors when comparing observed brown dwarf luminosities or effective temperatures to evolutionary cooling models. Although age information alone does not uniquely determine brown dwarf masses due to the inherent age–mass degeneracy, it significantly restricts the range of allowed solutions. We use bolometric luminosity values from the Ultracool Sheet\footnote{\url{https://zenodo.org/records/15802304}} in conjunction with our [Y/Mg] derived ages to place constraints on model dependent-mass values for the brown dwarf companions~\citep{Burrows2001, Baraffe2015, Baraffe2002,Saumon2008}. We use the \texttt{SPLAT} code~\citep{BurgasserSPLAT2017}, where the evolutionary model parameters are derived by log-linear interpolation of the original model grid. When bolometric luminosity measurements are not available in the literature, we use the reported spectral type of the brown dwarf companion to determine the effective temperature using the \texttt{SEDA} code (Suarez et al. in submitted), which follows the \cite{Filippazzo2015} spectral-type temperature relationship. For some our FGK + brown dwarf systems we find that the determined mass from adopting our [Y/Mg] derived ages are within 1 -- 2$\sigma$ of literature reported mass value. These results prove promising in continuing to evaluate the use of the [Y/Mg] age-diagnostic for FGK stars and age-dating in particular field brown dwarfs. We provide our findings in Table \ref{tab:Y_Mg_mass_vs_Literature}.

\begin{deluxetable*}{l|cccc}[h!]
\tablewidth{\columnwidth}
\tablecaption{Estimated of brown dwarf companion mass from spectroscopic [Y/Mg] age compared to brown dwarf evolutionary models from the literature}
\label{tab:Y_Mg_mass_vs_Literature}
\tablehead{
\colhead{Companion} & \colhead{L$_\mathrm{Bol}$/T$_\mathrm{eff}$} &  \colhead{M$_\mathrm{Jup}$ [Y/Mg]\tablenotemark{a}} &  \colhead{M$_\mathrm{Jup}$ (Literature)} & \colhead{References}}
\startdata
HD 89744 B & $-3.61$ $\pm$ 0.03 & 80.05 $\pm$ 10.02 & 82.92 $\pm$ 0.61 & UltraCoolSheet\\
HD 46588 B & $-4.68$ $\pm$ 0.05 & 70.18 $\pm$ 14.11&  64.0 $\pm$ 8.0 & UltraCoolSheet \\
2MASS J00250365+4759191 & $-3.33$ $\pm$ 0.04 &91.18 $\pm$ 3.95& 84 -- 87 & \cite{Faherty_2010}
\\
2MASS J12173646+1427119 & $-3.64$ $\pm$ 0.02 & 55.45 $\pm$ 14.25& 77.64 $\pm$ 8.51 & UltraCoolSheet\\
HN Peg B & $-4.83$ $\pm$ 0.05 & 49.20 $\pm$ 11.52 &  21.00$\pm$ 7.00 & \cite{Suarez2021} \\
BD+01 2920B & $-4.43$ $\pm$ 0.15 & 70.60 $\pm$ 7.72,& 38.41 $\pm$ 24.24 & UltraCoolSheet\\
GJ 417BC & $-4.12$ $\pm$ 0.04 & 70.73 $\pm$ 11.97 &99 $\pm$ 3\tablenotemark{b} & \citet{Dupuy2014}\\
2MASS J06324849+5053351& $-3.71$ $\pm$ 0.07& 75.51 $\pm$ 10.25& 79.43 $\pm$ 11.73 & UltraCoolSheet\\
CWISE J065752.45+163350.2\tablenotemark{c} & 1482 $\pm$ 100 & 67.61 $\pm$ 10.08 &62.8 & \cite{Rothermich2024}\\
HD 8291 B\tablenotemark{c} &1625 $\pm$ 100 & 51.43 $\pm$ 11.04 & 74.6$^{+1.3}_{-29.0}$ & \cite{Baig2024}\\
2MASS J14165987+5006258 & $-4.28$ $\pm$ 0.02 & 72.27 $\pm$ 5.38 & 74.29 $\pm$ 0.27 & UltraCoolSheet \\
HD 37216 B  & $-3.84$ $\pm$ 0.02 & 78.17 $\pm$ 4.11 & 80.13 $\pm$ 0.95 & UltraCoolSheet\\ 
2MASS J06135342+1514062 &$-3.26$ $\pm$ 0.02 & 
92.91 $\pm$ 5.92 &77.5 $\pm$ 31.41 & UltraCoolSheet  \\
CWISE J132539.70+022309.4\tablenotemark{c}& 2394 $\pm$ 100 & \nodata & $\approx$84&\cite{Rothermich2024}\\
HIP 9269B& $-4.42$ $\pm$ 0.03 & \nodata & 72.25 $\pm$ 1.06 & UltraCoolSheet \\
HD 3651 B  & $-5.60$ $\pm$ 0.06& 43.45 $\pm$ 7.58 & 45.96 $\pm$ 4.05 & UltraCoolSheet \\
HD 202030 B & $-4.99$ $\pm$	0.10 &54.55 $\pm$ 15.53 &$33.52^{+9.43}_{-12.57}$  & \cite{Metchev_2006}\\
2MASS J13204427+0409045 & $-3.92$ $\pm$ 0.03 & \nodata & 78.07 $\pm$ 0.44 & UltraCoolSheet\\
WISE J124332.17+600126.6 & $-4.42 \pm 0.10$ & 67.78 $\pm$ 11.03 & $13.47 \pm 5.67$& \cite{Phillips2024}\\
CWISE J214129.80-024623.6\tablenotemark{c} &1694 $\pm$ 100& 72.46 $\pm$ 6.74&$\approx$84 &\cite{Rothermich2024} \\
2MASS J00302476+2244492  & $-3.66 \pm 0.02$ & 83.69 $\pm$ 1.05&81.92 $\pm$ 3.91 & UltraCoolSheet\\
CWISE J005635.48–240401.9\tablenotemark{c}& 1336 $\pm$ 100& 64.93 $\pm$ 10.74& 52.36& \cite{Rothermich2024} \\
CWISE J165325.10+490439.7\tablenotemark{c}  & 1581 $\pm$ 100 & 69.93 $\pm$ 8.39&62.84&\citet{Rothermich2024} \\
ULAS J103131.49+123736.4\tablenotemark{c} & 1821 $\pm$ 100 &  73.08 $\pm$ 8.33& $\approx$84 &\cite{Rothermich2024} \\
HD 16270 B & $-3.57 \pm 0.03$ &  81.65 $\pm$ 8.55& 82.0 $\pm$ 2.67 & UltraCoolSheet\\
CWISE J174509.03+380733.2\tablenotemark{c} & 1694 $\pm$ 100 & 74.64 $\pm$ 0.89 & 62.84&\citet{Rothermich2024} \\
WIS 395 B&$-4.31 \pm 0.03$ & 52.25 $\pm$ 10.37&  11.36 $\pm$ 0.35 & UltraCoolSheet  \\
2MASS J13005061+4214473 \tablenotemark{c} &2102 $\pm$ 100 & \nodata & \nodata & \cite{Filippazzo2015} \\
CWISE J210640.16+250729.0\tablenotemark{c}&1184 $\pm$ 100&62.01 $\pm$ 9.58&$\approx$ 31.42&\citet{Rothermich2024}\\
2MASS J00193275+4018576  & $-3.66 \pm 0.02$ & \nodata &81.92 $\pm$ 3.91 & UltraCoolSheet\\
ULAS J014016.91+015054.7\tablenotemark{c}& 1149 $\pm$ 30 & 58.78 $\pm$ 13.09& \nodata & \citet{Burgasser_2025} \\
ULAS J150457.65+053800.8\tablenotemark{c} & 914 $\pm$ 100 & \nodata & \nodata & \citet{Burgasser_2025}\\
\enddata 
\tablecomments{Several brown dwarfs have physically implausible old ages ($>12$ Gyr), as a result we cannot use evolutionary models to determine a mass. }
\tablenotetext{a}{Mass of companion if [Y/Mg] age adopted. Mass not reported if [Y/Mg] derived age is implausible or if only upper limit for [Y/Mg]}
\tablenotetext{b}{Dynamical mass measurement for the brown dwarf binary GJ 417BC}
\tablenotetext{c}{$T_\mathrm{eff}$ determined using the \texttt{SEDA} based \cite{Filippazzo2015}  spectral type-temperature relationships, an error of 100K is adopted. 
}

\end{deluxetable*}

\section{Conclusions}
\label{sec:conclusion}

In this paper, we present a uniform abundance analysis of 32 FGK stars that host benchmark brown dwarfs. We utilize the high-resolution (R = 50,000 and R = 130,000) PEPSI spectrograph on the Large Binocular Telescope to obtain high signal-to-noise (SNR $\geq$ 200)  data. A fraction of our targets  are of particular interest to the brown dwarf and substellar science community, as the companions have current and forthcoming JWST data with NIRSpec/PRISM and MIRI/LRS data~\citep{SinkingSilicates, Zhang2025}.

\par

We use the \bacchus~framework to determine precise spectroscopic parameters and abundance ratios of our sample, with a focus on elements used to determine the following ratios: C/O, Mg/Si, Ca/Al, S/N, and [Y/Mg]. We summarize the primary conclusions derived from this work:

\begin{enumerate}
    \item We achieve a typical precision of 42 K in in $\teff$ and $\sim$0.03 dex in [Fe/H] for our sample. We have a similar precision of $\sim$0.03 dex for our R = 50,000 and R = 130,000 spectra.
    
    \item In line with previous works, we experience challenges in abundance determination for our coolest K-dwarf targets, which often resulted in upper limits for key elements and ratios, like [O/H] and C/O.
    
    \item We use the theoretical chemical frameworks from \cite{Calamari_2024} to predict cloud species in our brown dwarf companions based on host star Mg/Si ratios. We find that most of our systems predict enstatite and forsterite clouds with five systems also predicting quartz clouds (Table \ref{tab:cloud_species}) Future and current retrieval results of the companion with JWST will assist in determining best fit cloud species to compare to predictions from this work. 

    \item We compare the C/O of brown dwarf host star sample with the wide orbit directly-imaged population from \cite{Baburaj2025a, Baburaj2025b} and find that our sample exhibits similar chemical diversity for stellar hosts of brown dwarf companions

    \item We use the Si, Mg, Ca, Al, O, and Ti abundances to determine relative oxygen sinks (or sequestering) for our systems~\cite{Calamari_2024}. This has implications for determining the bulk vs retrieved C/O ratio in the brown dwarf companions. We find a median oxygen sink of 21.40 $\pm$ 5.3\%, which is higher than the local solar neighborhood value of 17.8$^{+1.7}_{-2.3}$\%

    \item We continue to test and investigate the use of the [Y/Mg] stellar clock for FGK stars and in particular those that host brown dwarf companions with a wide age estimate (Section \ref{sec:age_YtoMg}). We find for a number of systems,  our derived ages from [Y/Mg] in combination with fundamental parameters (L$_\mathrm{Bol}$ or T$_\mathrm{eff}$) provide mass estimates within 1$\sigma$ of previous mass estimates (Table \ref{tab:Y_Mg_mass_vs_Literature})
\end{enumerate}

\facility{Large Binocular Telescope}
\software{\texttt{EXOFASTv2}~\citep{exofast1}, \texttt{BACCHUS}~\citep{Masseron2013}, \texttt{seda}~(Suarez et al. submitted), \texttt{SPLAT}~\citep{BurgasserSPLAT2017}, numpy~\citep{Numpy2020}, astropy~\citep{Astropy1,astropy2,astropy3}}

%{\textemdash}Acknowledgements

\clearpage
\begin{acknowledgments}

\textit{This work is dedicated to my late mom, Janet Leigh Phillips (1961 $-$ 2024). I began this work at the Center for Computational Astrophysics as a CCA Predoc. Less than three weeks later, I lost her.
My mom was my foundation and the reason I became an astronomer. She saw the first astronomer in me when I was a young child. I hope this work serves as a foundation to help guide my colleagues in grounding and interpreting their analyses of companion brown dwarfs, just as my mom grounded and guided me.
Her favorite color was purple, and to honor her, I incorporated purple into all the figures so that she is always with me. I carry my mom’s love and care with me in everything I do.
“The cosmos is within us. We are made of star-stuff. We are a way for the universe to know itself.” My mom has returned to the cosmos, but her love is still here with me. I love and miss you, momma, until we meet again to exchange a hug amongst the stars.}

We thank the anonymous reviewer for the feedback, which
improved this paper. C.L.P thanks Yayaati Chachan for the helpful discussions regarding
N/O, C/N and C/N ratios and their implications in regards to brown dwarfs and Adam Wheeler and Andy Casey for a helpful start to this project. C.L.P would also like to thank John Brewer for helpful discussions on carbon and oxygen abundances.

CM is supported by the NSF Astronomy and Astrophysics Fellowship award number AST-2401638. JW acknowledges the support by the National Science Foundation under Grant No. 2143400.

Support for this work was provided by NASA through the NASA Hubble Fellowship grant
HST-HF2-51585.001-A awarded by the Space Telescope Science Institute, which is operated by the
Association of Universities for Research in Astronomy, Inc., for NASA, under contract
NAS5-26555.
C.L.P would like to thank the Center for Computational Astrophysics (CCA) at the Flatiron Institute for their support of this research during her time as a CCA Predoc. The Flatiron Institute is a division of the Simons Foundation. C.L.P would also like to thank the many people who participated in observational runs and collected data for this project.

\par
E.J.G. acknowledges support for this work provided by NASA through the NASA Hubble Fellowship Program grant No. HST-HF2-51576.001-A awarded by the Space Telescope Science Institute, which is operated by the Association of Universities for Research in Astronomy, Inc., for NASA, under the contract NAS 5-26555.
\par

The research shown here acknowledges use of the Hypatia Catalog, an online compilation of stellar abundance data as described in \cite{Hinkel2014}.

\par
This work is based on observations made with the Large Binocular Telescope. The LBT is an international collaboration among institutions in the United States, Italy and Germany. LBT Corporation partners are: The University of Arizona on behalf of the Arizona Board of Regents; Istituto Nazionale di Astrofisica, Italy; LBT Beteiligungsgesellschaft, Germany, representing the Max- Planck Society, The Leibniz Institute for Astrophysics Potsdam, and Heidelberg University; The Ohio State University, representing OSU, University of Notre Dame, University of Minnesota and University of Virginia.

\par
This work presents results from the European Space Agency (ESA) space mission Gaia. Gaia data are being processed by the Gaia Data Processing and Analysis Consortium (DPAC). Funding for the DPAC is provided by national institutions, in particular the institutions participating in the Gaia MultiLateral Agreement (MLA). The Gaia mission website is \url{https://www.cosmos.esa.int/gaia}. The Gaia archive website is \url{https://archives.esac.esa.int/gaia}

This work has benefited from The UltracoolSheet at \url{http://bit.ly/UltracoolSheet}, maintained by Will Best, Trent Dupuy, Michael Liu, Aniket Sanghi, Rob Siverd, and Zhoujian Zhang, and developed from compilations by  Dupuy \& Liu (2012), Dupuy \& Kraus (2013), Deacon et al. (2014),  Liu et al. (2016), Best et al. (2018), Best et al. (2021), Sanghi et al. (2023), and Schneider et al. (2023).
\end{acknowledgments}

\par

This paper makes use of EXOFAST~\citep{exofast1} as provided by the NASA Exoplanet Archive, which is operated by the California Institute of Technology, under contract with the National Aeronautics and Space Administration under the Exoplanet Exploration Program.

\par
This research has made use of the SIMBAD database, operated at CDS, Strasbourg, France~\citep{Wenger2000}.

The land in which this research was finalized is the unceded territory of the Awaswas-speaking Uypi Tribe. The Amah Mutsun Tribal Band, comprised of the descendants of indigenous people taken to missions Santa Cruz and San Juan Bautista during Spanish colonization of the Central Coast, is today working hard to restore traditional stewardship practices on these lands and heal from historical trauma.
\clearpage
%{\textemdash}Appendix
\appendix

%{\textemdash}Sun Fits
\section{Solar Fits with \bacchus}

Here we show examples of \bacchus~fits for sulfur, nitrogen, oxygen, yttrium, calcium, and carbon for the PEPSI spectra of the Sun from \cite{Strassmeier2015}.

\begin{figure*}[h!]
    \includegraphics[width=\linewidth]{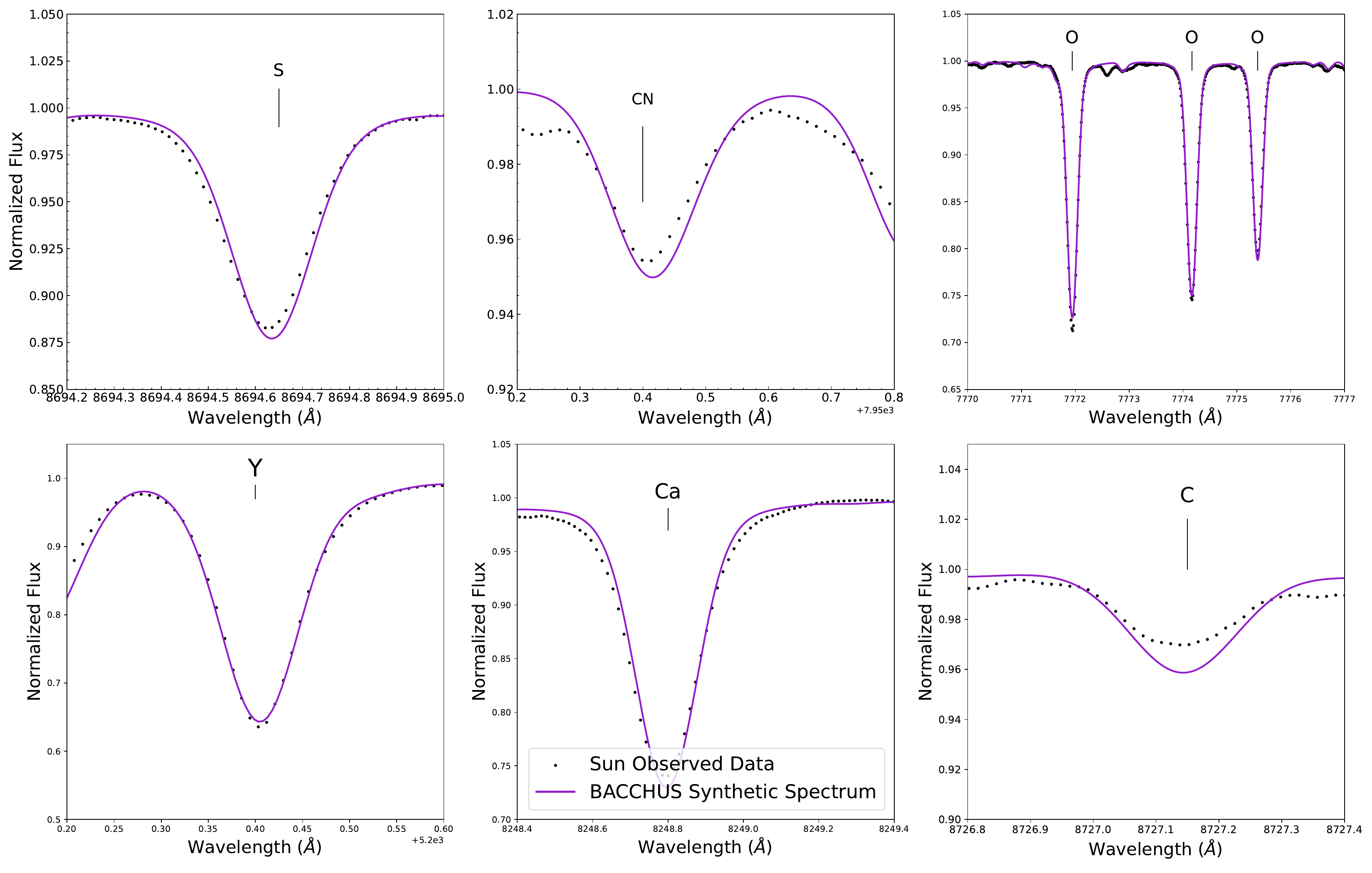}
    \caption{\bacchus~fits to the Sun as observed from PEPSI. We show the fits to key features, sulfur, nitrogen, oxygen, yttrium, calcium and carbon. 
    % \cm{we are still diagnosing some of these fits so this figure might change a smidge but the concept of the figure will remain!}.
    }
    \label{fig:solar_fits_bacchus}
\end{figure*}

%{\textemdash}-Commenting out until after the referee stage and working on Carbon for SktM-1526 and quant
% \section{Upper Limit Fits with  \bacchus}

% Here we show examples of BACCHUS fits for NLTT 1011, BD+24 4329, and StKM 1-1526 for upper limits on some elements such as yttrium, calcium, oxygen, and carbon as described in \S \ref{sec:upperlimits}.

% \begin{figure*}[h!]
% \centering
%     \includegraphics[width=\linewidth]{figures/Upper_Limit_FITS.pdf}
%     \caption{\bacchus~fits for NLTT 1011, BD+24 4329, StKM 1-1526 for upper limits for yttrium, calcium, oxygen, and carbon. }
%     \label{fig:upper_limit_fits}
% \end{figure*}

\clearpage
%{\textemdash}Oxygen Corrections
\section{Oxygen NLTE Corrections}

We provide \ion{O}{1} corrections used to calculate the NLTE \ion{O}{1} that is used for our C/O calculations.

\begin{table}[h!]
    \label{tab:oxy_correction}
    \caption{Oxygen correction values for sample based on spectroscopic stellar parameters from Table \ref{tab:params}}
    \centering
    \begin{tabular}{l|ccc}
    \hline
        Star & \ion{O}{1} [7771.9] & \ion{O}{1} [7774.2] & \ion{O}{1} [7775.4] \\ \hline
        HD 89744  & $-$0.326 & $-$0.301 & $-$0.242 \\ \hline
        HD 46588 & $-$0.119 & $-$0.102 & $-$0.075 \\ \hline
        $*$ 54 Psc & $-$0.035 & $-$0.031 & $-$0.024 \\ \hline
       V$^{*}$ HN Peg & $-$0.105 & $-$0.092 & $-$0.068 \\ \hline
        HD 126053 & $-$0.044 & $-$0.037 & $-$0.027 \\ \hline
        GJ 417 & $-$0.084 & $-$0.073 & $-$0.055 \\ \hline
        HIP 9269 & $-$0.060 & $-$0.053 & $-$0.040 \\ \hline
        G171-58  & $-$0.125 & $-$0.111 & $-$0.084 \\ \hline
        HIP 26653 & $-$0.040 & $-$0.034 & $-$0.026 \\ \hline
        HD 106888 & $-$0.090 & $-$0.079 & $-$0.059 \\ \hline
        HD 51400 & $-$0.040 & $-$0.034 & $-$0.026 \\ \hline
        HD 203030 & $-$0.042 & $-$0.036 & $-$0.027 \\ \hline
        HD 8291 & $-$0.047 & $-$0.040 & $-$0.030 \\ \hline
        HD 16270 & $-$0.020 & $-$0.018 & $-$0.014 \\ \hline
        HD 116012 & $-$0.015 & $-$0.013 & $-$0.010 \\ \hline
        BD+21 55  & $-$0.014 & $-$0.012 & $-$0.009 \\ \hline
       BD+60 1417 & $-$0.014 & $-$0.012 & $-$0.009 \\ \hline
        StKM 2-1777 & $-$0.020 & $-$0.017 & $-$0.013 \\ \hline
        HD 253662 & $-$0.040 & $-$0.037 & $-$0.029 \\ \hline
        CD-24 407 & $-$0.021 & $-$0.018 & $-$0.014 \\ \hline
        BD+13 2269 & $-$0.030 & $-$0.026 & $-$0.020 \\ \hline
        BD+01 299 & $-$0.011 & $-$0.010 & $-$0.008 \\ \hline
        StKM 1-1526 & $-$0.032 & $-$0.029 & $-$0.022 \\ \hline
        TYC 5213-545-1 & $-$0.010 & $-$0.009 & $-$0.007 \\ \hline
        LP 617-58 & $-$0.019 & $-$0.017 & $-$0.013 \\ \hline
        LSPM J0632+5053 & $-$0.035 & $-$0.031 & $-$0.024 \\ \hline
        NLTT 1011 & $-$0.018 & $-$0.016 & $-$0.012 \\ \hline
        BD+49 2561 & $-$0.013 & $-$0.011 & $-$0.009 \\ \hline
        HIP 63506 & $-$0.016 & $-$0.014 & $-$0.011 \\ \hline
        HD 125141 & $-$0.061 & $-$0.052 & $-$0.039 \\ \hline
        BD+24 4329 & $-$0.014 & $-$0.012 & $-$0.009 \\ \hline
    \end{tabular}
\end{table}

%% For this sample we use BibTeX plus aasjournalv7.bst to generate the
%% the bibliography. The sample7.bib file was populated from ADS. To
%% get the citations to show in the compiled file do the following:
%%
%% pdflatex sample7.tex
%% bibtext sample7
%% pdflatex sample7.tex
%% pdflatex sample7.tex

%{\textemdash}Acknowledgments

\clearpage

\bibliography{sample701}{}
\bibliographystyle{aasjournalv7}

%% This command is needed to show the entire author+affiliation list when
%% the collaboration and author truncation commands are used.  It has to
%% go at the end of the manuscript.
%\allauthors

%% Include this line if you are using the \added, \replaced, \deleted
%% commands to see a summary list of all changes at the end of the article.
%\listofchanges

\end{document}